\documentclass[9pt,twocolumn,twoside]{neutral}
\usepackage{xcolor}
\usetikzlibrary{calc, arrows.meta}
\newif\ifnotes\notesfalse

\makeatletter

\newcommand{\listofnotes}{\ifnotes\section*{Pending Notes}\@starttoc{not}\fi}
\makeatother
\usepackage{soul}      %
\usepackage{cleveref}  %
\crefname{figure}{Fig.}{Figs.}\crefname{equation}{Eq.}{Eqs.}

\usepackage{glossaries-extra}
\setabbreviationstyle[acronym]{long-short}
\glssetcategoryattribute{acronym}{nohyperfirst}{true}

\glsdisablehyper

\newacronym{cmip6}{CMIP6}{Coupled Model Intercomparison Project phase 6}
\newacronym{v}{V}{internal variability}
\newacronym{m}{M}{model uncertainty}
\newacronym{s}{S}{scenario uncertainty}
\newacronym{era5}{ERA5}{ECMWF Reanalysis v5}
\newacronym{esm}{ESM}{Earth System Model}
\newacronym{crps}{CRPS}{continuous ranked probability score}
\newacronym{rmse}{RMSE}{root mean squared error}
\newacronym{etccdi}{ETCCDI}{Expert Team on Climate Change Detection and Indices}
\newacronym{kde}{KDE}{kernel density estimation}
\newacronym{qq}{Q-Q}{quantile-quantile}
\newacronym{ri}{RI}{reliability index}
\newacronym{co2e}{CO$_2$e}{carbon dioxide equivalent}
\newacronym{ssp}{SSP}{Shared Socioeconomic Pathway}
\newacronym{ssp245}{SSP2-4.5}{SSP2-4.5}
\newacronym{ssp370}{SSP3-7.0}{SSP3-7.0}
\newacronym{ssp585}{SSP5-8.5}{SSP5-8.5}
\newacronym{245}{2-4.5}{2-4.5}
\newacronym{370}{3-7.0}{3-7.0}
\newacronym{585}{5-8.5}{5-8.5}
\newacronym{loocv}{LOOCV}{leave-one-out cross-validation}
\newacronym{ipcc}{IPCC}{Intergovernmental Panel on Climate Change}
\newacronym{ar6-wg1}{AR6 WGI}{Sixth Assessment Report of the Working Group I}
\newacronym{mvt}{MVT}{mean--variance--trend}
\newacronym{pp}{pp}{percentage points}
\newacronym{edm}{EDM}{elucidated diffusion model}
\newacronym{kl}{KL}{Kullback--Leibler}
\newacronym{tv}{TV}{total variation}
\newacronym{qdm}{QDM}{quantile delta mapping}
\newacronym{ecs}{ECS}{equilibrium climate sensitivity}
\newacronym{pca}{PCA}{principal component analysis}
\newacronym{eemd}{EEMD}{ensemble empirical mode decomposition}
\newacronym{ode}{ODE}{ordinary differential equation}

\newcommand{\degC}{^{\circ}\mathrm{C}}
\newcommand{\tgt}{\operatorname{tgt}}
\newcommand{\src}{\operatorname{src}}

\newcommand{\EE}{\mathbb{E}}
\newcommand{\KL}{\mathrm{KL}}
\newcommand{\TV}{\mathrm{TV}}
\newcommand{\Acc}{\mathrm{Acc}}

\usepackage[figuresright]{rotating}
\newcommand{\SIFIG}{figures/supplementary-information}
\begin{document}

\newcommand{\papertitle}{Guiding future global temperature distributions with simulations and observations}
 \title{\papertitle}

\author[a,b,1,2]{Francesco Immorlano}
\author[a,1]{Elijah Tavares}
\author[a]{Felix Draxler}
\author[b,c]{Juan Nathaniel}
\author[d]{Violette Launeau}
\author[b,e]{Duncan Watson-Parris}
\author[a]{Padhraic Smyth}
\author[b,c,d]{Pierre Gentine}
\author[a,b,2]{Stephan Mandt}

\affil[a]{Department of Computer Science, University of California Irvine, Irvine, CA 92697}
\affil[b]{Learning the Earth with AI and Physics, New York, NY 10027}
\affil[c]{Department of Earth and Environmental Engineering, Columbia University, New York, NY 10027}
\affil[d]{Celest Science, Paris, France}
\affil[e]{Scripps Institution of Oceanography, University of California San Diego, San Diego, CA 92093}

\significancestatement{Climate risks depend not only on extreme temperatures at individual locations, but also on their simultaneous occurrence across large regions. Estimating such risks requires large ensembles of daily temperature fields, yet climate simulations are expensive and biased. Observations can help correct these biases, but relying on them too strongly risks suppressing information about future warming from climate models. We develop a generative framework that combines simulated and observed temperature distributions, using withheld simulated futures to select their relative influence. We find that the selected observational influence decreases as projections extend into the future. The resulting ensembles enable analysis of rare, geographically widespread heat extremes without additional climate simulations.
}

\authorcontributions{Author contributions: F.I., E.T., F.D., and S.M. designed research; F.I., E.T., and J.N. performed research; F.I., E.T., F.D., J.N., V.L., and S.M. analyzed data; F.I., E.T., and S.M. wrote the paper; and F.D., J.N., V.L., P.G., D.W.-P., and P.S. reviewed and edited the paper.}
\authordeclaration{The authors declare no competing interest.}
\equalauthors{\textsuperscript{1}F.I. contributed equally to this work with E.T.}
\correspondingauthor{\textsuperscript{2}To whom correspondence may be addressed. E-mail: \mbox{fimmorla@uci.edu} or \mbox{mandt@uci.edu}}


\begin{abstract}
Assessing climate risks requires projections of daily temperature distributions, including rare and spatially widespread extremes. Climate simulations are costly and biased, while correcting them toward observations risks suppressing inter-model differences. We train a single machine-learning emulator, a conditional diffusion model, on seventeen CMIP6 models and ERA5 reanalysis to generate spatially coherent daily temperature fields conditioned on model identity, date, and CO$_2$-equivalent concentration. At sampling time, diffusion guidance reduces model biases by steering each emulated climate-model distribution toward observations, with its strength selected through perfect-model experiments using withheld simulated futures. Observational influence decreases with projection horizon and forcing. Across seventeen synthetic targets, guidance more than halves median biases in the 1st and 99th percentiles of daily temperature, improves probabilistic scores and reliability, and reduces median annual global-mean temperature RMSE by $32$--$57\%$ across three emission scenarios. In a detailed benchmark, guidance also outperforms established bias corrections on daily-field probabilistic scores. Efficient sampling generates large ensembles without further climate simulations. For example, our framework corrects CMIP6's roughly threefold historical overestimate of extreme-heat coverage, yet projects that concurrent extremes will affect nearly half of global land on a typical day in the 2080s under SSP3-7.0.
\end{abstract}

\maketitle
\thispagestyle{firststyle}
\ifthenelse{\boolean{shortarticle}}{\ifthenelse{\boolean{singlecolumn}}{\abscontentformatted}{\abscontent}}{}

\Firstpage

Climate models provide the foundation for understanding and adapting to a changing climate \citep{ipcc-2021-wg1-ar6}. Yet they differ in their representation of small-scale processes (e.g., clouds or convection) that strongly affect the response of the planet to greenhouse gases, leading to systematic biases relative to observations \citep{eyring-2016-cmip6}. Even models that reproduce historical climate relatively well can project markedly different future warming, making inter-model differences a major source of long-term uncertainty \citep{hawkins-sutton-2009-the-potential,lehner-2020-partitioning,zelinka-2020-causes}. Correcting model outputs against observations can mitigate these biases, but overcorrection risks erasing the spread that reflects genuine structural uncertainty and muting internal climate variability. Two questions follow: how strongly should observations constrain climate projections, and how can that choice be validated without future observations?

These questions extend beyond mean warming to full distributions of variables such as daily temperature. For example, the probability of local hot days depends on the tails of the distribution, whereas the total area affected is determined by concurrent heat extremes across regions \citep{seneviratne-2021-weather}. Accurately estimating these distributions, especially for rare events, requires extensive sampling \citep{mahesh-2025-huge-ensembles,lancelin-2026-ai-boosted}, but climate simulation archives provide few realizations for each model and forcing scenario. Although large ensembles can increase these sample sizes \citep{deser-2020-insights,rodgers-2021-cesm-lens2}, each additional realization requires another computationally expensive simulation \citep{mankin-2020-value}. Generative emulators \citep{tebaldi-2025-emulators} instead learn distributions from existing simulations and efficiently sample new fields \Parasplit without further integrations \citep{bassetti-2024-diffesm,ruhling-cachay-2024-dyffusion,bouabid-2026-score,clyne-2026-forced,brenowitz-2025-cbottle}. Their larger ensembles reduce sampling error but inherit the biases of the training simulations.

Existing methods use historical observations to weight climate models \citep{brunner-2020-reduced} or reduce biases in their output \citep{cannon-2015-qdm,xu-2021-mvt,pan-2021-learning,hess-2022-physically,mcgibbon-2024-global,aich-2026-conditional}. For a generative emulator, such adjustments operate directly on probability distributions, raising the question of how strongly generated fields should be steered toward observations. Optimizing a distribution to match historical records does not imply accurate projections of unobserved future conditions, because only physics-based models simulate the climate response to future forcing. Steering generated fields toward observations therefore requires explicit control over the relative influence of climate models and observations, and an empirical framework to evaluate this balance against synthetic target futures.

\begin{figure*}[t!]
\centering
\scalebox{0.9656}{%
\graphicspath{{figures/main-text/fig1_assets/}{figures/fig1_assets/}{figures/}{fig1_assets/}{./}}%
\definecolor{ink}{HTML}{2C3E50}%
\sffamily%
\begin{tikzpicture}[x=1cm, y=1cm, font=\footnotesize]
\node[anchor=north west, inner sep=0pt, outer sep=0pt] (strip) at (0, 0)
  {\includegraphics[width=17.6cm]{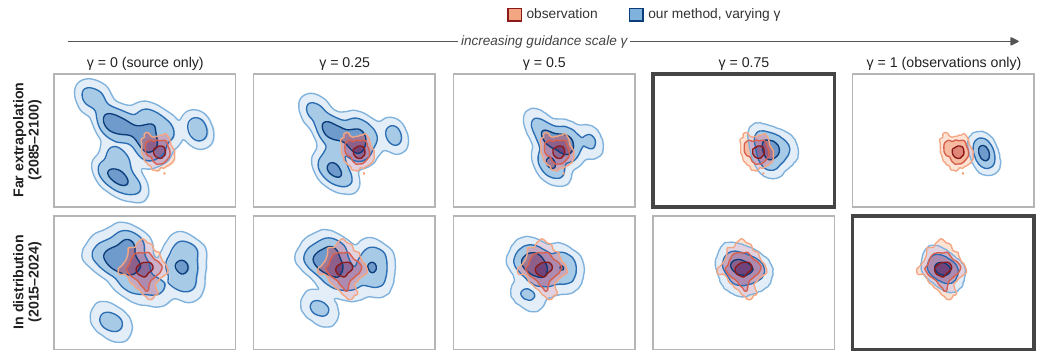}};
\node[anchor=west, ink, inner sep=0pt, font=\footnotesize] at (0, -0.25)
  {{\small\bfseries A.}\quad The right guidance scale is not necessarily pure observation emulation};
\end{tikzpicture}%
}
\captionsetup{skip=6pt}
\caption{\textbf{Guidance balances observational fidelity and model-based extrapolation.} Daily temperature fields from a synthetic observation (MPI-ESM1-2-LR, withheld from training after 2024; red) and from the diffusion emulator conditioned on the remaining sixteen \glsentryshort{cmip6} models (blue), projected onto the first two principal components. Columns increase the guidance scale $\gamma$ from source-model emulation ($\gamma=0$) to target-only emulation ($\gamma=1$). Within the observational period (2015--2024; bottom), target-only emulation best reproduces the target distribution. Far beyond the observed period (2085--2100; top), intermediate guidance better overlaps the target, indicating that extrapolation benefits from retaining information from the physics-based model ensemble. Bold frames indicate the guidance scale closest to the selected value for each period.}
\label{fig:schematic}
\end{figure*}

To address these challenges, we develop a generative modeling framework, starting with a single emulator trained on daily temperature data. The data comprise simulations from seventeen \gls{cmip6} models \citep{eyring-2016-cmip6} under three \gls{ssp} scenarios \citep{oneill-2016-scenariomip} and the \gls{era5} \citep{hersbach-2020-era5} through 2024. The emulator, a generative diffusion model \citep{karras-2022-elucidating}, learns the distribution of global daily near-surface temperature fields conditioned on model identity, calendar day and year, and \gls{co2e} concentration (Materials and Methods). Modeling individual days rather than sequences suffices for the risk measures above, which depend only on the distribution of daily fields. 

We denote by $p_i(\mathbf{x})$ the learned distribution of daily fields $\mathbf{x}$ for climate model $i$, and by $p_{\mathrm{tgt}}(\mathbf{x})$ that of the \gls{era5} target. We align each $p_i$ with $p_{\mathrm{tgt}}$ by sampling their geometric mean weighted by $\gamma$. Unlike an additive mixture, which would draw each field from one distribution or the other, this product favors fields likely under both:
\begin{align*}
    p_{\gamma,i}(\mathbf{x}) \propto p_i(\mathbf{x})^{1-\gamma}\, p_{\mathrm{tgt}}(\mathbf{x})^{\gamma}. \numberthis \label{eq:geomean}
\end{align*}
This \emph{condition-to-condition guidance} builds on guidance methods for diffusion models \citep{dhariwal-nichol-2021-diffusion-beats-gans,ho-2021-classifier-free,zhong-2025-domain-guidance} and samples \cref{eq:geomean} by following the weighted sum of the two learned scores (gradients of the log-densities), $(1-\gamma)\nabla_{\mathbf{x}}\log p_i + \gamma\nabla_{\mathbf{x}}\log p_{\mathrm{tgt}}$ (Materials and Methods). The guidance scale $\gamma \in [0,1]$, set at inference time without further training, moves $p_{\gamma,i}$ from the climate model ($\gamma=0$) to the observations ($\gamma=1$; \cref{fig:schematic}). Guiding each of the seventeen models produces an ensemble that we use to evaluate how guidance strength affects both the daily temperature distribution and inter-model spread.

We evaluate this framework through perfect-model experiments, or \gls{loocv}, treating each of the seventeen \gls{cmip6} models as a synthetic observational reference whose future, unlike that of \gls{era5}, is known. In each experiment, we train the emulator jointly on the target's simulation up to 2024, except 2001--2010, and full simulations of the remaining sixteen models. We use the target's withheld 2025--2100 future to quantify projection errors and determine the optimal guidance strength. In the experiment with MIROC6 as the target, we benchmark against three methods: the bias corrections \gls{mvt} \citep{xu-2021-mvt} and \gls{qdm} \citep{cannon-2015-qdm}, and classifier guidance \citep{dhariwal-nichol-2021-diffusion-beats-gans}.

Perfect-model experiments show that the selected observational influence decreases with projection horizon and forcing. Guidance lowers 1st- and 99th-percentile biases, averaged over the three scenarios, for all seventeen synthetic targets, outperforms established bias corrections for MIROC6 on most metrics beyond their calibration period, and reduces the overestimated historical extent of concurrent heat extremes. Our framework shows how generative climate emulation and observations can be combined in a statistically principled manner.

\section*{Results}
In perfect-model experiments, where a held-out \gls{cmip6} model serves as a synthetic observation with a known future, we fit how strongly guidance should weight the observations, evaluate the resulting schedule on all seventeen targets, and benchmark it against established bias corrections. We then turn to \gls{era5}: we verify the emulator on a withheld observed decade, project global warming with the schedule, and analyze local and regional heat extremes.

We first determine how strongly guidance should weight the observations. No observations of the future exist to tune this weight, so we tune it in perfect-model experiments, whose withheld futures are known. In \cref{fig:schematic}, target-only emulation ($\gamma=1$) best replicates the synthetic target in 2015--2024, while intermediate guidance better captures its withheld 2085--2100 distribution. The same pattern emerges in all seventeen perfect-model experiments when we select the guidance scale that minimizes 1st- and 99th-percentile daily temperature biases. As projections extend farther beyond the historical record, the fitted schedule shifts weight away from the target, decreasing from $\gamma=0.94$ in 2025 to $0.84$, $0.80$, and $0.77$ in 2100 under \glsentryshort{ssp245}, \glsentryshort{ssp370}, and \glsentryshort{ssp585} (Materials and Methods; \emph{Supporting Information}, Fig.~\ref{fig:si-gamma-schedule}). The guided projections stay anchored to observations near the present and rely increasingly on simulated dynamics as forcing and lead times grow.

We next ask whether the fitted schedule, selected on daily temperature tails, also improves the projected warming trajectory of each withheld target. For each target and scenario, we calculate the \gls{rmse} between the target's annual global-mean trajectory and the ensemble mean over 2025--2100 (\emph{Supporting Information}, \emph{Evaluation Metrics}). Across scenarios, median \gls{rmse} drops from $0.47$--$0.56\,\degC$ unguided to $0.24$--$0.38\,\degC$ guided. Averaged over the three scenarios, guidance also reduces area-weighted 1st- and 99th-percentile daily temperature biases for every target model and improves broader distributional metrics (\emph{Supporting Information}, Tables~\ref{tab:loocv-skill} and \ref{tab:loocv-metrics} and Figs.~\ref{fig:si-folds-mean-370}--\ref{fig:si-folds-rank-370}). These scores are in-sample, but the schedule has only three parameters.

\begin{figure}[t!]  
\centering
\includegraphics[width=\columnwidth]{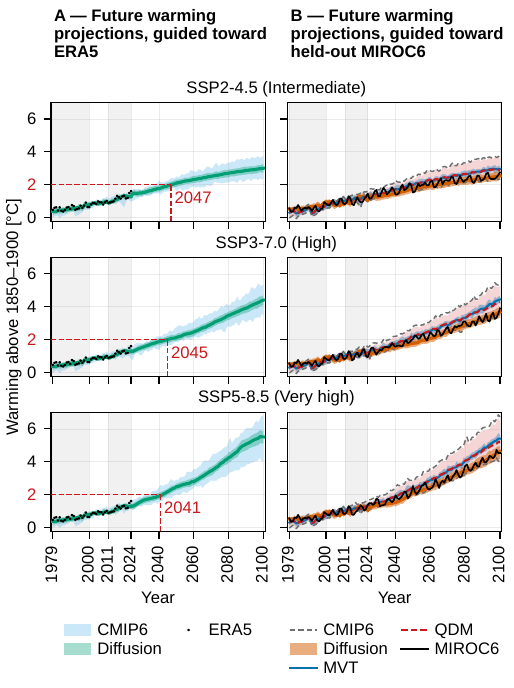}
\caption{\textbf{Guided projections constrain future warming and recover a known synthetic future.} Annual global-mean warming above 1850--1900 under \glsentryshort{ssp245}, \glsentryshort{ssp370}, and \glsentryshort{ssp585}; shaded bands show the 5--95\% range of each ensemble. (\emph{A}) Projections guided toward \glsentryshort{era5}: the emulator reproduces the observational record through 2024 and yields a narrower range of future warming than raw \glsentryshort{cmip6} while retaining inter-model spread. (\emph{B}) Corresponding perfect-model experiment with MIROC6 as a synthetic observational target, whose future, withheld after 2024, serves as ground truth: guided projections follow the MIROC6 trajectory more closely than raw \glsentryshort{cmip6} and the \glsentryshort{mvt} and \glsentryshort{qdm} bias corrections. Gray shading marks the training period.}
\label{fig:projections}
\end{figure}

Statistical bias corrections are the established alternative, but they are calibrated on the historical period and applied unchanged to the future. The MIROC6 experiment (\cref{fig:projections}\emph{B}) tests both approaches beyond this calibration period. Guidance reduces the error of the raw \gls{cmip6} trajectory by $41$--$62\%$, whereas \gls{mvt} \citep{xu-2021-mvt} and \gls{qdm} \citep{cannon-2015-qdm} reduce it by $8$--$16\%$ (\emph{Supporting Information}, Table~\ref{tab:loocv-miroc6-baselines}). Classifier guidance (run on a separate emulator trained without MIROC6) underperforms our approach in trajectory error and all daily-field metrics. Over 2081--2100, guided fields have the most uniform rank histograms and lower mean absolute bias than either correction, while \gls{mvt} better captures local standard deviation and skewness and, applied to the emulator output, has a lower 99th-percentile bias (\emph{Supporting Information}, Table~\ref{tab:loocv-miroc6-baselines} and Figs.~\ref{fig:si-m6-maps-raw-370}--\ref{fig:si-m6-cycle-diff}).

Projections must also convey their uncertainty, and here our ensembles remain somewhat under-dispersed: the guided 5--95\% warming interval covers the withheld MIROC6 trajectory in $62$--$72\%$ of years (nominal $90\%$; \emph{Supporting Information}, Table~\ref{tab:loocv-miroc6-baselines}). For a second target, MPI-ESM1-2-LR, the guided ensemble underestimates warming after mid-century (\emph{Supporting Information}, \emph{Held-Out MPI-ESM1-2-LR}). In these experiments, guidance refines forced responses and distributions but leaves residual under-dispersion and target-specific biases in long-term warming.

Before projecting from \gls{era5}, we check the emulator against real observations on the decade withheld from training, where the corrections can be scored as well. Over 2001--2010, our emulator replicating \gls{era5} ($\gamma=1$) leaves a spatially smoother residual bias than \gls{mvt} and \gls{qdm} and better preserves the observed temperature differences between neighboring cells, which these per-cell corrections distort, although they better match marginal distribution tails and day-to-day variability (\emph{Supporting Information}, Tables~\ref{tab:baselines} and \ref{tab:spatial-consistency} and Figs.~\ref{fig:si-baseline-maps}--\ref{fig:si-clouds-eur}). Beyond 2024, we apply the fitted schedule. Guidance narrows the range of projected global warming while preserving structural differences among the seventeen \gls{cmip6} models (\cref{fig:projections}\emph{A}). Over 2081--2100, the guided ensemble retains $20.5$--$22.9\%$ of the raw \gls{cmip6} 5--95\% inter-model range, across scenarios (\emph{Supporting Information}, Table~\ref{tab:retention}). The guided ensemble mean first exceeds $1.5\,\degC$ in 2028--2029 and $2\,\degC$ in 2041--2047 across \glspl{ssp} (\emph{Supporting Information}, Table~\ref{tab:gwl-crossings}).

\begin{figure}[t!]
\centering
\includegraphics[width=\columnwidth]{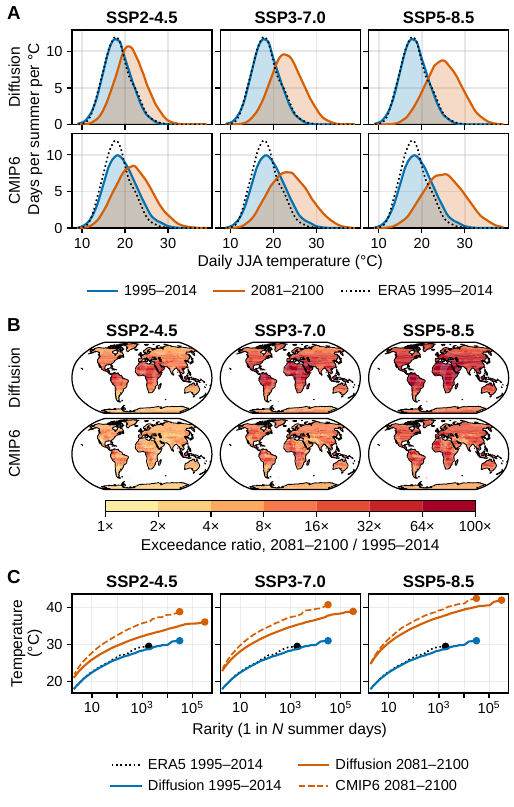}
\caption{\textbf{Spatially resolved analysis of extreme heat.} Daily summer temperature statistics for 1995--2014 and 2081--2100 under \glsentryshort{ssp245}, \glsentryshort{ssp370}, and \glsentryshort{ssp585}. (\emph{A}) Distribution of June--August daily mean temperature at the Paris grid cell for the guided diffusion emulator and raw \glsentryshort{cmip6}, with \glsentryshort{era5} as the historical reference. (\emph{B}) Ratio of the probability of exceeding each ensemble's local 1995--2014 99th-percentile summer temperature in 2081--2100 versus 1995--2014. Values greater than one indicate more frequent future exceedance of historical extremes. (\emph{C}) Paris temperature exceeded on average once in $N$ summer days; dots mark the hottest day of each curve, whose extent reflects the number of available samples.}
\label{fig:distributions}

\end{figure}

Our emulator's computational efficiency overcomes the sample-size limitations of standard climate archives: ten daily samples for each guided model produce $312{,}800$ summer days over 2081--2100, against $31{,}280$ in the \gls{cmip6} archive analyzed here. This ten-fold increase resolves rare tail events that raw archives undersample. Observational constraints also reshape the local distributions behind extreme-heat probabilities. At Paris, the emulator replicating \gls{era5} over 1995--2014 (half in the training set) reduces most of the raw \gls{cmip6} historical bias, particularly in the warm tail (\cref{fig:distributions}\emph{A}). Guidance and raw \gls{cmip6} identify broadly similar land regions of strongest growth in summer extremes, but guided projections show greater relative increases in exceedance frequency of local historical 99th-percentile thresholds across most land areas, a divergence that grows with forcing (\cref{fig:distributions}\emph{B}). However, because guidance corrects the historical warm bias, late-century guided extremes remain lower in absolute temperature than raw \gls{cmip6} at equal return periods (\cref{fig:distributions}\emph{C}). Risk estimates that depend on historical thresholds therefore change once the historical bias is removed.

\begin{figure*}[t!]
\centering
\includegraphics[width=\textwidth]{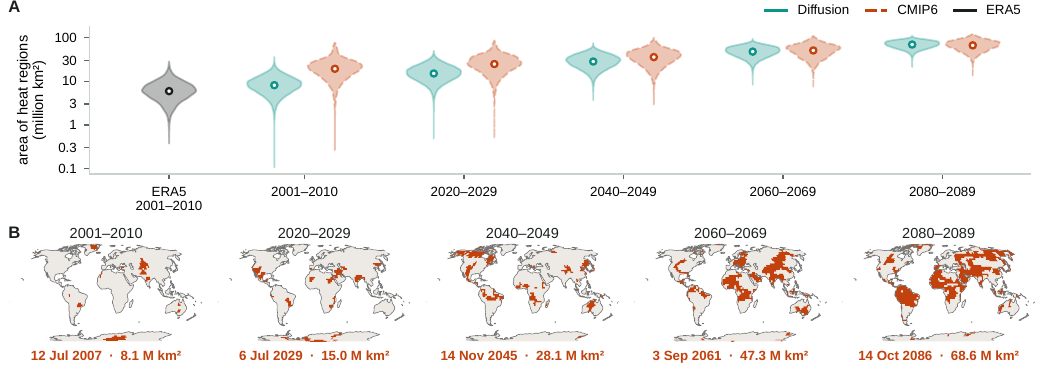}
\caption{\textbf{Extreme heat affects an increasing share of global land.} Total land area contained in contiguous regions of extreme heat under \glsentryshort{ssp370}. A land grid cell is extreme when its daily mean temperature exceeds the \glsentryshort{era5} 1979--2024 95th percentile for that day of year; the same thresholds are applied to all datasets, and only regions of at least two adjacent cells are counted. (\emph{A}) Distribution of daily affected area for \glsentryshort{era5}, the guided diffusion ensemble, and raw \glsentryshort{cmip6}; circles mark the medians. In the 2001--2010 validation decade, the emulator, which replicates \glsentryshort{era5} before 2025, substantially reduces the \glsentryshort{cmip6} overestimate of observed extreme-heat area; the projected distribution then shifts toward much larger affected areas through the century. (\emph{B}) Representative guided fields for each decade, chosen to have total affected area closest to the median in panel \emph{A}, each labeled with its date and total area.}
\label{fig:heat-region}
\end{figure*}

Finally, concurrent heat extremes across regions determine the total area affected, and measuring them requires spatially complete daily fields, which our emulator provides in large numbers. Against local day-of-year 95th-percentile \gls{era5} thresholds, raw \gls{cmip6} models overestimate the extreme-heat land area in the withheld 2001--2010 decade, with a median extent roughly three times the observed value. The emulator replicating \gls{era5} instead aligns closely with observations, with a median coverage of $8.1$ million km$^2$ ($5.5\%$ of global land; \cref{fig:heat-region}\emph{B}). Correcting this historical bias does not suppress the projected response to warming. Under \glsentryshort{ssp370}, the median affected area grows steadily each decade to $68.6$ million km$^2$, or $46.9\%$ of global land, in the 2080s, converging near the raw \gls{cmip6} baseline (\cref{fig:heat-region}\emph{A} and \emph{B}).

\section*{Discussion}
Our experiments show that a single conditional diffusion emulator can expand the scope of statistical questions addressed using a fixed archive of climate simulations and observational data. Its additional, spatially coherent daily fields enable the analysis of quantities often difficult to estimate from standard ensembles, particularly simultaneous regional extremes and sparsely sampled distribution tails. Initial-condition large ensembles offer comparable sample sizes but remain computationally prohibitive for most climate models \citep{deser-2020-insights,rodgers-2021-cesm-lens2}. Our emulator bypasses this bottleneck by sampling from the joint distribution learned across existing climate archives, so custom analyses need no new simulations, provided the training data span the target climate variability and the emulator learns those distributions. 

Condition-to-condition guidance empirically extrapolates historical climate distributions into the future. Reanalyses capture historical variability but contain no signals of future forcing. \gls{cmip6} simulations cover the projection period but carry systematic biases relative to observations. Our perfect-model experiments test this trade-off by replacing the unknown future of \gls{era5} with held-out model projections. As lead time and forcing grow, the selected scale shifts weight toward the climate models. These scales ($\gamma < 1$) and the reduced trajectory and distributional errors across held-out models confirm that combining observations with model dynamics outperforms either alone. Nonetheless, tuning guidance strength is distinct from fully calibrating the predictive distribution. Daily diagnostics and rank histograms indicate improved distributional agreement, and annual uncertainty bands contain the withheld MIROC6 trajectory in most years, but coverage remains below the nominal $90\%$. We view guidance as a practical framework for adjusting projections to observational reference data. Fully correcting the remaining distributional errors would require test-time calibration algorithms, which trade computational cost for statistical accuracy \citep{geyfman-2026-calibrated-guidance}.

Our extreme-heat analyses illustrate how guidance separates historical bias from projected warming trends. In Paris, guiding toward \gls{era5} removes the historical warm-tail bias of raw \gls{cmip6}, shifting the local extreme-heat threshold. The guided ensemble therefore projects a larger relative increase in exceedance frequency, yet lower absolute late-century temperatures than raw \gls{cmip6}. This distinction matters whenever future changes are evaluated against model-specific historical thresholds. The extreme-heat area analysis illustrates the same mechanism spatially: guidance reduces the historical \gls{cmip6} bias without suppressing projected warming. Beyond these applications, our guided ensembles can support downstream evaluations, including population exposure under changing heat thresholds, spatially resolved return levels, or regional climate responses at fixed global warming levels.

We acknowledge several limitations. Additional samples reduce Monte Carlo sampling error within the learned distribution but cannot correct errors in the emulator, guidance scheme, or simulations used for training. Because our perfect-model experiments evaluate out-of-model generalization only within the \gls{cmip6} archive, they cannot reveal biases shared across models or confirm calibration against observations. Because our emulator draws days independently, it cannot directly estimate multi-day heatwave duration, event sequencing, or temporally compound risks. Our analysis is limited to near-surface temperature at moderate spatial resolution; within this scope, guided generative emulation resolves long-term forced warming, daily extreme quantiles, and spatial heat coverage across multi-model archives. Extending this framework to temporally continuous, multivariate, and higher-resolution fields would enable broader impact applications, though with higher computational and validation demands.

These limitations suggest future directions. Additional computing power would enable calibration-aware guidance, refined guidance schedules, and more evaluations of tail statistics and spatial dependencies. Because the guidance scale is set at inference time, a single pre-trained emulator can adapt to diverse user requirements, allowing region- and time-specific scales that can be tested in our perfect-model framework. We introduced this methodology to complement physics-based climate modeling: dynamical simulations provide forced responses, reanalyses capture historical variability, and our emulator combines both into a probabilistic ensemble. Climate model archives can now function as continuous, observationally based probability distributions, enabling direct queries for spatial patterns, extreme quantiles, and joint risk statistics.

\matmethods{

    \subsection*{Data}\label{methods:data}
    We use daily near-surface temperature fields from seventeen \gls{cmip6} models (\emph{Supporting Information}, Table~\ref{tab:cmip6-models}) over 1979--2100 under \glspl{ssp} \glsentryshort{245}, \glsentryshort{370}, and \glsentryshort{585}, and the \gls{era5} reanalysis~\citep{hersbach-2020-era5} over 1979--2024 as observational reference. We conservatively remap all fields~\citep{taylor-2024-conservative-remapping} to a $64 \times 128$ grid (${\sim}2.8^{\circ}$) and pair each with the annual global-mean \gls{co2e} concentration~\citep{tsutsui-2022-rcm} of its year and scenario (\glsentryshort{ssp245} for \gls{era5}). We train on \gls{era5} and withhold the decade 2001--2010 for validation (\emph{Supporting Information}, \emph{Data Split}).
    
    \subsection*{Conditional Diffusion Emulator}\label{methods:diffusion}
    We emulate the seventeen \gls{cmip6} models and \gls{era5} with a single conditional diffusion model. This network learns the distribution $p_\theta(\mathbf{x} \mid m, d, c, y)$ of a daily temperature field $\mathbf{x}$, conditioned jointly on the model identity $m \in \{\mathrm{CMIP6}_1, \ldots, \mathrm{CMIP6}_{17}, \mathrm{ERA5}\}$, the day of year $d$, the annual global-mean \gls{co2e} concentration $c$, and the calendar year $y$. We train on the three \glspl{ssp} jointly and without a scenario index, so the emulator sees the forcing pathway only through $c$ and $y$. Because the scenarios reach a given \gls{co2e} concentration in different years, $c$ alone does not identify this pathway, and the climate response depends on the pathway through the lagged ocean response, not only on the current concentration. Conditioning also on $y$ resolves this ambiguity.

    We train a UNet denoiser with \gls{edm} preconditioning and loss~\citep{karras-2022-elucidating} on every daily field of the training data and generate fields with its deterministic Heun sampler (\emph{Supporting Information}, \emph{Machine Learning Setup Details}). Since a single network represents all eighteen conditional distributions, the score of any source model and that of \gls{era5} are available from the same denoiser at every sampling step.

    \subsection*{Condition-to-Condition Guidance}\label{methods:ctc-guidance}
    We introduce condition-to-condition guidance, a central contribution of this work, which merges the emulated distributions of a climate model and of the observations at sampling time, without retraining. For a given day, \gls{co2e} concentration and year, let $p_i(\mathbf{x}) = p_\theta(\mathbf{x} \mid \mathrm{CMIP6}_i, d, c, y)$ be the emulated distribution of climate model $i$, the source, and $p_{\mathrm{tgt}}(\mathbf{x}) = p_\theta(\mathbf{x} \mid \mathrm{ERA5}, d, c, y)$ that of the observations, the target. Guidance from source $i$ toward the \gls{era5} target samples their weighted geometric mean $p_{\gamma,i}$ of \cref{eq:geomean}, where the guidance scale $\gamma \in [0,1]$ sets the weight of the target. \Cref{eq:geomean} assigns high probability only to fields that are likely under both the climate model and the observations. The mixture $(1-\gamma)\,p_i + \gamma\,p_{\mathrm{tgt}}$, in contrast, would return fields from one or the other (\emph{Supporting Information}, \emph{Interpolating between Two Distributions}). Finally, we pool the guided distributions with equal weight,
    \begin{align*}
        p_{\gamma}(\mathbf{x}) = \frac{1}{|\mathcal{I}|} \sum_{i \in \mathcal{I}} p_{\gamma,i}(\mathbf{x}), \numberthis \label{eq:ensemble}
    \end{align*}

    where $\mathcal{I}$ contains the seventeen \gls{cmip6} models or, in a perfect-model test, the sixteen source models guided toward the held-out one. At $\gamma = 0$, \cref{eq:ensemble} is the emulated climate-model ensemble; at $\gamma = 1$, every component becomes $p_{\mathrm{tgt}}$. In between, each component moves toward the target and keeps part of what distinguishes its source model. For Gaussian $p_i$ and $p_{\mathrm{tgt}}$ with means $\boldsymbol{\mu}_i$ and $\boldsymbol{\mu}_{\mathrm{tgt}}$ and a common covariance, $p_{\gamma,i}$ is Gaussian with that covariance and mean $(1-\gamma)\,\boldsymbol{\mu}_i + \gamma\,\boldsymbol{\mu}_{\mathrm{tgt}}$, and the spread of the model means around the target shrinks by a factor $1-\gamma$.

    To draw a single sample from \cref{eq:ensemble}, we fix a source model $i \in \mathcal I$ and perform diffusion sampling by integrating the probability flow ordinary differential equation~\citep{song2021scorebased} from noise to data with the interpolated score function
    \begin{align*}
        \nabla_{\mathbf{x}_t} \log p_{\gamma,i}(\mathbf{x}_t) \approx
           (1-\gamma)  & \nabla_{\mathbf{x}_t}\log p_\theta(\mathbf{x}_t \mid \text{CMIP6}_i, d, c, y) \quad + \\
            \gamma & \nabla_{\mathbf{x}_t}\log p_\theta(\mathbf{x}_t \mid \text{ERA5}, d, c, y), \numberthis \label{eq:guidance}
    \end{align*}
    where $\mathbf{x}_t$ is the partially denoised field at noise level $t$.

    Sampling starts from random noise and removes it step by step, moving the field along the weighted sum of the two scores (i.e., the directions in which the field becomes more likely under the source model and under \gls{era5}). \cref{eq:guidance} approximates the score of \cref{eq:geomean}~\citep{geyfman-2026-calibrated-guidance}, since the geometric mean of two noised distributions is not the noised geometric mean; the approximation is exact in the Gaussian example above and also underlies classifier-free guidance~\citep{ho-2021-classifier-free}. Unlike classifier guidance~\citep{dhariwal-nichol-2021-diffusion-beats-gans}, which requires a separate classifier, condition-to-condition guidance evaluates both scores with the same parameters $\theta$. Domain Guidance~\citep{zhong-2025-domain-guidance} uses the same construction for a single source domain; we generalize it to a pool of sources sharing one target, and the ensemble of \cref{eq:ensemble} keeps their differences.

    \subsection*{Selecting the Guidance Strength}\label{methods:gamma-selection}
    We select $\gamma$ values with the perfect-model experiments. We divide 2025--2100 into 32 consecutive blocks and, for each fold, \gls{ssp} and block, select among nine scales the one whose guided ensemble has the smallest absolute 1st- and 99th-percentile daily temperature bias against the held-out model, averaged over grid cells and both tails (\emph{Supporting Information}, \emph{The Criterion Used to Select the Guidance Scale}). Because the selections vary between folds and blocks, we fit one guidance schedule, linear in calendar year and \gls{co2e} concentration, to all selections pooled over folds and \glspl{ssp}, and clip it to $[0,1]$ (\emph{Supporting Information}, \emph{Fitting the Guidance Schedule}). We apply this schedule to projections toward \gls{era5}, assuming it transfers because these tests replicate the real-world situation of a target known up to 2024 and projected beyond. Before 2025, where the target is known, we do not extrapolate. Our diffusion model is conditioned on the target only ($\gamma = 1$), so it replicates the known record with many realizations. For simplicity, we also refer to this ensemble as guided.

    \subsection*{Annual-Variance Rescaling}\label{methods:variance-rescaling}
    Independent daily sampling makes emulated annual means average away almost all the year-to-year variability that climate models and observations keep through correlated daily anomalies. We therefore fit a cubic polynomial in time to every emulated annual global-mean series (one per sample of each source model) and multiply the residuals by a factor $k \geq 1$, the ratio of the year-to-year standard deviation of the target's annual global mean over 1979--2024 (\gls{era5}, or the held-out model of a perfect-model experiment) to that of the emulator (\emph{Supporting Information}, \emph{Estimating the Rescaling Factor}). We apply one $k$ to all series, years and \glspl{ssp} of an ensemble, assuming stationary interannual variability and a sampler under-dispersion independent of $\gamma$. We rescale the annual global-mean series of every emulated ensemble, guided or not, including classifier guidance.
    
}

\showmatmethods{} %

\dataavail{The datasets used in this study are freely accessible, under their respective licenses, from the following public repositories: \href{https://leap-stc.github.io}{LEAP Pangeo} for \gls{cmip6} simulations; \href{https://cds.climate.copernicus.eu/datasets/reanalysis-era5-single-levels?tab=download}{Climate Data Store} for \gls{era5} reanalysis~\citep{hersbach-2020-era5,hersbach-2023-era5-single-levels}. The code will be made available at \url{https://github.com/mandt-lab/gdso}. The \emph{Supporting Information} figures for the other scenarios and for 2030--2050 will be made available at \url{https://github.com/mandt-lab/gdso-figures}.}

\acknow{We acknowledge funding from National Science Foundation (NSF) through Learning the Earth with Artificial Intelligence and Physics (LEAP) Science and Technology Center (STC) (Award \#2019625). Stephan Mandt acknowledges funding from NSF through an NSF CAREER Award IIS-2047418, IIS-2611549, and funding from the Hasso Plattner Research Center at UCI. Pierre Gentine acknowledges funding from the European Research Council Synergy Grant ``Understanding and modeling the Earth System with Machine Learning'' under the Horizon 2020 research and innovation programme (Grant Agreement No. 855187). Duncan Watson-Parris acknowledges funding from NSF award 2441832. Padhraic Smyth acknowledges funding from NSF award CAIG-2425932.}

\showacknow{} %

\bibsplit[18]

\SIauthors{Francesco Immorlano, Elijah Tavares, Felix Draxler, Juan Nathaniel, Violette Launeau, Duncan Watson-Parris, Padhraic Smyth, Pierre Gentine and Stephan Mandt}
\supplement

\SItext
\section*{Data}\label{sec:si:data} 
This section details the data description in \emph{Materials and Methods}, \emph{Data}.

\paragraph*{Temperature Fields}
We use \gls{cmip6} and \gls{era5}~\cite{si:hersbach-2023-era5-single-levels} records from 1979 onwards, when the transition to the satellite era improved the quality and spatial completeness of global observations~\cite{si:hersbach-2020-era5}. For each \gls{cmip6} model we use one initial-condition member, \emph{r1i1p1f1}, the primary member used in the \gls{ipcc} \gls{ar6-wg1}~\cite{si:ipcc-2021-wg1-ar6}. We drop leap days, so that every year has 365 days. The common grid is the $64 \times 128$ Gaussian grid of CanESM5, the coarsest of the seventeen models used in this work.

\paragraph*{\glsentryshort{co2e} Forcing}
We obtain the annual global-mean \gls{co2e} concentration for 1979--2100 under the three \glspl{ssp} from the Minimal CMIP Emulator v1.2~\citep{si:tsutsui-2022-rcm}, a reduced-complexity model whose effective radiative forcing of greenhouse gases and aerosols we convert to \gls{co2e} by inverting the CO$_2$ forcing expression of Meinshausen et al.~\citep{si:meinshausen-2020-ssp}. This variable encodes the combined greenhouse-gas and aerosol forcing that drives long-term climate response. Annual resolution is sufficient for our purposes because well-mixed greenhouse gases have atmospheric lifetimes longer than several years~\citep{si:ipcc-2021-wg1-ar6}, and the day-of-year condition carries the seasonal cycle of temperature.

\section*{Machine Learning Setup Details}
This section details the emulator of \emph{Materials and Methods}, \emph{Conditional Diffusion Emulator}: training data, the scaling applied, conditioning, denoiser, noise schedule, training, and sampling. We train eighteen such emulators: one for each of the seventeen perfect-model experiments, or \gls{loocv} folds, and one (the \gls{era5} projection model) trained on all seventeen \gls{cmip6} models and \gls{era5} used as the actual observation.

\paragraph*{Data Split}
We jointly train the \gls{era5} projection model on the full \gls{cmip6} simulation record (1979--2100, three \glspl{ssp}), so that it is exposed to the whole range of forced responses across scenarios, and on \gls{era5} (1979--2000 and 2011--2024, reserving 2001--2010 for validation).
The years 2025--2100 are dedicated to projection, where the \gls{era5} endpoint of guidance requires extrapolation. Because the \glspl{ssp} diverge after 2014, we pair the \gls{era5} temperature fields with \gls{co2e} values from the \glsentryshort{ssp245} trajectory, following the \gls{cmip6} convention of extending the historical period with \glsentryshort{ssp245}~\citep{si:gillett-2016-damip,si:oneill-2016-scenariomip}. Our emulator identifies the scenario only through the combination of \gls{co2e} and year, and the \gls{co2e} trajectories of the three \glspl{ssp} barely diverge over 2015--2024.

In each of the seventeen \gls{loocv} folds,  we hold out a single \gls{cmip6} model from the pool and use it as a
synthetic observational target for which we also have ground truth in the future. This target model plays the role of \gls{era5}. We train the fold-specific emulator on the daily temperature fields simulated by the remaining sixteen \gls{cmip6} models over 1979--2100. Only the periods 1979--2000 and 2011--2024 of the held-out model enter training.
We hold out 2001--2010 to evaluate the emulation quality closer to the present day while keeping the start of each \gls{ssp} (2015--2024) in the training data. This choice also lets the emulator learn how different \glspl{ssp} begin to diverge, which is useful for future projections. We withhold consecutive years rather than random days because daily temperature is autocorrelated and random days would leak information from training into validation. 
During data loading, we divide the daily fields into the historical period (1979--2014) and \glspl{ssp} \glsentryshort{245}, \glsentryshort{370}, and \glsentryshort{585}, because the three scenarios share the historical period up to 2014. In this way, the emulator sees each daily field once per epoch, in random order.

\paragraph*{Data Scaling}
We compute a scalar mean and standard deviation over all grid cells and days of the training data, separately for each fold, to apply z-score normalization~\cite{si:han-2012-zscore}. We standardize \gls{co2e} in a similar manner. Because \gls{co2e} is a function of year and \gls{ssp} in our setup, its statistics include each possible value once across the full historical and \gls{ssp} periods, independent of any fold.

\paragraph*{Conditioning}
Of the four conditioning inputs of \emph{Materials and Methods}, \emph{Conditional Diffusion Emulator}, two are categorical: the model identity $m$ (from a fixed list of \gls{cmip6} models and \gls{era5}) and the day of year $d$. We encode each by its integer index and pass it through its own sinusoidal positional embedding $\mathbf{e}_m$ and $\mathbf{e}_d$, respectively. A learned feed-forward network embeds the z-scored annual global-mean \gls{co2e} concentration $c$ as $\mathbf{e}_c$. We rescale the year $y$ linearly to $[-1,1]$ and embed it with its own learned feed-forward network as $\mathbf{e}_y$. We concatenate these embeddings and project them to the timestep-embedding dimension (128) to obtain the joint conditioning vector $\mathbf{c}_{\mathrm{comb}}$:
\begin{align*}
    \mathbf{c}_{\mathrm{comb}} = \mathrm{FFN}\bigl([\mathbf{e}_m;\,\mathbf{e}_c;\,\mathbf{e}_d;\,\mathbf{e}_y]\bigr) \numberthis \label{eq:conditioning}
\end{align*}
 The vector is added to the timestep embedding throughout the network. 

\paragraph*{Denoiser}
The UNet backbone expands its channels progressively (32$\to$64$\to$128 channels, multipliers [1,2,4]), with ResNet blocks at each resolution level and self-attention at the coarsest scale. It also uses \gls{edm} preconditioning~\citep{si:karras-2022-elucidating} with $\sigma_{\mathrm{data}} = 1$.

\paragraph*{Noise Schedule}
We use a log-uniform noise schedule over the noise level $\sigma$ (the diffusion time $t$ of Eq.~\ref{eq:guidance} of the main text; $\sigma(t) = t$ in \gls{edm}) with $\sigma_{\min} = 0.002$ and $\sigma_{\max} = 200$, where the large $\sigma_{\max}$ ensures the forward process fully destroys the climate signal to prevent low-frequency structure from leaking across noise levels~\cite{si:brenowitz-2025-cbottle}.

\paragraph*{Training}
We train for 150 epochs with AdamW~\cite{si:loshchilov-2018-adamw} (one-cycle learning rate schedule peaking at $10^{-3}$, batch size 192, 16-bit bfloat mixed-precision training) and maintain an exponential moving average of the network parameters, which we use for all evaluation. We average the \gls{edm} loss~\citep{si:karras-2022-elucidating} over grid cells with weights proportional to their true area, so that each region contributes in proportion to its size, following standard practice in climate emulation~\citep{si:nguyen-2023-climatelearn,si:willard-2025-analyzing,si:aich-2026-conditional}.

\paragraph*{Sampling}
We sample with the deterministic Heun sampler of Karras et al.~\cite{si:karras-2022-elucidating}, which integrates the probability flow \gls{ode} and enables log-likelihood computation.

\begin{figure}[H]
\centering
\includegraphics[width=\textwidth]{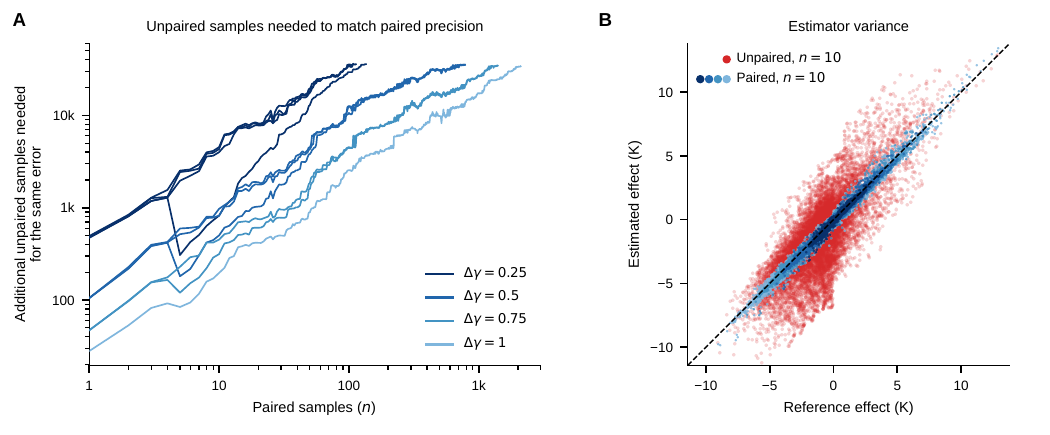}
\caption{\textbf{Sharing the initial noise across guidance scales makes the effect of guidance cheaper to estimate.} We sample one condition of the MPI-ESM1-2-LR fold (source model NorESM2-LM, SSP5-8.5, day 195 of 2090) at $\gamma = 0$, $0.25$, $0.5$, $0.75$ and $1$. For a pair of scales the effect of guidance is the difference between the mean fields at the two scales at each grid cell. In paired estimates both scales start from the same initial noise tensors while in unpaired estimates they start from different ones. A reference effect is computed from $37{,}498$ samples per scale and the paired and unpaired estimates are drawn from separate samples so that no initial noise is used twice. Accuracy is the area-weighted root-mean-square error (RMSE) of an estimated effect map against the reference. (\emph{A}) For each of the ten scale pairs ($\binom{5}{2} = 10$), the number of additional samples per scale an unpaired estimate needs to be as accurate as a paired estimate from $n$ samples per scale is plotted. Each curve ends where the available unpaired samples are no longer enough. Colors mark the gap between the two scales, $\Delta\gamma$: four pairs share $\Delta\gamma = 0.25$, three share $0.5$, two share $0.75$ and one has $\Delta\gamma = 1$. (\emph{B}) Estimates from only $n = 10$ samples per scale against the reference, one point per grid cell, for the four pairs that compare $\gamma = 0$ with a guided scale. Red points are unpaired estimates. Blue points are paired estimates shaded by $\Delta\gamma$ as in \emph{A}. Points on the dashed line indicate agreement with the reference.}
\label{fig:si-shared-noise}
\end{figure}

\paragraph*{Shared Initial Noise Sampling}
Since the sampler integrates the probability flow \gls{ode} deterministically, a sample is a deterministic function of its initial noise, its conditioning inputs and the guidance scale. As in our earlier work~\citep{si:immorlano-2025-technical}, we pass the same initial noise tensors to every guidance scale for a given conditioning and sample index $s$, which separates the effect of the guidance scale from sampling variability (Fig.~\ref{fig:si-shared-noise}). Each noise tensor is still sampled independently for each sample index, source model identity, \gls{co2e} value, day, year, and \gls{loocv} fold.

\section*{Interpolating between Two Distributions}\label{sec:si:paths}
The two endpoints of a path are the two fixed distributions that it connects: that of a climate model and that of the observations. Both are learned by our conditional diffusion model, $p_\theta$, which returns a different distribution of daily temperature fields for each condition (\emph{Materials and Methods}, \emph{Conditional Diffusion Emulator}). The identity condition selects whose distribution it returns. As in the main text, we write the two endpoints as $p_i=p_\theta(\cdot\mid\mathrm{CMIP6}_i,d,c,y)$ for the source (\gls{cmip6} model $i$, $i \in \{1, \dots, 17\}$ in our case) and $p_{\mathrm{tgt}}=p_\theta(\cdot\mid\mathrm{ERA5},d,c,y)$ for the target, so $p_i$ is the emulated climate model $i$ and $p_{\mathrm{tgt}}$ the emulated \gls{era5}. In a perfect-model test, the held-out model plays the role of \gls{era5} and the sixteen remaining models are the sources. A diffusion sampler does not need to evaluate $p_\theta$ itself: at each noise level, it follows the score $\nabla_{\mathbf{x}} \log p_\theta$, the direction in which a field $\mathbf{x}$ becomes more likely, which the denoiser estimates. Eq.~\ref{eq:guidance} of the main text averages the scores of $p_i$ and $p_{\mathrm{tgt}}$ with weights $1-\gamma$ and $\gamma$; the paths below show which distribution this average is the score of.

\paragraph*{Optimal Interpolating Paths}
We call any distribution $q$ of daily temperature fields a \emph{candidate} for the interpolating distribution, and let $D(p,q)$ be a measure of distance between two distributions. We look for the interpolating distribution, the candidate that is simultaneously close to both endpoints. For each $\gamma \in [0,1]$, it minimizes the weighted objective:

\begin{align*}
J_\gamma(q)=(1-\gamma)\,D(p_i,q)+\gamma\,D(p_{\mathrm{tgt}},q), \numberthis \label{eq:Jgamma}
\end{align*}
where the weight $\gamma$ sets how much closeness to the target counts relative to closeness to the source. When $D$ is convex in $q$, the minimizers span, as $\gamma$ varies, the Pareto frontier of $\big(D(p_i,q),D(p_{\mathrm{tgt}},q)\big)$ (which is the set of best compromises between the two distances): a minimizer cannot be moved closer to one endpoint without moving farther from the other. The path traced by varying $\gamma$ depends on $D$, and we consider three choices of $D$ below.

\paragraph*{Forward Kullback--Leibler Divergence Gives the Mixture Path}
We first use the forward \gls{kl} divergence, which compares an endpoint $p$ ($p_i$ or $p_{\mathrm{tgt}}$) with the candidate $q$:
\begin{align*}  
    \KL(p \,\|\, q) = \EE_{p}\!\left[\log \frac{p(\mathbf{x})}{q(\mathbf{x})}\right] ,
\end{align*}
where the expectation $\EE_{p}$ is computed under the endpoint (average over fields $\mathbf{x}$ that the endpoint produces, the emulated climate model for $p_i$ or the emulated \gls{era5} for $p_{\mathrm{tgt}}$). The divergence is large when the candidate gives low probability to fields that the endpoint produces often. To be close to both endpoints, a candidate must therefore cover the fields of both. Dropping terms that do not involve $q$,

\begin{align*}
    J_\gamma(q)
    &= -\int \bigl[(1-\gamma)\,p_i(\mathbf{x}) + \gamma\,p_{\mathrm{tgt}}(\mathbf{x})\bigr] \log q(\mathbf{x})\,\mathrm{d}\mathbf{x} + \mathrm{const} \\
    &= \KL(p_{\gamma,i}^{\mathrm{mix}} \,\|\, q) + \mathrm{const}, \numberthis \label{eq:forward-derivation}
\end{align*}

with
\begin{align*}
p_{\gamma,i}^{\mathrm{mix}} = (1-\gamma)\,p_i + \gamma\,p_{\mathrm{tgt}} , \numberthis \label{eq:mixture}
\end{align*}

where the second line of Eq.~\ref{eq:forward-derivation} adds and subtracts $\int p_{\gamma,i}^{\mathrm{mix}}(\mathbf{x})\log p_{\gamma,i}^{\mathrm{mix}}(\mathbf{x})\,\mathrm{d}\mathbf{x}$, a term that does not depend on $q$ and joins the constant. Because the \gls{kl} divergence is zero when its two arguments coincide and positive otherwise, the minimizer is $q=p_{\gamma,i}^{\mathrm{mix}}$ of Eq.~\ref{eq:mixture}. This is the arithmetic mixture, which covers the fields of both endpoints, known as the \emph{mixture path} or \emph{m-geodesic}. The optimal value is the skewed Jensen--Shannon divergence $(1-\gamma)\KL(p_i\|p_{\gamma,i}^{\mathrm{mix}}) + \gamma\KL(p_{\mathrm{tgt}}\|p_{\gamma,i}^{\mathrm{mix}})$.

\paragraph*{Reverse Kullback--Leibler Divergence Gives the Geometric Path}
The reverse \gls{kl} divergence swaps the roles of the endpoint and the candidate:

\begin{align*}
\KL(q \,\|\, p) = \EE_{q}\!\left[\log \frac{q(\mathbf{x})}{p(\mathbf{x})}\right],
\end{align*}

where the expectation $\EE_q$ is now computed under the candidate (average over fields that the candidate $q$ itself produces). The divergence is large when the candidate produces fields that the endpoint finds unlikely. To be close to both endpoints, a candidate must therefore keep only fields that are plausible under both. Combining the two logarithms,

\begin{align*}
    J_\gamma(q)
    &= (1 - \gamma) \int q(\mathbf{x}) \log \frac{q(\mathbf{x})}{p_i(\mathbf{x})}\,\mathrm{d}\mathbf{x} + \gamma \int q(\mathbf{x}) \log \frac{q(\mathbf{x})}{p_{\mathrm{tgt}}(\mathbf{x})}\,\mathrm{d}\mathbf{x} \\
    &= \int q(\mathbf{x}) \log \frac{q(\mathbf{x})}{p_i(\mathbf{x})^{1-\gamma} p_{\mathrm{tgt}}(\mathbf{x})^{\gamma}}\,\mathrm{d}\mathbf{x} \\
    &=: \KL(q \,\|\, p_{\gamma,i}) - \log Z_\gamma, \numberthis \label{eq:reverse-derivation}
\end{align*}

with

\begin{align*}
    p_{\gamma,i}=\frac{p_i(\mathbf{x})^{\,1-\gamma}p_{\mathrm{tgt}}(\mathbf{x})^{\,\gamma}}{Z_\gamma},
    \quad
    Z_\gamma=\int p_i(\mathbf{x})^{\,1-\gamma}p_{\mathrm{tgt}}(\mathbf{x})^{\,\gamma}\,\mathrm{d}\mathbf{x}. \numberthis \label{eq:geometric}
\end{align*}

The normalization constant $Z_\gamma$ does not depend on $q$ and so does not affect the minimization: minimizing $J_\gamma$ corresponds to minimizing $\KL(q \,\|\, p_{\gamma,i})$, which, as for the forward divergence, is zero only when $q = p_{\gamma,i}$. The minimizer is therefore $q = p_{\gamma,i}$ of Eq.~\ref{eq:geometric}, the weighted geometric mean of Eq.~\ref{eq:geomean} of the main text, known as the \emph{geometric path} or \emph{e-geodesic}.
\newline
\newline
\noindent The names of the two paths refer to the coordinate in which each is linear: the mixture path interpolates linearly in the densities, the geometric path in the log-densities \citep{si:amari-2016-information}. Condition-to-condition guidance follows the geometric path, because at each noise level Eq.~\ref{eq:guidance} of the main text is the score of the weighted geometric mean of the two endpoints $p_i$ and $p_{\mathrm{tgt}}$:
\begin{align*}
( 1-\gamma)\,\nabla\!\log p_i+\gamma\,\nabla\!\log p_{\mathrm{tgt}} & =  \nabla\Bigl[(1-\gamma)\log p_i+\gamma\log p_{\mathrm{tgt}}\Bigr] \\
& = \nabla\log\Bigl[p_i^{\,1-\gamma}p_{\mathrm{tgt}}^{\,\gamma}\Bigr] \\
& = \nabla\log p_{\gamma,i}, \numberthis \label{eq:score-to-geomean}
\end{align*}
with the last equality holding because $Z_\gamma$ is constant in $\mathbf{x}$ and vanishes under the gradient. The score used in Eq.~\ref{eq:guidance} of the main text is therefore the score of

\begin{align*}
p_{\gamma,i}(\mathbf{x}_t)\;\propto\;p_\theta(\mathbf{x}_t \mid \mathrm{CMIP6}_i, d, c, y)^{\,1-\gamma}\;p_\theta(\mathbf{x}_t \mid \mathrm{ERA5}, d, c, y)^{\,\gamma}. \numberthis \label{eq:si-geomean}
\end{align*}
Eq.~\ref{eq:score-to-geomean} involves no approximation: at a given noise level, averaging scores and computing a geometric mean of densities correspond to the same operation. Across noise levels, the reverse diffusion driven by these scores samples the geometric mean of the noise-free conditionals only approximately, because adding noise to a geometric mean of two distributions does not give the geometric mean of the noised distributions (\emph{Materials and Methods}, \emph{Condition-to-Condition Guidance}).

\paragraph*{Under Total Variation, the Mixture Path Has a Linear Tradeoff}
A third choice of $D$, \gls{tv}, has a simple meaning: it measures how well the best possible classifier can tell the candidate $q$ from an endpoint $p$.
For fields drawn from $\mathbf{x} \sim \frac12(q(\mathbf{x}) + p(\mathbf{x}))$, that is, from $q$ or $p$ with equal probability, the accuracy of this best possible (Bayes-optimal) classifier between $q$ and $p$ is
\begin{align*}
    \Acc(p, q) = 1 - \tfrac12 \int \min\bigl(p(\mathbf{x}), q(\mathbf{x})\bigr)\,\mathrm{d}\mathbf{x}
    = \tfrac12\bigl(1 + \TV(p, q)\bigr),
\end{align*}
where $\TV(p,q) = \tfrac12 \int |p(\mathbf{x})-q(\mathbf{x})|\,\mathrm{d}\mathbf{x}$ is the total variation and the second equality uses $\min(a, b) = (a + b - |a - b|) / 2$. The larger the total variation, the better the classifier tells the two distributions apart.

For the mixture of Eq.~\ref{eq:mixture} as candidate, the differences from the two endpoints are proportional to one another,

\begin{align*}
p_{\gamma,i}^{\mathrm{mix}} - p_i = \gamma (p_{\mathrm{tgt}} - p_i),
\quad
p_{\gamma,i}^{\mathrm{mix}} - p_{\mathrm{tgt}} = (\gamma-1)(p_{\mathrm{tgt}} - p_i),
\end{align*}

and therefore

\begin{align*}
& \TV(p_i, p_{\gamma,i}^{\mathrm{mix}}) = \gamma\, \TV(p_i, p_{\mathrm{tgt}}), \\
& \TV(p_{\mathrm{tgt}}, p_{\gamma,i}^{\mathrm{mix}}) = (1 - \gamma)\, \TV(p_i, p_{\mathrm{tgt}}).
\end{align*}

Total variation is a metric, so it satisfies the triangle inequality:
\begin{align*}
    \TV(p_i, p_{\mathrm{tgt}}) \leq \TV(p_i, q) + \TV(p_{\mathrm{tgt}}, q).
\end{align*}
For $q = p_{\gamma,i}^{\mathrm{mix}}$ the equality holds, meaning that the sum of the two $\TV$ terms reaches its minimum:

\begin{align*}
    \TV(p_i, p_{\gamma,i}^{\mathrm{mix}}) + \TV(p_{\mathrm{tgt}}, p_{\gamma,i}^{\mathrm{mix}}) = & [\gamma + 1 - \gamma] \TV(p_i, p_{\mathrm{tgt}}) \\
    = & \TV(p_i, p_{\mathrm{tgt}}).
\end{align*}

The mixture therefore lies on the linear path between $p_i$ and $p_{\mathrm{tgt}}$ in total variation and in accuracy: its distance to $p_i$ grows, and its distance to $p_{\mathrm{tgt}}$ shrinks, linearly in $\gamma$.

\begin{figure}[H]
\centering
\includegraphics[width=\textwidth]{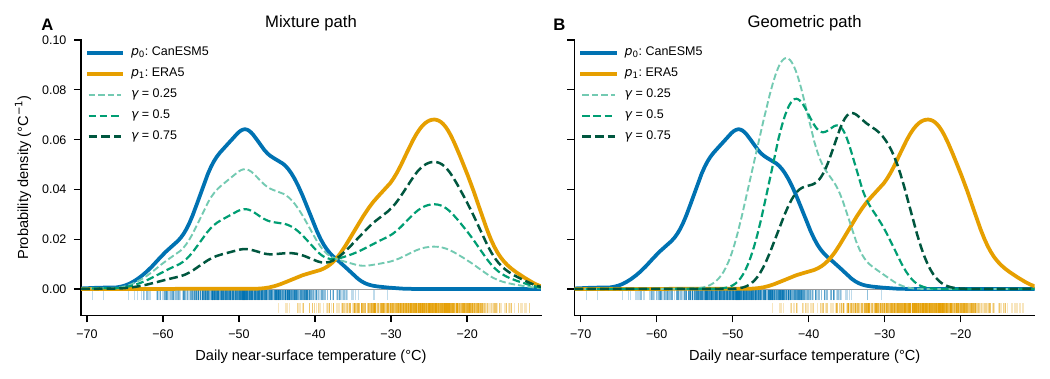}
\caption{\textbf{Mixture and geometric paths between CanESM5 and ERA5 at one Antarctic grid cell ($85.1^{\circ}$S, $140.6^{\circ}$W), March to May 2001--2010.} (\emph{A}) The mixture path $(1-\gamma)\,p_i+\gamma\,p_{\mathrm{tgt}}$ and (\emph{B}) the geometric path $p_i^{\,1-\gamma}p_{\mathrm{tgt}}^{\,\gamma}/Z_\gamma$ between $p_i$, the density of daily near-surface temperature of CanESM5 (blue), and $p_{\mathrm{tgt}}$, that of ERA5 (orange), each from $920$ daily values. Dashed curves are the paths at $\gamma = 0.25$, $0.5$ and $0.75$, darker for larger $\gamma$. Blue and orange ticks below zero mark the daily values of CanESM5 (upper row) and ERA5 (lower row). The paths are defined in \emph{Interpolating between Two Distributions}.}
\label{fig:si-paths-example}
\end{figure}

\paragraph*{Which Path Bias Correction Requires}
The mixture of Eq.~\ref{eq:mixture} is sampled by drawing from $p_{\mathrm{tgt}}$ with probability $\gamma$ and from $p_i$ otherwise. Every draw comes from one endpoint or the other, so for well-separated endpoints the mixture is bimodal, with two separate peaks, and no sample lies between them. The geometric mean of Eq.~\ref{eq:si-geomean} instead assigns high density only where both factors are large, since a field that is unlikely under either endpoint brings the product close to zero. The mixture is a disjunction over the two endpoints: a field is likely if it is likely under one endpoint or the other. The geometric mean is a conjunction: a field is likely only if it is likely under both. Bias correction requires the conjunction: a corrected field should be one that the climate model and the observations could both have produced, in line with the geometric mean in Eq.~\ref{eq:si-geomean}. Fig.~\ref{fig:si-paths-example} shows the mixture path and the geometric path side by side for a one-dimensional example built from \gls{cmip6} and \gls{era5} data.

\section*{Fitting the Guidance Schedule}\label{sec:si:schedule}
\begin{figure}[H]
\centering
\begin{minipage}[t]{0.49\textwidth}\textsf{\textbf{A}}\\
\includegraphics[width=\linewidth]{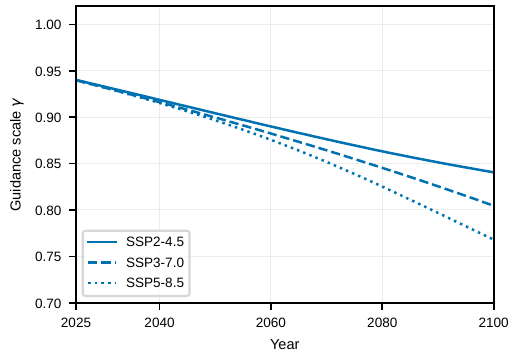}\end{minipage}\hfill
\begin{minipage}[t]{0.49\textwidth}\textsf{\textbf{B}}\\
\includegraphics[width=\linewidth]{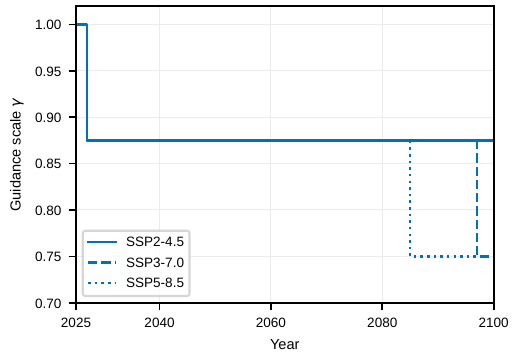}\end{minipage}
\caption{\textbf{The fitted guidance schedule, 2025--2100.} (\emph{A}) The guidance scale $\gamma(y, c)$ of Eq.~\ref{eq:gamma-schedule} for each scenario, the continuous values supplied directly to the sampler in the scheduled run and the projections toward ERA5, which sample every day of the year. (\emph{B}) The same schedule rounded to the nearest scale of the nine-scale sweep (\emph{Sampling Designs}): $\gamma = 1$ through 2026, then $0.875$, then $0.75$ from 2097 under \glsentryshort{ssp370} and from 2085 under \glsentryshort{ssp585}; the three scenarios coincide before these dates.}
\label{fig:si-gamma-schedule}
\end{figure}

The selections described in \emph{Materials and Methods}, \emph{Selecting the Guidance Strength}, give one $\gamma$ for each of the seventeen folds, three \glspl{ssp} and 32 blocks, $17 \times 3 \times 32 = 1{,}632$ values in total (the criterion is defined in \emph{Evaluation Metrics}, \emph{The Criterion Used to Select the Guidance Scale}, and the samples in \emph{Sampling Designs}, \emph{The Scale Sweep}). Each selection is based on only the twelve days per year of its block, one sample per source model and nine discrete scales, and the selected values vary between folds and blocks. For the projections toward \gls{era5}, no selection is possible, because the future of the target is unknown. The schedule must therefore depend only on quantities known for any future year: the calendar year and the \gls{co2e} concentration of the scenario. To obtain a schedule for all years, we fit one linear function of the calendar year and the \gls{co2e} concentration to all values pooled over folds and \glspl{ssp}:

\begin{align*}
    \gamma(y, c) = 3.0516 - 1.0138\times10^{-3}\,y - 1.3605\times10^{-4}\,c , \numberthis \label{eq:gamma-schedule}
\end{align*}

where $y$ is the midpoint year of a block and $c$ its mean \gls{co2e} concentration in ppm, and we clip the result to $[0,1]$. Evaluated along the \gls{co2e} pathway of each \gls{ssp}, the schedule declines from $\gamma=0.940$ in 2025 to $0.841$, $0.805$ and $0.768$ in 2100 under \glsentryshort{ssp245}, \glsentryshort{ssp370} and \glsentryshort{ssp585}, that is $-0.0135$, $-0.0180$ and $-0.0231$ per decade (Fig.~\ref{fig:si-gamma-schedule}\emph{A}). We pool over folds so that the schedule is not tuned to any one held-out model, in line with the assumption that it transfers to \gls{era5} (\emph{Materials and Methods}, \emph{Selecting the Guidance Strength}). We pool over \glspl{ssp} and include the \gls{co2e} term so that the dependence on the scenario enters through the forcing and one function serves all three scenarios. A linear function is the simplest form that lets the scale change gradually with year and \gls{co2e}. Fig.~\ref{fig:si-gamma-schedule}\emph{B} shows the same schedule rounded to the nearest of the nine scales of the scale sweep. We use this rounded schedule only to read scheduled-guidance results from the sweep, which covers all seventeen folds (\emph{Sampling Designs}, \emph{The Scale Sweep}); the scheduled run and the projections toward \gls{era5}, which sample every day, use the continuous schedule of Fig.~\ref{fig:si-gamma-schedule}\emph{A}.

\section*{Estimating the Rescaling Factor}\label{sec:si:rescaling}
Because our emulator samples each day independently, its annual global means keep almost none of the year-to-year variability that climate models and observations show, and we correct this by rescaling (\emph{Materials and Methods}, \emph{Annual-Variance Rescaling}). The rescaling acts on each emulated annual global-mean \emph{series}: the annual global means $g_y$ of one sample of one source model, one value for each year $y$. It replaces each series $g_y$ by
\begin{align*}
    g'_y = \tilde{g}_y + k\,\bigl(g_y - \tilde{g}_y\bigr), \qquad k = \max\bigl(\sigma_{\mathrm{tgt}}/\sigma_{\mathrm{emulator}},\, 1\bigr), \numberthis \label{eq:variance-k}
\end{align*}
where $\tilde{g}_y$ is the cubic polynomial in time fitted to $g_y$ by least squares. The cubic keeps the forced trajectory, the departure of the source model from the others and the constant offset of the sample, so the residual $g_y - \tilde{g}_y$ is the year-to-year fluctuation. Eq.~\ref{eq:variance-k} thus keeps the trend $\tilde{g}_y$ unchanged and stretches the fluctuations around it by the factor $k$, which compares how much the annual mean varies from year to year in the target and in the emulator and is never smaller than 1.

For the target, we use its 46 annual global means $G_y$ of 1979--2024 (for a held-out \gls{cmip6} model, its historical simulation followed by \glsentryshort{ssp245} from 2015). Both the expected temperature of a year and the spread of the annual mean around it can change over time, so we estimate both as functions of the year. Following Guillaumin and Zanna~\cite{si:guillaumin-2021-stochastic}, a small neural network learns from these values the expected temperature of each year, $\mu_y$, and then, with $\mu_y$ held fixed, the standard deviation around it, $s_y$, the year-to-year variability of the target in year $y$. The typical deviation over the whole period, $\sigma_{\mathrm{tgt}}$, is:

\begin{align*}
    \sigma_{\mathrm{tgt}} = \Bigl(\frac{1}{46}\sum_{y=1979}^{2024} s_y^{2}\Bigr)^{1/2}.
\end{align*}

For the emulator, we use its samples replicating the target ($\gamma = 1$) over the same years, 1979--2024. We use them on the same days of each year as the series we rescale: every day for the designs that sample every day, and the twelve days per year of the scale sweep for the sweep (\emph{Sampling Designs}). We fit the cubic trend $\tilde{g}_y$ of each of these series, compute the standard deviation of its residuals $g_y - \tilde{g}_y$, and use the median of these standard deviations over the series as $\sigma_{\mathrm{emulator}}$. Two cases need special treatment. The sweep samples only from 2025 and its pattern of days repeats every 31 years, so over 1979--2024 we continue the same pattern backward from 2025. The emulator behind classifier guidance never learned the held-out model and so cannot replicate it; for classifier guidance, we use the series we rescale, over 2025--2100.

\section*{Additional Details for Figures}
\paragraph*{Figure~\ref{fig:schematic}}

Figure~\ref{fig:schematic} of the main text shows the \gls{loocv} fold in which MPI-ESM1-2-LR is held out as the synthetic observation, under SSP5-8.5, over two periods: 2015--2024 (in-distribution emulation) and 2085--2100 (long-term extrapolation). We build it in four steps: sampling, removal of the seasonal cycle, projection with a \gls{pca} and density estimation. Sampling: for each period, we draw ten samples each day, for $\gamma \in \{0, 0.25, 0.5, 0.75, 1\}$, using the same initial noise for every guidance scale (\emph{Machine Learning Setup Details}, \emph{Shared Initial Noise Sampling}). Removal of the seasonal cycle: before fitting the \gls{pca} and projecting the data, we remove the seasonal cycle to make the impact of guidance clearly visible. We remove it separately for each group of data: the synthetic observation, each source in the simulation data, and each source at each $\gamma$ in the generated data. For each group and grid cell, we fit the mean and first three harmonics of the annual cycle (sinusoids with periods of one year, six months and four months) to all days of the period by least squares and subtract the harmonics, similar to Wheeler and Hendon~\cite{si:wheeler-hendon-2004-rmm}. Projection: for each period, we fit a \gls{pca} only to the simulation data of that period, from the sixteen source models and from the held-out model serving as our synthetic observation. We then project onto its first two principal components the samples drawn above from the sixteen sources and, for the synthetic observation, its ground-truth simulation data.

Density estimation: these steps give point clouds in the projected \gls{pca} space, one for our method for each period and $\gamma$ and one for the synthetic observation for each period. To visualize the distributions and their densities, we compute a 2D \gls{kde} for each cloud, a smooth estimate of its density. We discretize the densities and show the highest-density regions, the smallest regions that contain 39.35\%, 86.47\% and 98.89\% of the distribution, shaded darker the smaller the region. These are the fractions inside the 1, 2 and 3 standard-deviation contours of a two-dimensional Gaussian distribution, $1 - e^{-n^2/2}$ for $n = 1, 2, 3$.

\begin{figure}[H]
\centering
\includegraphics[width=\textwidth]{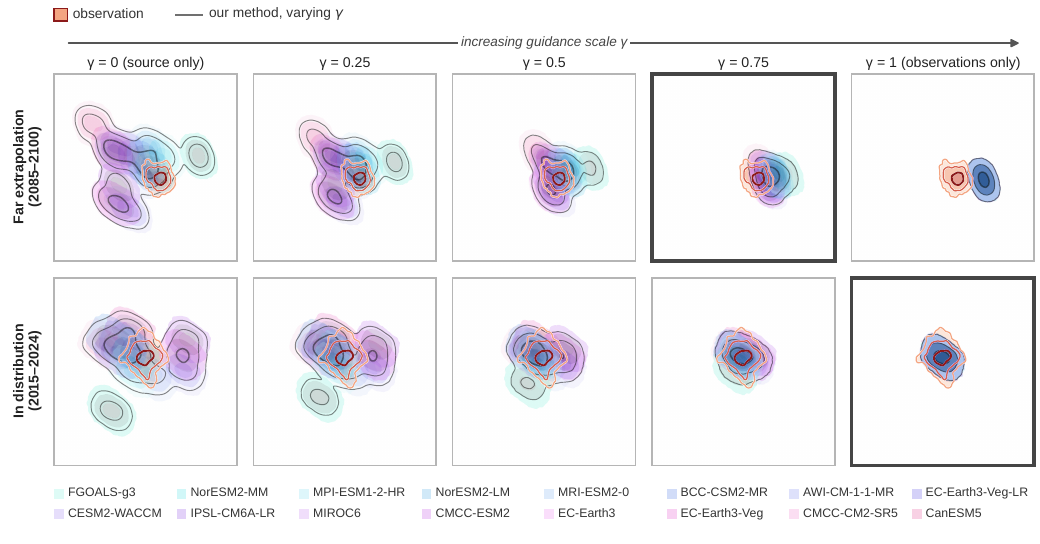}
\caption{\textbf{Figure~\ref{fig:schematic} of the main text decomposed by source model.} Same fold (MPI-ESM1-2-LR as synthetic observation), scenario (SSP5-8.5), windows, guidance scales, projection and density estimate as Fig.~\ref{fig:schematic}. Each of the sixteen source models is drawn as its own kernel density with the same three highest-density bands ($39.35$, $86.47$ and $98.89\%$), filled at $20\%$ opacity in the color given in the key; the gray outlines are the bands of the sixteen sources pooled, the blue distribution of Fig.~\ref{fig:schematic}, and the coral bands are the synthetic observation. The key shows each source's color as it appears on its own; where sources overlap their fills stack and the color deepens. The pooled distribution is a mixture of individually compact clouds, and every cloud contracts toward the emulator's version of the synthetic observation ($\gamma = 1$) as $\gamma$ increases.}
\label{fig:si-fig1-sources}
\end{figure}

The bold frames mark, for each period, the guidance scale closest to the selected value. For 2085--2100, the selected value is that of the continuous $\gamma$ schedule fitted over 2025--2100, which we round to the nearest scale shown; for 2015--2024, where the target is known, it is $\gamma = 1$ (\emph{Materials and Methods}, \emph{Selecting the Guidance Strength}). We show fixed scales, rather than the schedule, to display how $\gamma$ changes the sampled distribution. As $\gamma$ increases toward $1$, the sampled distribution changes from multimodal to the single target distribution. It is multimodal at small $\gamma$ because it pools the sixteen sources, each of which forms its own compact cloud (Fig.~\ref{fig:si-fig1-sources}).

\section*{Sampling Designs}
Every evaluation and projection in the main text and in this Supporting Information is based on one of the five sampling designs below. Fig.~\ref{fig:schematic} of the main text and Fig.~\ref{fig:si-fig1-sources}, which illustrate how guidance acts, use their own samples, described in \emph{Additional Details for Figures}.

\paragraph*{The Emulator Replicating ERA5}
The emulator conditioned on \gls{era5} ($\gamma = 1$) replicates the observed record with seventeen samples on every day of 1979--2024. Tables~\ref{tab:baselines} and \ref{tab:spatial-consistency} and Figs.~\ref{fig:si-baseline-maps}--\ref{fig:si-clouds-eur} use these samples over 2001--2010, the decade withheld from training and used for validation. Figs.~\ref{fig:projections}\emph{A}, \ref{fig:distributions} and \ref{fig:heat-region} of the main text use them before 2025.

\paragraph*{The Projections toward ERA5}
We guide each of the seventeen \gls{cmip6} models toward \gls{era5} with the continuous $\gamma$ schedule, with ten samples per source \gls{cmip6} model on every day of 2025--2100. Fig.~\ref{fig:projections}\emph{A}, the 2081--2100 half of Fig.~\ref{fig:distributions} and Tables~\ref{tab:retention} and \ref{tab:gwl-crossings} use all ten samples. Fig.~\ref{fig:heat-region} uses one sample per source model and day, so that the guided and raw \gls{cmip6} distributions are based on the same number of maps.

\paragraph*{The Scheduled Run}
Sampling every day with ten samples per source model is computationally expensive, so we use this design only for two \gls{loocv} folds, as case studies: those with MIROC6 and MPI-ESM1-2-LR as the synthetic observation. In each, we guide the sixteen remaining models toward the held-out one with the continuous $\gamma$ schedule, with ten samples per source model on every day of 2025--2100. For MIROC6, we also sample, at the same density, the diffusion emulator without guidance ($\gamma=0$) and classifier guidance. Fig.~\ref{fig:projections}\emph{B} of the main text and Fig.~\ref{fig:si-mpi-warming} use all ten samples. The comparisons of methods (Tables~\ref{tab:loocv-miroc6-baselines} and \ref{tab:loocv-mpi}) and the maps and distributions of the daily fields (Figs.~\ref{fig:si-m6-maps-raw-370}--\ref{fig:si-m6-cycle-diff} and \ref{fig:si-mpi-maps-370}) use one of the ten samples per source model and day, so that every ensemble has sixteen members, one per source model, as in the raw \gls{cmip6} archive used in this work. The guided panels of Figs.~\ref{fig:si-mpi-kde}--\ref{fig:si-mpi-rank} use one sample per source model on only the twelve days per year of the scale sweep (the same days as their unguided panels; see \emph{The Scale Sweep}).

\paragraph*{The Scale Sweep}
The design used to select the guidance strength covers nine discrete scales $\gamma \in \{0, 0.125, \ldots, 0.875, 1\}$, all seventeen folds of the \gls{loocv} and all three \glspl{ssp} over 2025--2100. We draw one sample per source \gls{cmip6} model, day, and scale on twelve days per year, about a month apart and shifted from year to year so that every calendar day is used over the period. Results with the scheduled guidance strength that we read from the sweep require no further sampling: for each year, we use the samples of the sweep at the scale nearest the value of the $\gamma$ schedule of Eq.~\ref{eq:gamma-schedule} for that year. This rounding gives $\gamma = 1$ in 2025 and 2026, $0.875$ afterward, and $0.75$ from 2097 under SSP3-7.0 and from 2085 under SSP5-8.5; under SSP2-4.5, the scale stays at $0.875$ through 2100 (Fig.~\ref{fig:si-gamma-schedule}\emph{B}). The two designs that guide with the schedule on every day, \emph{The Scheduled Run} and \emph{The Projections toward ERA5}, use the continuous schedule instead. At $\gamma = 1$ the sweep holds ten samples of the emulator conditioned on the held-out model and no sixteen-member ensemble, so we score 2025 and 2026 with the $\gamma = 0.875$ samples. Sparse sampling makes all seventeen folds and every scale affordable, at the cost of depth: twelve days a year over a twenty-one-year window give 252 values per grid cell and member, so only about 2.5 of them lie below the 1st percentile of a single cell, and as many above the 99th. The selection criterion averages over grid cells and members, which makes this depth sufficient. Every seventeen-fold statistic in the main text comes from this design: the trajectory \gls{rmse}, the percentile biases and the distributional metrics. Because the $\gamma$ schedule is fitted to selections made on these folds, these statistics are in-sample for the choice of $\gamma$. Tables~\ref{tab:loocv-skill} and \ref{tab:loocv-metrics}, Figs.~\ref{fig:si-folds-mean-370}--\ref{fig:si-folds-rank-370} and the unguided panels ($\gamma=0$) of Figs.~\ref{fig:si-mpi-kde}--\ref{fig:si-mpi-rank} use this design.

\paragraph*{The Emulator Conditioned on Each Model Identity}
To evaluate the two endpoints between which guidance interpolates, we sample the emulator conditioned on each of the seventeen \gls{cmip6} models without guidance ($\gamma = 0$), with ten samples on every day of 2080--2100 (and of 2030--2050 for the figures at \url{https://github.com/mandt-lab/gdso-figures}), and conditioned on \gls{era5} ($\gamma = 1$), with ten samples on every day of 2001--2010. Figs.~\ref{fig:si-prior-mean-370}--\ref{fig:si-prior-cycle-370} use this design.
\section*{Baseline Methods}\label{sec:si:baselines} This section describes the three methods against which the main text benchmarks condition-to-condition guidance: two statistical bias corrections, \gls{mvt} and \gls{qdm}, and classifier guidance. We evaluate all three in the held-out MIROC6 fold, and the two bias corrections also against \gls{era5} over 2001--2010.

\subsection*{Mean--Variance--Trend Transform (MVT)}
We implement the \gls{mvt} correction of Xu et al.~\cite{si:xu-2021-mvt} independently at each grid cell. Let $x_{\src}(\tau)$ be the daily temperature of a source \gls{cmip6} model on date $\tau$, $T_{\src}(\tau)$ its nonlinear monthly trend, and $C_{\src}(d)$ its climatology after removing that trend, for calendar day $d$ of $\tau$. The source weather anomaly is
\begin{align*}
    a_{\src}(\tau)
    =
    x_{\src}(\tau)-T_{\src}(\tau)-C_{\src}(d).
    \numberthis
    \label{eq:mvt-anomaly}
\end{align*}
We estimate the nonlinear trends from monthly temperature with \gls{eemd}. The corrected temperature is
\begin{align*}
    \widehat{x}_{\mathrm{MVT}}(\tau)
    =
    C_{\tgt}(d)
    +
    a_{\src}(\tau)\frac{\sigma_{\tgt}(d)}{\sigma_{\src}(d)}
    +
    T_{\mathrm{MME}}(\tau)
    -
    B_{m},
    \numberthis
    \label{eq:mvt-correction}
\end{align*}
where $T_{\tgt}$ and $C_{\tgt}$ are the target's nonlinear monthly trend and its climatology after removing that trend, $\sigma_{\tgt}(d)$ and $\sigma_{\src}(d)$, the standard deviations of the target and source anomalies on calendar day $d$, rescale the daily variability, and $T_{\mathrm{MME}}$ is the nonlinear trend of the raw-\gls{cmip6} multi-model ensemble mean. The monthly trend correction is
\begin{align*}
    B_m
    =
    \overline{T_{\mathrm{MME}}}^{\,H}_{m}
    -
    \overline{T_{\tgt}}^{\,H}_{m},
    \numberthis
    \label{eq:mvt-trend-bias}
\end{align*}
where the overline denotes the average over the calibration years $H$ in calendar month $m$, the month of $\tau$. Thus, \gls{mvt} combines the target climatology and variability with the projected nonlinear trend of the source-model ensemble. We use five \gls{eemd} trials with a noise width of $0.2$ and recompute the daily standard deviations after excluding anomalies larger than three times their initial standard deviation. Applied to the unguided emulator ($\gamma=0$), \gls{mvt} estimates the monthly source trend and the calibration statistics jointly from all samples. 

\paragraph*{ERA5 as the Target} We compute $T_{\mathrm{MME}}$ from all seventeen raw \gls{cmip6} models. Historical calibration uses 1979--2000 and 2011--2014, and each \gls{ssp} adds 2015--2024, so the calibration years are 1979--2000 and 2011--2024. The decade 2001--2010 is excluded from all calibration statistics and retained for validation.

\paragraph*{MIROC6 as the Target}
MIROC6 replaces \gls{era5} as the target and is not corrected, and we recompute $T_{\mathrm{MME}}$ from the remaining sixteen raw \gls{cmip6} models. We use the same calibration years and withhold the MIROC6 simulation over 2001--2010 and 2025--2100 for validation. Because \gls{eemd} requires a continuous monthly series, we fill the missing target values over 2001--2010 by linear interpolation, separately for each calendar month, only to estimate $T_{\tgt}$. The target climatology, variability and trend-bias statistics use only the calibration years.

\subsection*{Quantile Delta Mapping (QDM)}
We implement additive \gls{qdm} following Cannon et al.~\cite{si:cannon-2015-qdm}, independently at each grid cell. For a source temperature $x_{\src}(\tau)$, its percentile in the local projected source distribution is
\begin{align*}
    p(\tau)
    =
    F_{\src,\mathrm{proj}(\tau)}
    \left[x_{\src}(\tau)\right].
\end{align*}

The corrected temperature is
\begin{align*}
    \widehat{x}_{\mathrm{QDM}}(\tau)
    =
    F_{\tgt,H}^{-1}\left[p(\tau)\right]
    +
    x_{\src}(\tau)
    -
    F_{\src,H}^{-1}\left[p(\tau)\right],
    \numberthis
    \label{eq:qdm-correction}
\end{align*}
where $F_{\tgt,H}$ and $F_{\src,H}$ are the empirical target and source distributions over the calibration years. The additive delta term preserves the source's modeled temperature change at percentile $p(\tau)$. We estimate the distributions separately for each grid cell and calendar month, pooling each month with its two adjacent months, and estimate $F_{\src,\mathrm{proj}(\tau)}$ from a centered 30-year window, shifted at the beginning and end of the record to keep 30 complete years. Applied to the unguided emulator, \gls{qdm} pools all samples to estimate the source distributions.

\paragraph*{ERA5 as the Target}
We estimate the target distributions from \gls{era5} and correct each of the seventeen \gls{cmip6} sources independently. Historical calibration uses 1979--2000 and 2011--2014, and each \gls{ssp} adds 2015--2024. The decade 2001--2010 is excluded from the calibration distributions and retained for validation.

\paragraph*{MIROC6 as the Target}
MIROC6 supplies the target distributions over the calibration years and the ground truth over 2001--2010 and 2025--2100, and is not a source. The remaining sixteen models are the sources, and we correct both their raw simulations and the corresponding unguided emulator samples.

\subsection*{Classifier Guidance}
We compare condition-to-condition guidance with classifier guidance \citep{si:dhariwal-nichol-2021-diffusion-beats-gans} in the MIROC6 fold of our \gls{loocv} framework. In this fold, MIROC6 over 2001--2010 and 2025--2100 is the withheld synthetic observation, and the other sixteen \gls{cmip6} models are the sources. With classifier guidance, the correction toward MIROC6 must come from the classifier, not from information already contained in the emulator. We therefore train a second diffusion emulator on the sixteen source models over 1979--2100, with all MIROC6 simulations (historical and future) held out and all other hyperparameters unchanged. The classifier keeps the encoder and conditioning embeddings of this second emulator and adds an attention head. We train it with a cross-entropy loss to identify which of the seventeen \gls{cmip6} models a noised field comes from, over the same noise levels as the diffusion model. Its training data contain MIROC6 only over 1979--2000 and 2011--2024.

We find that classifier guidance underperforms condition-to-condition guidance. This is likely because the classifier is trained to classify rather than to provide accurate gradients, so its corrections are less precise. In contrast, condition-to-condition guidance adds scores that the emulator has learned directly.

Classifier guidance has two scales, $\eta$ for the pull toward the target and $\kappa$ for the push away from the source model:

\begin{align*}
    \nabla_{\mathbf{x}_t} \log \tilde p_i(\mathbf{x}_t)
    = \;& \nabla_{\mathbf{x}_t} \log p_{\theta}(\mathbf{x}_t \mid \text{CMIP6}_i, d, c, y) \\
      & + \nabla_{\mathbf{x}_t} \bigl[\, \eta \log p_\phi(\text{MIROC6} \mid \mathbf{x}_t, d, c, y)
        - \kappa \log p_\phi(\text{CMIP6}_i \mid \mathbf{x}_t, d, c, y) \,\bigr],
    \numberthis \label{eq:cg-guidance}
\end{align*}
where $\tilde p_i$ is the distribution sampled by classifier guidance from source $i$, $p_\theta$ is the second diffusion emulator and $p_\phi$ is the classifier.

We apply guidance only at noise levels $\sigma < 10$ with $\eta = \kappa = 2.5$. We select this threshold and tune both scales together ($\eta = \kappa$) by sampling within the held-out 2001--2010 period and comparing to the MIROC6 ground truth.

\section*{Evaluation Metrics}\label{sec:si:metrics}

This section defines every metric used to evaluate the emulator, both in the perfect-model experiments and against \gls{era5}, and every quantity used to describe its projections.

\paragraph*{Notation}
A \emph{source model} is a \gls{cmip6} model on which the emulator is conditioned and from which guidance starts (i.e., the $\gamma = 0$ endpoint of Eq.~\ref{eq:guidance} of the main text). In the perfect-model setting, one \gls{cmip6} model is held out and treated as the truth, and the remaining $N_m = 16$ models are the source models. In the projections toward \gls{era5}, all $N_m = 17$ models are source models. For a given setting (no guidance, $\gamma = 0$; guidance at scale $\gamma$; or a bias correction), let
$$
\hat{x}_{y,d,s,m,j,k}, \qquad y = 1,\ldots,N_y,\ d = 1,\ldots,N_d,\ s = 1,\ldots,N_s,\ m = 1,\ldots,N_m,
$$
denote the $s$-th sample of daily near-surface temperature generated from source model $m$ at latitude index $j$, longitude index $k$, on day $d$ of year $y$. The truth is the held-out model's
own simulation, a single realization,
$$
x^{\mathrm{truth}}_{y,d,j,k},
$$
and is replaced by $x^{\mathrm{ERA5}}_{y,d,j,k}$ wherever the evaluation is against observations. The grid is $64 \times 128$, global, and all temperatures are in $\degC$. The sampling designs of \emph{Sampling Designs} differ in $N_d$ and $N_s$: the scale sweep has $N_d = 12$ and $N_s = 1$, and the other designs sample every day ($N_d = 365$) and each states its $N_s$. Raw \gls{cmip6} and the \gls{mvt} and \gls{qdm} bias corrections have $N_s = 1$. At $\gamma = 1$ the source model plays no role: the ensemble then consists of $N_s$ independent samples of the emulator conditioned on the target (\gls{era5} or the held-out model), and $N_m = 1$. Each table states the window of its metrics, and $N_y$ is the number of years in that window.

Throughout, an \emph{ensemble member} is one source model together with its samples. Unless a table states otherwise, we use the pooled reduction, in which the index pair $(s,m)$ is flattened into a single sample axis of length $N_s N_m$ before any statistic is computed.

\paragraph*{Area-Weighted Average}
Every spatial average in this work is area-weighted, and the $64\times128$ grid is Gaussian in latitude. Let $A_{j,k}$ be the true area of cell $(j,k)$ supplied with the grid; summed over the grid, these areas recover the surface area of the Earth. The normalized weight is
$$
w_{j,k} \;=\; \frac{A_{j,k}}{\sum_{j',k'} A_{j',k'}}, \qquad \sum_{j,k} w_{j,k} = 1 ,
$$
so that the area-weighted spatial mean of any field $f_{j,k}$ is
$\langle f \rangle = \sum_{j,k} w_{j,k} f_{j,k}$. Most metrics below reduce a bias map to a
scalar through $\langle\,\cdot\,\rangle$ applied to its absolute value, so that errors of
opposite sign in different regions cannot cancel.

\subsection*{Trajectory Metrics}

These two metrics score the projected annual global-mean temperature, one value per year, against that of the ground truth. The trajectory \gls{rmse} measures how closely the ensemble mean follows the true trajectory, and the band coverage how often the ensemble's spread contains it.

\paragraph*{Trajectory RMSE}
Let $g_{y,m} = \langle \frac{1}{N_d N_s}\sum_{d,s} \hat{x}_{y,d,s,m,\cdot,\cdot} \rangle$ be the annual global-mean temperature of member $m$ in year $y$, rescaled where this applies (\emph{Materials and Methods}, \emph{Annual-Variance Rescaling}), and $g^{\mathrm{truth}}_{y}$ the same quantity for the held-out model. With $\bar{g}_y = \frac{1}{N_m}\sum_m g_{y,m}$ the ensemble mean, the trajectory \gls{rmse} over the $N_y$ years of a period is
$$
\mathrm{RMSE}_{\mathrm{traj}} \;=\; \sqrt{\frac{1}{N_y}\sum_{y=1}^{N_y}\bigl(\bar{g}_y - g^{\mathrm{truth}}_y\bigr)^{2}} .
$$ 
Tables~\ref{tab:loocv-skill}, \ref{tab:loocv-miroc6-baselines} and \ref{tab:loocv-mpi} report it.

\paragraph*{Band Coverage}
Each sample of each source model gives one annual series, $g_{y,s,m} = \langle \frac{1}{N_d}\sum_{d} \hat{x}_{y,d,s,m,\cdot,\cdot} \rangle$, rescaled where this applies (\emph{Materials and Methods}, \emph{Annual-Variance Rescaling}). In each year $y$, the 5--95\% band is the interval between the empirical 5th and 95th percentiles of the $N_s N_m$ values $g_{y,s,m}$. The \emph{band coverage} is the fraction of the $N_y$ years in which $g^{\mathrm{truth}}_y$ falls inside the band. The band is nominally $90\%$, so $0.90$ is the target, and a lower coverage means that the ensemble is under-dispersive. Tables~\ref{tab:loocv-skill}, \ref{tab:loocv-miroc6-baselines} and \ref{tab:loocv-mpi} report it.

Tables~\ref{tab:loocv-miroc6-baselines} and \ref{tab:loocv-mpi} compute both trajectory metrics on the warming above 1850--1900 (\emph{Warming-Level Crossing Year}), as Fig.~\ref{fig:projections}\emph{B} of the main text shows it, instead of on the annual global-mean temperature.

\subsection*{Distributional Metrics}

These metrics compare, at every grid cell, the daily temperatures of an ensemble with those of the truth on the same days. Together they test the mean, the spread, the shape and the tails of the daily distribution, the calibration of the ensemble and the spatial texture of its mean bias.

Write $\overline{\hat{x}}_{j,k} = \frac{1}{N_y N_d N_s N_m}\sum_{y,d,s,m} \hat{x}_{y,d,s,m,j,k}$ for the climatological-mean field of the ensemble and $\overline{x}^{\mathrm{truth}}_{j,k}$ for the corresponding truth climatology.

\paragraph*{Mean Absolute Bias and RMSE}
The pointwise mean-bias map and its two scalar summaries, the mean absolute bias and the \gls{rmse}, are
$$
B^{\mu}_{j,k} = \overline{\hat{x}}_{j,k} - \overline{x}^{\mathrm{truth}}_{j,k}, \qquad
\langle |B^{\mu}| \rangle, \qquad
\mathrm{RMSE}_\mu = \sqrt{\textstyle\sum_{j,k} w_{j,k}\bigl(B^{\mu}_{j,k}\bigr)^{2}} .
$$
Tables~\ref{tab:loocv-metrics}, \ref{tab:loocv-miroc6-baselines}, \ref{tab:baselines} and \ref{tab:loocv-mpi} report the \gls{rmse}, and Tables~\ref{tab:loocv-miroc6-baselines}, \ref{tab:baselines} and \ref{tab:loocv-mpi} the mean absolute bias.

\paragraph*{Standard-Deviation Ratio Deviation}
With $\sigma^{\mathrm{pred}}_{j,k}$ and $\sigma^{\mathrm{truth}}_{j,k}$ the standard deviations at cell $(j,k)$ across $(y,d,s,m)$ and $(y,d)$ respectively, the reported scalar is
$\langle |\sigma^{\mathrm{pred}}_{j,k}/\sigma^{\mathrm{truth}}_{j,k} - 1| \rangle$. It is zero
when the daily spread matches exactly and penalizes over- and under-dispersion equally. Tables~\ref{tab:loocv-metrics}, \ref{tab:loocv-miroc6-baselines}, \ref{tab:baselines} and \ref{tab:loocv-mpi} report it.

\paragraph*{Skewness and Excess-Kurtosis Bias}
The pointwise standardized third and fourth central moments are
$$
\mathrm{Skew}^{\mathrm{pred}}_{j,k} = \mathbb{E}\Bigl[\bigl(\tfrac{\hat{x}_{\cdot,j,k} - \overline{\hat{x}}_{j,k}}{\sigma^{\mathrm{pred}}_{j,k}}\bigr)^{3}\Bigr], \qquad
\mathrm{Kurt}^{\mathrm{pred}}_{j,k} = \mathbb{E}\Bigl[\bigl(\tfrac{\hat{x}_{\cdot,j,k} - \overline{\hat{x}}_{j,k}}{\sigma^{\mathrm{pred}}_{j,k}}\bigr)^{4}\Bigr] - 3 ,
$$
the expectation computed empirically over $(y,d,s,m)$, and analogously for the truth. The reported scalars are $\langle |\mathrm{Skew}^{\mathrm{pred}} - \mathrm{Skew}^{\mathrm{truth}}| \rangle$ and the same for kurtosis: the skewness bias and the excess-kurtosis bias. They diagnose shape discrepancies that the mean and variance miss. Tables~\ref{tab:loocv-metrics}, \ref{tab:loocv-miroc6-baselines} and \ref{tab:loocv-mpi} report both.

\paragraph*{Percentile Biases}
Let $q^{\mathrm{pred}}_{\alpha,j,k}$ and $q^{\mathrm{truth}}_{\alpha,j,k}$ denote the empirical $\alpha$-quantiles at cell $(j,k)$, computed over $(y,d,s,m)$ and $(y,d)$ respectively. The quantile-bias map is $B^{q_\alpha}_{j,k} = q^{\mathrm{pred}}_{\alpha,j,k} - q^{\mathrm{truth}}_{\alpha,j,k}$ and the reported scalar is $\langle |B^{q_\alpha}| \rangle$, the 1st-percentile bias for $\alpha = 0.01$ and the 99th-percentile bias for $\alpha = 0.99$. Tables~\ref{tab:loocv-metrics}, \ref{tab:loocv-miroc6-baselines}, \ref{tab:baselines} and \ref{tab:loocv-mpi} report both.

\paragraph*{TX90p and TN10p Biases}
The names follow two indices of the \gls{etccdi} \citep{si:zhang-2011-indices}: TX90p, the percentage of days on which the daily maximum temperature exceeds its 90th percentile, and TN10p, the percentage of days on which the daily minimum temperature falls below its 10th percentile. We apply them to daily mean temperature, with one threshold per grid cell: the 90th or 10th percentile of the truth's daily values in the window (\emph{Percentile Biases}). The warm-day exceedance frequency of the ensemble is
$$
F^{\mathrm{TX90p}}_{j,k} = \frac{1}{N_y N_d N_s N_m}\sum_{y,d,s,m} \mathbf{1}\bigl[\hat{x}_{y,d,s,m,j,k} > q^{\mathrm{truth}}_{0.9,j,k}\bigr] .
$$
The bias map is $F^{\mathrm{TX90p}}_{j,k} - F^{\mathrm{TX90p,truth}}_{j,k}$, where the truth's own exceedance frequency is $0.1$ up to the discreteness of the empirical percentile, so that an unbiased ensemble scores zero. It is reported in percentage points, and the scalar is its area-weighted absolute mean. TN10p is defined analogously with $\mathbf{1}[\hat{x} < q^{\mathrm{truth}}_{0.1,j,k}]$. Tables~\ref{tab:loocv-metrics}, \ref{tab:loocv-miroc6-baselines} and \ref{tab:loocv-mpi} report both, and Table~\ref{tab:baselines} the TX90p bias.

\paragraph*{Wasserstein Distance}
At each cell the 1-Wasserstein distance between the daily distributions of the ensemble and the truth is the mean absolute difference of their quantile functions,
$$
W_{1,j,k} \;=\; \int_0^1 \bigl| q^{\mathrm{pred}}_{u,j,k} - q^{\mathrm{truth}}_{u,j,k} \bigr| \, \mathrm{d}u ,
$$
and the scalar is $\langle W_1 \rangle$. Tables~\ref{tab:loocv-metrics}, \ref{tab:loocv-miroc6-baselines}, \ref{tab:baselines} and \ref{tab:loocv-mpi} report it.

\paragraph*{Continuous Ranked Probability Score}
\Gls{crps} is a proper scoring rule that penalizes jointly the distance from the ensemble to the truth and the miscalibration of its spread. At each grid cell and day, the $S = N_s N_m$ pooled values $\hat{x}_1,\ldots,\hat{x}_S$ of the ensemble are scored against the truth $x$ of that cell and day with the fair estimator
$$
\mathrm{CRPS} \;=\; \frac{1}{S}\sum_{s} |\hat{x}_s - x| \;-\; \frac{1}{2S(S-1)}\sum_{s,s'} |\hat{x}_s - \hat{x}_{s'}| ,
$$
averaged over the $N_y N_d$ days at each cell and then over cells with area weights. Tables~\ref{tab:loocv-metrics}, \ref{tab:loocv-miroc6-baselines} and \ref{tab:loocv-mpi} report it.

\paragraph*{Rank Histogram and Reliability Index}
For each $(y,d,j,k)$ let $r$ be the number of the $S$ pooled ensemble values strictly below the truth, so that $r \in \{0,\ldots,S\}$ and a calibrated ensemble gives ranks uniform over the $K = S+1$ bins. The rank histograms of the figures show the area-weighted frequency of each rank over all cells and days. The \gls{ri} is computed at each grid cell from the frequencies $f_{\ell,j,k}$ of the ranks over the $T = N_y N_d$ days at that cell,
$$
\mathrm{RI}_{j,k} \;=\; \sum_{\ell=0}^{S}\Bigl(f_{\ell,j,k} - \frac{1}{K}\Bigr)^{2} ,
$$
zero at perfect calibration, and the reported scalar is $\langle \mathrm{RI} \rangle$. Computing it per cell prevents miscalibrations of opposite sign in different regions from canceling, as they would in a histogram pooled over cells. A finite sample of $T$ ranks per cell gives even a perfect ensemble a positive floor $\mathrm{RI}_0 = (K-1)/(TK)$, which we report alongside the value. Tables~\ref{tab:loocv-metrics}, \ref{tab:loocv-miroc6-baselines}, \ref{tab:baselines} and \ref{tab:loocv-mpi} report it as \gls{ri}, per cell.

\paragraph*{Laplacian Roughness}
Small-scale checkerboard structure in a bias field indicates that the spatial coherence of the underlying maps has not been preserved. For the mean-bias field the discrete Laplacian at an interior cell is
$$
(\nabla^{2} B^{\mu})_{j,k} = B^{\mu}_{j+1,k} + B^{\mu}_{j-1,k} + B^{\mu}_{j,k+1} + B^{\mu}_{j,k-1} - 4 B^{\mu}_{j,k} ,
$$
with the longitude index wrapping around the globe. The reported scalar is the area-weighted mean of its absolute value over the cells $\mathcal{C}$ of every latitude row except the northernmost and the southernmost, where the stencil is incomplete:

$$
\mathrm{LapRough} = \frac{\sum_{(j,k)\in\mathcal{C}} w_{j,k}\,|(\nabla^{2} B^{\mu})_{j,k}|}{\sum_{(j,k)\in\mathcal{C}} w_{j,k}} .
$$
The bias field is averaged over $(y,d,s,m)$ before the Laplacian is applied, so this measures systematic small-scale structure in the climatological bias rather than in any single sample. Tables~\ref{tab:loocv-metrics}, \ref{tab:loocv-miroc6-baselines}, \ref{tab:baselines} and \ref{tab:loocv-mpi} report it.

\subsection*{Spatial-Consistency Metrics}
A per-cell correction can repair every cell's marginal distribution and still leave the relation between neighboring cells wrong. The two quantities of Table~\ref{tab:spatial-consistency} test that relation inside a regional box, the first at the pair of north--south adjacent cells whose \gls{mvt} mean biases differ most. Write $x_A$ and $x_B$ for the daily temperatures at the pair's southern and northern cell, over all sampled days and members.

\paragraph*{Displacement}
For \gls{era5} we compute the \emph{principal axis} of the joint distribution of $(x_A^{\mathrm{ERA5}}, x_B^{\mathrm{ERA5}})$ over all sampled days: the line through its mean along the leading eigenvector of the $2\times2$ covariance matrix. This line is the reference, rather than the diagonal $x_B = x_A$, because the two cells have different climates and a perfect emulator's points would lie along it. For each method we then compute the signed perpendicular distance from that line to the mean of its own joint distribution,
$$
D \;=\; \bigl(\overline{\mathbf{x}} - \overline{\mathbf{x}}^{\mathrm{ERA5}}\bigr)\cdot \mathbf{n} ,
\qquad \mathbf{n} \propto (-\hat{m},\, 1),
$$
with $\overline{\mathbf{x}} = (\overline{x}_A, \overline{x}_B)$ and $\overline{\mathbf{x}}^{\mathrm{ERA5}}$ the means of the joint distributions of the method and of \gls{era5}, $\hat{m}$ the slope of the principal axis and $\mathbf{n}$ the unit normal to it. $D = 0$ means the method reproduces the observed relation between the two cells; a negative $D$ means its points sit below that relation, i.e.\ the contrast between the cells is too weak. Table~\ref{tab:spatial-consistency} reports $|D|$ as the displacement.

\paragraph*{Roughness Ratio}
The mean absolute five-point Laplacian of a \emph{daily} field, over the interior cells of the regional box,
$$
\rho \;=\; \frac{1}{|\mathcal{C}|}\sum_{(j,k)\in\mathcal{C}} \bigl| (\nabla^2 x)_{j,k} \bigr| ,
$$
averaged over days and members and divided by the same quantity for \gls{era5}. A value of $1$ matches the observed spatial texture; below $1$ the fields are too smooth. This differs from the Laplacian roughness defined earlier, which is applied to the \emph{climatological bias field} rather than to individual daily fields. Table~\ref{tab:spatial-consistency} reports it.

\subsection*{Projection and Extreme-Heat Metrics}

These quantities describe the projections of the main text. Except where stated, they are not scored against a held-out truth.

\paragraph*{Retained Inter-Model Spread}
This metric measures how much of the disagreement between the climate models about late-century warming is preserved by guidance toward \gls{era5}. For each scenario, write $w_{y,s,m}$ for the warming above 1850--1900 (\emph{Warming-Level Crossing Year}) in year $y$ of sample $s$ of the projection guided from source model $m$, after the annual-variance rescaling (\emph{Materials and Methods}, \emph{Annual-Variance Rescaling}; $N_s = 10$, $N_m = 17$), and $w^{\mathrm{raw}}_{y,m}$ for that of raw \gls{cmip6} model $m$, both as drawn in Fig.~\ref{fig:projections}\emph{A} of the main text. Each model is reduced to its mean warming over 2081--2100, averaged over its samples,
$$
W_m = \frac{1}{20\,N_s}\sum_{y=2081}^{2100}\sum_{s=1}^{N_s} w_{y,s,m}, \qquad W^{\mathrm{raw}}_m = \frac{1}{20}\sum_{y=2081}^{2100} w^{\mathrm{raw}}_{y,m} .
$$
The 5--95\% inter-model range $R$ is the difference between the empirical 95th and 5th percentiles of the seventeen values $W_m$, and $R^{\mathrm{raw}}$ is the same range of the $W^{\mathrm{raw}}_m$. The retained inter-model spread is the fraction $R/R^{\mathrm{raw}}$. Averaging over ten samples and twenty years removes nearly all year-to-year variability, so $R$ measures the spread between models. Table~\ref{tab:retention} reports $R^{\mathrm{raw}}$, $R$ and the retained inter-model spread.

\paragraph*{Exceedance-Probability Ratio}
Summer is June--August north of the equator and December--February south of it. At a grid cell, every daily summer temperature of a dataset over 1995--2014 is pooled across members, samples, days and years, and $q_{\alpha}$ is the $\alpha$ quantile of this pool. With $\hat{P}_{w}(x > q_{\alpha})$ the fraction of pooled values above $q_{\alpha}$ in window $w$,
$$
R_{\alpha} \;=\; \frac{\hat{P}_{2081\text{--}2100}(x > q_{\alpha})}{\hat{P}_{1995\text{--}2014}(x > q_{\alpha})} .
$$
Each dataset is referred to its own $q_{\alpha}$, so its 1995--2014 exceedance is $1-\alpha$ and $R_{\alpha}$ cannot exceed $1/(1-\alpha)$, that is $10$ at the 90th percentile and $100$ at the 99th. A constant offset added to every value of a dataset leaves $R_{\alpha}$ unchanged. Fig.~\ref{fig:distributions}\emph{B} of the main text shows it.

\paragraph*{Temperature Exceeded Once in $N$ Days}
The $n$ pooled daily values of a dataset and window are sorted in decreasing order, $x_{(1)} \ge x_{(2)} \ge \cdots \ge x_{(n)}$. The value of rank $i$ is exceeded on average once in $N_i = (n+1)/i$ days, and intermediate values of $N$ are interpolated linearly in $\log N$. A curve therefore ends at its hottest day, $N = n+1$, and each level up to $N = (n+1)/10$ is reached on at least ten sampled days. The emulator draws each day independently, whereas real summer temperatures persist from day to day, so $N$ counts days, not a return period in years. Fig.~\ref{fig:distributions}\emph{C} of the main text shows it.

\paragraph*{Area of Contiguous Extreme Heat}
A land cell is extreme on a given day when its daily-mean temperature exceeds the 95th percentile of \gls{era5} at that cell over 1979--2024, computed from all days within five days of the same calendar day. Extreme cells that share an edge form one region, with the grid wrapped across the date line. The daily statistic is the total area of all regions of at least two cells, summed from the true cell areas; the land surface of the model grid is $146.4$ million km$^2$. The statistic is computed on every daily map, one per member and day, and the value of a decade is the median over all of its maps: $3{,}650$ for \gls{era5}, and $62{,}050$ for the guided emulator and for raw \gls{cmip6}, each with seventeen members (\emph{Sampling Designs}). Fig.~\ref{fig:heat-region} of the main text shows it.

\paragraph*{Warming-Level Crossing Year}
We express every series in Fig.~\ref{fig:projections} of the main text as warming above 1850--1900: we reference it to its own 1995--2014 mean and add the warming of $0.85\,\degC$ from 1850--1900 to 1995--2014 assessed by \gls{ipcc} \gls{ar6-wg1}~\cite{si:ipcc-2021-wg1-ar6}. The guided projections start in 2025, so we reference them to the 1995--2014 mean of the emulator that replicates the target ($\gamma = 1$) before 2025. The year in which the guided ensemble mean crosses $1.5$ or $2\,\degC$ is the midpoint of the first 20-year period whose mean warming exceeds the level~\cite{si:ipcc-2021-wg1-ar6}, rounded down to a calendar year. Table~\ref{tab:gwl-crossings} reports the crossing year and its 5--95\% range.

\subsection*{The Criterion Used to Select the Guidance Scale}
We select the guidance scale that minimizes the bias of the tails of the daily distribution. For a given fold, scenario and block (\emph{Materials and Methods}, \emph{Selecting the Guidance Strength}), we compute at every grid cell the 1st and 99th percentiles of daily temperature for the held-out model and, separately, for each of the $N_s N_m$ samples of the ensemble. The ensemble percentile is the mean of the sample percentiles, unlike the pooled percentiles of \emph{Percentile Biases}. The bias maps $B^{q_{0.01}}(\gamma)$ and $B^{q_{0.99}}(\gamma)$ are the ensemble percentiles minus those of the held-out model. We then select
$$
\gamma^{\star} \;=\; \arg\min_{\gamma}\ \tfrac{1}{2}\Bigl(\bigl\langle |B^{q_{0.01}}(\gamma)| \bigr\rangle + \bigl\langle |B^{q_{0.99}}(\gamma)| \bigr\rangle\Bigr)
$$
over the nine sampled scales (\emph{Sampling Designs}, \emph{The Scale Sweep}).

\section*{Fidelity of the Unguided Emulator}

Before assessing guidance, we check the two endpoints it interpolates between: the emulator without guidance, conditioned on each of the eighteen model identities in turn (\emph{Sampling Designs}, \emph{The Emulator Conditioned on Each Model Identity}). The $\gamma = 0$ endpoint is the emulator conditioned on one of the seventeen \gls{cmip6} models, evaluated over 2080--2100 under \glsentryshort{ssp370}, as a representative case for the long-term future. The $\gamma = 1$ endpoint is the emulator conditioned on \gls{era5}, evaluated over 2001--2010, the decade withheld from its observational training record. Figures~\ref{fig:si-prior-mean-370}--\ref{fig:si-prior-q99-370} map the biases of both endpoints, and Figs.~\ref{fig:si-prior-rank-370}--\ref{fig:si-prior-cycle-370} show the rank histograms, density, and seasonal cycle of the daily global-mean temperature.

Conditioned on a \gls{cmip6} model, the emulator reproduces that model's daily temperature distribution closely: under SSP3-7.0 the mean, 1st-percentile and 99th-percentile biases are small for all seventeen models (Figs.~\ref{fig:si-prior-mean-370}--\ref{fig:si-prior-q99-370}, panels \emph{A}--\emph{Q}). The observational endpoint (panel \emph{R}), evaluated on a decade withheld from training, has larger residuals than most model endpoints.

\section*{Perfect-Model Validation with a Scheduled Guidance Strength} This section details the perfect-model evaluation of the guidance schedule summarized in the \emph{Results} of the main text: first in all seventeen folds, then in the folds with MIROC6 and MPI-ESM1-2-LR held out (\emph{Sampling Designs} names the design of each table and figure).

\paragraph*{All Seventeen Folds under SSP3-7.0}
Guidance improves the calibration of the emulated ensemble against every held-out model, but for several of them the guided ensemble has a bias of one sign over most of the globe. Figures~\ref{fig:si-folds-mean-370}--\ref{fig:si-folds-rank-370} show each of the seventeen folds under SSP3-7.0 over 2080--2100, without and with guidance, with the samples of the scale sweep. Guidance lowers the per-cell reliability index for every held-out model, most for CanESM5 and BCC-CSM2-MR, the two with the highest index without guidance (Fig.~\ref{fig:si-folds-rank-370}). Over most of the globe the guided ensemble is warmer than some held-out models, such as FGOALS-g3 and the two NorESM2 models, and colder than others, such as MPI-ESM1-2-LR, CMCC-ESM2 and EC-Earth3-Veg (Figs.~\ref{fig:si-folds-mean-370} and \ref{fig:si-folds-kde-370}).

\begin{table}[p]
\centering
\caption{\textbf{Perfect-model skill for each of the seventeen \glsentryshort{loocv} iterations.} For the unguided and guided case, the trajectory RMSE and the band coverage over 2025--2100. The row of each held-out model gives the mean over the three scenarios. The final rows give the median over the seventeen held-out models for each scenario. Bold marks a guided trajectory RMSE smaller than the unguided one. \emph{Unguided} is the emulator conditioned on the sixteen source models with no guidance. \emph{Guided} uses the fitted $\gamma$ schedule rounded to the nearest of the nine sampled scales (\emph{Sampling Designs}). The metrics are defined in \emph{Evaluation Metrics}, \emph{Trajectory Metrics}.}\label{tab:loocv-skill}
\footnotesize
\setlength{\tabcolsep}{4pt}
\begin{tabular}{lrrrr}
Held-out model & \multicolumn{2}{c}{Trajectory RMSE [$\degC$]} & \multicolumn{2}{c}{Band coverage} \\
\cmidrule(lr){2-3}\cmidrule(lr){4-5}
 & unguided & guided & unguided & guided \\
\midrule
AWI-CM-1-1-MR & 0.16 & 0.35 & 1.00 & 0.44 \\
BCC-CSM2-MR & 0.70 & \textbf{0.19} & 0.71 & 0.65 \\
CanESM5 & 1.18 & \textbf{0.24} & 0.32 & 0.56 \\
CESM2-WACCM & 0.41 & \textbf{0.19} & 0.98 & 0.70 \\
CMCC-CM2-SR5 & 0.64 & \textbf{0.27} & 0.89 & 0.67 \\
CMCC-ESM2 & 0.50 & 0.65 & 0.95 & 0.18 \\
EC-Earth3 & 0.43 & 0.46 & 1.00 & 0.31 \\
EC-Earth3-Veg & 0.42 & \textbf{0.28} & 0.99 & 0.62 \\
EC-Earth3-Veg-LR & 0.32 & \textbf{0.20} & 1.00 & 0.67 \\
FGOALS-g3 & 1.51 & \textbf{0.80} & 0.00 & 0.17 \\
IPSL-CM6A-LR & 0.54 & \textbf{0.18} & 0.89 & 0.83 \\
MIROC6 & 0.83 & \textbf{0.24} & 0.53 & 0.64 \\
MPI-ESM1-2-HR & 0.39 & 0.39 & 1.00 & 0.24 \\
MPI-ESM1-2-LR & 0.82 & \textbf{0.35} & 0.84 & 0.37 \\
MRI-ESM2-0 & 0.28 & \textbf{0.13} & 1.00 & 0.87 \\
NorESM2-LM & 0.29 & 0.54 & 1.00 & 0.23 \\
NorESM2-MM & 0.78 & \textbf{0.58} & 0.90 & 0.26 \\
\midrule
Median, SSP2-4.5 & 0.56 & \textbf{0.24} & 0.92 & 0.58 \\
Median, SSP3-7.0 & 0.47 & \textbf{0.28} & 1.00 & 0.54 \\
Median, SSP5-8.5 & 0.56 & \textbf{0.38} & 0.95 & 0.50 \\
\bottomrule
\end{tabular}

\end{table}

\begin{sidewaystable}
\centering
\caption{\textbf{Full distributional evaluation for each of the seventeen LOOCV iterations.} Every metric is computed at each grid cell against the held-out model's own daily fields over 2080--2100 and reduced to one scalar by area weighting. The row of each held-out model gives the mean over the three scenarios. The median rows give the median over the seventeen held-out models for each scenario. Bold marks a guided value smaller than the unguided one. The RI floor is $0.0037$. \emph{Unguided} is the emulator conditioned on the sixteen source models with no guidance. \emph{Guided} uses the fitted $\gamma$ schedule rounded to the nearest of the nine sampled scales (\emph{Sampling Designs}). The metrics are defined in \emph{Evaluation Metrics}, \emph{Distributional Metrics}.}\label{tab:loocv-metrics}
\footnotesize
\setlength{\tabcolsep}{3.5pt}
\renewcommand{\arraystretch}{0.93}
\begin{tabular}{lrrrrrrrrrrrr}
Held-out model & \shortstack{RMSE\\{}[$\degC$]} & \shortstack{Standard-\\{}deviation ratio\\{}deviation} & \shortstack{Skewness\\{}bias} & \shortstack{Excess-kurtosis\\{}bias} & \shortstack{TX90p bias\\{}[percentage points]} & \shortstack{TN10p bias\\{}[percentage points]} & \shortstack{1st-percentile\\{}bias [$\degC$]} & \shortstack{99th-percentile\\{}bias [$\degC$]} & \shortstack{CRPS\\{}[$\degC$]} & \shortstack{RI,\\{}per cell} & \shortstack{Laplacian\\{}roughness [$\degC$]} & \shortstack{Wasserstein\\{}distance [$\degC$]} \\
\midrule
\multicolumn{13}{l}{\textit{Unguided ($\gamma = 0$)}} \\
AWI-CM-1-1-MR & 1.128 & 0.196 & 0.209 & 0.504 & 8.60 & 6.16 & 1.749 & 2.032 & 1.223 & 0.0266 & 1.159 & 1.018 \\
BCC-CSM2-MR & 1.657 & 0.408 & 0.286 & 0.757 & 7.61 & 20.75 & 3.189 & 2.075 & 1.487 & 0.0913 & 1.008 & 1.578 \\
CanESM5 & 2.466 & 0.242 & 0.203 & 0.555 & 7.65 & 24.99 & 3.597 & 1.696 & 1.821 & 0.1575 & 1.244 & 2.096 \\
CESM2-WACCM & 1.146 & 0.227 & 0.231 & 0.502 & 4.77 & 12.06 & 2.252 & 1.608 & 1.234 & 0.0347 & 0.614 & 1.090 \\
CMCC-CM2-SR5 & 1.650 & 0.196 & 0.223 & 0.535 & 5.18 & 9.31 & 2.451 & 1.391 & 1.294 & 0.0301 & 0.748 & 1.221 \\
CMCC-ESM2 & 1.457 & 0.259 & 0.227 & 0.520 & 4.89 & 9.88 & 2.535 & 1.564 & 1.226 & 0.0263 & 0.753 & 1.132 \\
EC-Earth3 & 1.656 & 0.158 & 0.198 & 0.599 & 8.34 & 8.81 & 2.185 & 1.504 & 1.372 & 0.0353 & 0.980 & 1.311 \\
EC-Earth3-Veg & 1.706 & 0.160 & 0.196 & 0.594 & 8.05 & 9.22 & 2.235 & 1.494 & 1.377 & 0.0378 & 0.970 & 1.326 \\
EC-Earth3-Veg-LR & 1.654 & 0.154 & 0.198 & 0.605 & 12.36 & 8.61 & 1.831 & 1.843 & 1.454 & 0.0502 & 0.616 & 1.448 \\
FGOALS-g3 & 2.997 & 0.333 & 0.293 & 0.525 & 18.60 & 6.45 & 2.120 & 3.665 & 1.805 & 0.0684 & 1.016 & 2.077 \\
IPSL-CM6A-LR & 1.588 & 0.199 & 0.259 & 0.562 & 6.40 & 6.35 & 1.674 & 1.939 & 1.328 & 0.0261 & 0.855 & 1.170 \\
MIROC6 & 2.031 & 0.350 & 0.222 & 0.591 & 13.39 & 9.49 & 2.638 & 2.462 & 1.476 & 0.0577 & 1.293 & 1.595 \\
MPI-ESM1-2-HR & 1.163 & 0.204 & 0.217 & 0.541 & 8.42 & 6.50 & 1.560 & 2.350 & 1.219 & 0.0278 & 1.129 & 1.066 \\
MPI-ESM1-2-LR & 1.491 & 0.238 & 0.224 & 0.577 & 14.75 & 5.18 & 1.709 & 2.941 & 1.316 & 0.0437 & 1.214 & 1.348 \\
MRI-ESM2-0 & 1.151 & 0.229 & 0.221 & 0.508 & 7.38 & 7.79 & 1.578 & 2.113 & 1.264 & 0.0283 & 0.774 & 1.057 \\
NorESM2-LM & 1.347 & 0.218 & 0.233 & 0.475 & 8.37 & 6.06 & 1.532 & 2.238 & 1.275 & 0.0249 & 1.124 & 1.076 \\
NorESM2-MM & 1.528 & 0.233 & 0.240 & 0.456 & 12.68 & 4.68 & 1.148 & 2.691 & 1.332 & 0.0309 & 0.622 & 1.202 \\
\addlinespace
Median, SSP2-4.5 & 1.590 & 0.221 & 0.222 & 0.549 & 8.02 & 8.22 & 2.025 & 1.966 & 1.344 & 0.0346 & 0.998 & 1.284 \\
Median, SSP3-7.0 & 1.507 & 0.225 & 0.223 & 0.531 & 7.83 & 7.85 & 2.069 & 1.938 & 1.311 & 0.0320 & 0.975 & 1.231 \\
Median, SSP5-8.5 & 1.609 & 0.231 & 0.226 & 0.542 & 7.71 & 8.63 & 2.114 & 2.152 & 1.339 & 0.0359 & 0.966 & 1.295 \\
\midrule
\multicolumn{13}{l}{\textit{Guided}} \\
AWI-CM-1-1-MR & \textbf{0.768} & \textbf{0.070} & \textbf{0.133} & \textbf{0.316} & \textbf{7.77} & \textbf{4.02} & \textbf{0.776} & \textbf{1.014} & \textbf{1.133} & \textbf{0.0239} & \textbf{0.470} & \textbf{0.705} \\
BCC-CSM2-MR & \textbf{0.694} & \textbf{0.099} & \textbf{0.173} & \textbf{0.370} & \textbf{4.52} & \textbf{5.65} & \textbf{1.100} & \textbf{0.774} & \textbf{1.011} & \textbf{0.0198} & \textbf{0.455} & \textbf{0.589} \\
CanESM5 & \textbf{0.823} & \textbf{0.116} & \textbf{0.153} & \textbf{0.347} & \textbf{5.15} & \textbf{5.16} & \textbf{1.244} & \textbf{0.902} & \textbf{1.101} & \textbf{0.0150} & \textbf{0.588} & \textbf{0.634} \\
CESM2-WACCM & \textbf{0.782} & \textbf{0.067} & \textbf{0.127} & \textbf{0.255} & 4.84 & \textbf{4.42} & \textbf{0.871} & \textbf{0.814} & \textbf{1.091} & \textbf{0.0168} & \textbf{0.319} & \textbf{0.627} \\
CMCC-CM2-SR5 & \textbf{0.667} & \textbf{0.055} & \textbf{0.111} & \textbf{0.258} & \textbf{3.43} & \textbf{3.39} & \textbf{0.846} & \textbf{0.594} & \textbf{1.040} & \textbf{0.0122} & \textbf{0.283} & \textbf{0.533} \\
CMCC-ESM2 & \textbf{1.061} & \textbf{0.071} & \textbf{0.118} & \textbf{0.253} & \textbf{4.35} & \textbf{8.09} & \textbf{1.501} & \textbf{0.592} & \textbf{1.111} & 0.0268 & \textbf{0.298} & \textbf{0.837} \\
EC-Earth3 & \textbf{0.927} & \textbf{0.052} & \textbf{0.116} & \textbf{0.220} & \textbf{5.74} & \textbf{4.11} & \textbf{0.909} & \textbf{0.827} & \textbf{1.129} & \textbf{0.0180} & \textbf{0.307} & \textbf{0.739} \\
EC-Earth3-Veg & \textbf{0.587} & \textbf{0.046} & \textbf{0.117} & \textbf{0.279} & \textbf{3.21} & \textbf{3.17} & \textbf{0.858} & \textbf{0.576} & \textbf{1.049} & \textbf{0.0093} & \textbf{0.271} & \textbf{0.490} \\
EC-Earth3-Veg-LR & \textbf{0.520} & \textbf{0.058} & \textbf{0.115} & \textbf{0.214} & \textbf{4.12} & \textbf{3.37} & \textbf{0.727} & \textbf{0.656} & \textbf{1.091} & \textbf{0.0112} & \textbf{0.339} & \textbf{0.514} \\
FGOALS-g3 & \textbf{1.811} & \textbf{0.112} & \textbf{0.159} & \textbf{0.284} & \textbf{12.20} & \textbf{5.28} & \textbf{1.546} & \textbf{1.609} & \textbf{1.388} & \textbf{0.0503} & \textbf{0.499} & \textbf{1.287} \\
IPSL-CM6A-LR & \textbf{0.861} & \textbf{0.086} & \textbf{0.146} & \textbf{0.317} & \textbf{4.40} & \textbf{4.78} & \textbf{0.986} & \textbf{0.780} & \textbf{1.142} & \textbf{0.0169} & \textbf{0.402} & \textbf{0.696} \\
MIROC6 & \textbf{0.644} & \textbf{0.095} & \textbf{0.129} & \textbf{0.253} & \textbf{5.18} & \textbf{3.16} & \textbf{0.826} & \textbf{0.962} & \textbf{1.028} & \textbf{0.0110} & \textbf{0.524} & \textbf{0.553} \\
MPI-ESM1-2-HR & \textbf{0.688} & \textbf{0.082} & \textbf{0.134} & \textbf{0.293} & \textbf{6.61} & \textbf{3.16} & \textbf{0.752} & \textbf{1.124} & \textbf{1.084} & \textbf{0.0151} & \textbf{0.470} & \textbf{0.630} \\
MPI-ESM1-2-LR & \textbf{0.767} & \textbf{0.075} & \textbf{0.138} & \textbf{0.437} & \textbf{3.58} & 5.45 & \textbf{1.233} & \textbf{0.583} & \textbf{1.044} & \textbf{0.0129} & \textbf{0.419} & \textbf{0.618} \\
MRI-ESM2-0 & \textbf{0.722} & \textbf{0.078} & \textbf{0.135} & \textbf{0.287} & \textbf{3.85} & \textbf{5.94} & \textbf{1.015} & \textbf{0.733} & \textbf{1.134} & \textbf{0.0173} & \textbf{0.374} & \textbf{0.617} \\
NorESM2-LM & \textbf{1.001} & \textbf{0.076} & \textbf{0.142} & \textbf{0.268} & \textbf{7.07} & \textbf{4.43} & \textbf{1.039} & \textbf{1.017} & \textbf{1.183} & \textbf{0.0212} & \textbf{0.343} & \textbf{0.793} \\
NorESM2-MM & \textbf{0.970} & \textbf{0.076} & \textbf{0.147} & \textbf{0.294} & \textbf{9.49} & \textbf{4.41} & \textbf{0.980} & \textbf{1.084} & \textbf{1.205} & \textbf{0.0268} & \textbf{0.347} & \textbf{0.824} \\
\addlinespace
Median, SSP2-4.5 & \textbf{0.688} & \textbf{0.066} & \textbf{0.132} & \textbf{0.282} & \textbf{4.40} & \textbf{3.59} & \textbf{0.879} & \textbf{0.738} & \textbf{1.099} & \textbf{0.0133} & \textbf{0.340} & \textbf{0.567} \\
Median, SSP3-7.0 & \textbf{0.781} & \textbf{0.076} & \textbf{0.132} & \textbf{0.277} & \textbf{4.51} & \textbf{4.46} & \textbf{0.927} & \textbf{0.818} & \textbf{1.092} & \textbf{0.0163} & \textbf{0.375} & \textbf{0.628} \\
Median, SSP5-8.5 & \textbf{0.878} & \textbf{0.083} & \textbf{0.136} & \textbf{0.290} & \textbf{4.76} & \textbf{5.16} & \textbf{1.007} & \textbf{0.831} & \textbf{1.106} & \textbf{0.0185} & \textbf{0.406} & \textbf{0.696} \\
\bottomrule
\end{tabular}
\end{sidewaystable}

\paragraph*{Comparison with Two Bias Corrections and Classifier Guidance beyond the Calibration Period}
Guidance reproduces the daily temperature fields of the held-out MIROC6 over 2081--2100 more closely than two statistical bias corrections, \gls{mvt} \citep{si:xu-2021-mvt} and \gls{qdm} \citep{si:cannon-2015-qdm} (Table~\ref{tab:loocv-miroc6-baselines} and Figs.~\ref{fig:si-m6-maps-raw-370}--\ref{fig:si-m6-cycle-diff}). We calibrate both corrections toward MIROC6 over 1979--2000 and 2011--2024 and apply them unchanged to 2081--2100 (\emph{Baseline Methods}). We apply them to two inputs. The first, the raw output of the sixteen source \gls{cmip6} models, reproduces the pipeline that guidance would replace. The second, the output of the emulator conditioned on the same sixteen models without guidance ($\gamma = 0$), forms an ablation that keeps the emulator fixed and changes only how its output is adjusted toward MIROC6. Every ensemble has sixteen members, one per source model (\emph{Sampling Designs}).

Under all three \glspl{ssp} and with either input, guidance has the smallest error on eight metrics: the mean absolute bias, the \gls{rmse}, the 1st-percentile bias, the TX90p and TN10p biases, \gls{crps}, the reliability index and the Wasserstein distance. The guided rank histograms are also the closest to uniform (Figs.~\ref{fig:si-m6-rank-raw} and \ref{fig:si-m6-rank-diff}). Both corrections produce fields warmer than MIROC6 in every calendar month (Figs.~\ref{fig:si-m6-cycle-raw} and \ref{fig:si-m6-cycle-diff}), and in their rank histograms MIROC6 falls below all sixteen members more often than at any other rank.

The other five metrics favor one or both corrections in some of the six combinations of scenario and input. \Gls{mvt} has a smaller standard-deviation ratio deviation and a smaller skewness bias than guidance in all six. \Gls{mvt} applied to the emulator output also has a smaller 99th-percentile bias under all three scenarios, and \gls{qdm} applied to the emulator output has a smaller excess-kurtosis bias under SSP2-4.5. Both corrections, with either input, have a smaller Laplacian roughness than guidance under SSP3-7.0 and SSP5-8.5 and similar values under SSP2-4.5. This metric favors guidance against \gls{era5} over 2001--2010 (Table~\ref{tab:baselines}) but no longer beyond the calibration period. Applying the corrections to the emulator output instead of to raw \gls{cmip6} lowers most of their errors slightly.

Classifier guidance \citep{si:dhariwal-nichol-2021-diffusion-beats-gans} steers a diffusion model toward a target with the gradient of a separately trained classifier, and we compare it with condition-to-condition guidance in the same setting (\emph{Baseline Methods}, \emph{Classifier Guidance}). Like condition-to-condition guidance, it is end-to-end: a diffusion model trained on the raw \gls{cmip6} models generates the corrected fields directly. We train a second dedicated diffusion emulator on the sixteen source models over 1979--2100, without MIROC6. We also train a classifier that recognizes MIROC6 among the \gls{cmip6} models, using MIROC6 data only from 1979--2000 and 2011--2024. During sampling, we condition the trained emulator on each source model and steer it toward MIROC6 with the gradient of this classifier. We use a classifier scale of $2.5$, applied only at diffusion noise levels below $10$. Classifier guidance therefore uses MIROC6 over the same years as the corrections and our emulator. It differs from condition-to-condition guidance in the training set of the emulator and in the guidance mechanism, so the comparison measures the two methods as a whole, each with its own emulator. It appears in the rightmost column of Figs.~\ref{fig:si-m6-maps-raw-370}, \ref{fig:si-m6-rank-raw} and \ref{fig:si-m6-kde-raw} and as a separate curve in Fig.~\ref{fig:si-m6-cycle-raw}.

Condition-to-condition guidance has a smaller trajectory \gls{rmse} than classifier guidance and a smaller error on all thirteen metrics of the daily fields under all three scenarios (Table~\ref{tab:loocv-miroc6-baselines}). \Gls{qdm}, applied to raw \gls{cmip6} or to the emulator output, also has a smaller error than classifier guidance on every metric of the daily fields except the TX90p bias under SSP3-7.0 and SSP5-8.5. Unlike the two corrections, classifier guidance produces fields colder than MIROC6, in every calendar month under SSP2-4.5 and SSP3-7.0 and in most months under SSP5-8.5 (Fig.~\ref{fig:si-m6-cycle-raw}).

\paragraph*{Held-Out MPI-ESM1-2-LR}
Guidance does not reproduce every held-out model equally well. We repeat the scheduled run with MPI-ESM1-2-LR as the synthetic observation, with the same design and the same $\gamma$ schedule (Table~\ref{tab:loocv-mpi} and Figs.~\ref{fig:si-mpi-warming}--\ref{fig:si-mpi-rank}). Until the middle of the century the guided ensemble follows MPI-ESM1-2-LR closely: over 2030--2050, its distribution and seasonal cycle of the daily global-mean temperature match those of MPI-ESM1-2-LR, and its rank histograms are close to uniform (Figs.~\ref{fig:si-mpi-kde}--\ref{fig:si-mpi-rank}). Later it warms less than MPI-ESM1-2-LR (Fig.~\ref{fig:si-mpi-warming}). Over 2080--2100 it is colder than MPI-ESM1-2-LR in every calendar month, and MPI-ESM1-2-LR ranks high among its members, most of all under SSP3-7.0 and SSP5-8.5 (Figs.~\ref{fig:si-mpi-cycle} and \ref{fig:si-mpi-rank}). Under SSP3-7.0 the guided ensemble is colder than MPI-ESM1-2-LR over most of the globe, whereas the raw \gls{cmip6} ensemble is warmer there (Fig.~\ref{fig:si-mpi-maps-370}). Its trajectory \gls{rmse} stays below that of the raw ensemble mean under all three scenarios, but its 5--95\% band contains MPI-ESM1-2-LR in only 17--38\% of the years (Table~\ref{tab:loocv-mpi}). In the scale sweep, averaged over the three scenarios, guidance also lowers the trajectory \gls{rmse} of this fold relative to the unguided emulator, while its band coverage falls (Table~\ref{tab:loocv-skill}). Over 2081--2100 the guided ensemble has a smaller error than raw \gls{cmip6} on every metric of the daily fields except the TN10p bias under SSP3-7.0 and SSP5-8.5 (Table~\ref{tab:loocv-mpi}).

\begin{table}[p]
\centering
\footnotesize
\renewcommand{\arraystretch}{0.93}
\caption{\textbf{Held-out MIROC6: condition-to-condition guidance against MVT, QDM, and classifier guidance.} Upper table: annual global-mean warming over 2025--2100 (complement to Fig.~\ref{fig:projections}\emph{B} of the main text), with trajectory RMSE and band coverage. Lower table: daily fields over 2081--2100, each metric computed at each grid cell against MIROC6's own daily fields and reduced to one scalar by area weighting. \emph{Raw CMIP6} is the ensemble of the sixteen source simulations. \emph{Emulator $\gamma=0$} is the emulator conditioned on the same sixteen source models without guidance. \emph{Guided} is the emulator guided from the same sixteen models toward MIROC6 with the $\gamma$ schedule. MVT and QDM are applied either to the raw CMIP6 data (\emph{on raw CMIP6}) or to the output of the emulator without guidance (\emph{on emulator output}). \emph{Classifier guidance} is the second emulator described in the text. In both tables each column has sixteen members, one per source model, with one sample per member and day, and the lower table pools the members. The bands of the emulator without guidance, of the guided emulator and of classifier guidance are rescaled (\emph{Materials and Methods}), and those of raw CMIP6, MVT and QDM are not. The RI floor is $0.0001$. The smallest value of each metric except the band coverage is in bold. The metrics are defined in \emph{Evaluation Metrics}, \emph{Trajectory Metrics} and \emph{Distributional Metrics}.}\label{tab:loocv-miroc6-baselines}
\begin{tabular}{llrrrrrrrr}
 & & & & & \multicolumn{2}{c}{on raw CMIP6} & \multicolumn{2}{c}{on emulator output} & Classifier \\
\cmidrule(lr){6-7}\cmidrule(lr){8-9}
Metric & Scenario & Raw CMIP6 & Emulator $\gamma=0$ & Guided & MVT & QDM & MVT & QDM & guidance \\
\midrule
\multicolumn{10}{l}{\textit{Annual global-mean warming, 2025--2100}} \\
Trajectory RMSE [$\degC$] & SSP2-4.5 & 0.363 & 0.368 & \textbf{0.214} & 0.311 & 0.305 & 0.303 & 0.313 & 0.514 \\
 & SSP3-7.0 & 0.467 & 0.446 & \textbf{0.200} & 0.431 & 0.408 & 0.400 & 0.393 & 0.459 \\
 & SSP5-8.5 & 0.532 & 0.511 & \textbf{0.201} & 0.490 & 0.466 & 0.465 & 0.456 & 0.360 \\
\addlinespace[1pt]
Band coverage & SSP2-4.5 & 0.82 & 0.78 & 0.62 & 0.47 & 0.87 & 0.05 & 0.71 & 0.24 \\
 & SSP3-7.0 & 0.78 & 0.79 & 0.72 & 0.41 & 0.83 & 0.03 & 0.72 & 0.25 \\
 & SSP5-8.5 & 0.83 & 0.76 & 0.67 & 0.29 & 0.89 & 0.05 & 0.72 & 0.39 \\
\midrule
\multicolumn{10}{l}{\textit{Daily fields, 2081--2100}} \\
Mean absolute bias [$\degC$] & SSP2-4.5 & 1.388 & 1.394 & \textbf{0.373} & 0.619 & 0.599 & 0.594 & 0.560 & 0.824 \\
 & SSP3-7.0 & 1.363 & 1.372 & \textbf{0.412} & 0.838 & 0.770 & 0.754 & 0.703 & 0.824 \\
 & SSP5-8.5 & 1.404 & 1.426 & \textbf{0.515} & 0.947 & 0.867 & 0.866 & 0.797 & 0.900 \\
\addlinespace[1pt]
RMSE [$\degC$] & SSP2-4.5 & 2.075 & 2.071 & \textbf{0.593} & 0.810 & 0.794 & 0.791 & 0.757 & 1.225 \\
 & SSP3-7.0 & 1.956 & 1.973 & \textbf{0.586} & 1.001 & 0.933 & 0.930 & 0.871 & 1.203 \\
 & SSP5-8.5 & 1.978 & 2.010 & \textbf{0.730} & 1.157 & 1.082 & 1.097 & 1.024 & 1.281 \\
\addlinespace[1pt]
Standard-deviation ratio deviation & SSP2-4.5 & 0.331 & 0.339 & 0.075 & \textbf{0.052} & 0.059 & 0.053 & 0.057 & 0.177 \\
 & SSP3-7.0 & 0.341 & 0.343 & 0.092 & 0.072 & 0.087 & \textbf{0.065} & 0.083 & 0.168 \\
 & SSP5-8.5 & 0.349 & 0.355 & 0.108 & 0.081 & 0.123 & \textbf{0.071} & 0.114 & 0.168 \\
\addlinespace[1pt]
Skewness bias & SSP2-4.5 & 0.220 & 0.214 & 0.107 & 0.095 & 0.103 & 0.091 & \textbf{0.090} & 0.196 \\
 & SSP3-7.0 & 0.210 & 0.202 & 0.110 & 0.103 & 0.114 & \textbf{0.101} & 0.106 & 0.191 \\
 & SSP5-8.5 & 0.213 & 0.212 & 0.121 & \textbf{0.103} & 0.143 & \textbf{0.103} & 0.136 & 0.187 \\
\addlinespace[1pt]
Excess-kurtosis bias & SSP2-4.5 & 0.581 & 0.570 & 0.203 & 0.244 & 0.218 & 0.204 & \textbf{0.166} & 0.414 \\
 & SSP3-7.0 & 0.586 & 0.563 & \textbf{0.210} & 0.301 & 0.280 & 0.226 & 0.224 & 0.410 \\
 & SSP5-8.5 & 0.566 & 0.572 & \textbf{0.235} & 0.310 & 0.333 & 0.242 & 0.286 & 0.426 \\
\addlinespace[1pt]
TX90p bias [percentage points] & SSP2-4.5 & 12.41 & 12.54 & \textbf{4.24} & 8.42 & 8.08 & 8.04 & 7.56 & 9.28 \\
 & SSP3-7.0 & 13.74 & 13.64 & \textbf{5.31} & 12.51 & 11.32 & 11.21 & 10.27 & 9.90 \\
 & SSP5-8.5 & 14.44 & 14.26 & \textbf{6.06} & 14.07 & 12.76 & 12.86 & 11.71 & 11.31 \\
\addlinespace[1pt]
TN10p bias [percentage points] & SSP2-4.5 & 9.78 & 9.89 & \textbf{2.73} & 4.59 & 4.38 & 4.67 & 4.28 & 6.42 \\
 & SSP3-7.0 & 8.99 & 9.24 & \textbf{2.77} & 5.21 & 4.66 & 5.20 & 4.42 & 5.66 \\
 & SSP5-8.5 & 8.92 & 9.11 & \textbf{3.48} & 5.68 & 4.80 & 5.74 & 4.66 & 5.32 \\
\addlinespace[1pt]
1st-percentile bias [$\degC$] & SSP2-4.5 & 2.626 & 2.582 & \textbf{0.672} & 0.702 & 0.740 & 0.758 & 0.770 & 1.333 \\
 & SSP3-7.0 & 2.448 & 2.442 & \textbf{0.670} & 0.791 & 0.777 & 0.835 & 0.779 & 1.175 \\
 & SSP5-8.5 & 2.448 & 2.471 & \textbf{0.802} & 0.977 & 0.895 & 1.053 & 0.928 & 1.264 \\
\addlinespace[1pt]
99th-percentile bias [$\degC$] & SSP2-4.5 & 2.249 & 2.252 & 0.726 & 0.731 & 0.869 & \textbf{0.629} & 0.737 & 1.558 \\
 & SSP3-7.0 & 2.467 & 2.419 & 0.887 & 1.053 & 1.259 & \textbf{0.832} & 1.107 & 1.605 \\
 & SSP5-8.5 & 2.693 & 2.654 & 1.070 & 1.248 & 1.634 & \textbf{0.987} & 1.426 & 1.706 \\
\addlinespace[1pt]
CRPS [$\degC$] & SSP2-4.5 & 1.486 & 1.485 & \textbf{1.019} & 1.114 & 1.082 & 1.106 & 1.069 & 1.238 \\
 & SSP3-7.0 & 1.464 & 1.466 & \textbf{1.022} & 1.202 & 1.133 & 1.174 & 1.107 & 1.230 \\
 & SSP5-8.5 & 1.467 & 1.472 & \textbf{1.039} & 1.249 & 1.153 & 1.225 & 1.132 & 1.261 \\
\addlinespace[1pt]
RI, per cell & SSP2-4.5 & 0.0539 & 0.0528 & \textbf{0.0059} & 0.0362 & 0.0222 & 0.0370 & 0.0196 & 0.0329 \\
 & SSP3-7.0 & 0.0555 & 0.0545 & \textbf{0.0072} & 0.0621 & 0.0329 & 0.0590 & 0.0268 & 0.0351 \\
 & SSP5-8.5 & 0.0602 & 0.0586 & \textbf{0.0106} & 0.0819 & 0.0363 & 0.0795 & 0.0323 & 0.0457 \\
\addlinespace[1pt]
Laplacian roughness [$\degC$] & SSP2-4.5 & 1.253 & 1.285 & 0.461 & 0.460 & \textbf{0.455} & 0.463 & 0.457 & 1.159 \\
 & SSP3-7.0 & 1.248 & 1.281 & 0.517 & 0.485 & 0.471 & 0.489 & \textbf{0.470} & 1.170 \\
 & SSP5-8.5 & 1.270 & 1.300 & 0.590 & 0.530 & \textbf{0.512} & 0.535 & 0.513 & 1.193 \\
\addlinespace[1pt]
Wasserstein distance [$\degC$] & SSP2-4.5 & 1.559 & 1.571 & \textbf{0.466} & 0.668 & 0.650 & 0.647 & 0.615 & 0.961 \\
 & SSP3-7.0 & 1.544 & 1.554 & \textbf{0.500} & 0.881 & 0.818 & 0.802 & 0.754 & 0.947 \\
 & SSP5-8.5 & 1.595 & 1.611 & \textbf{0.605} & 0.996 & 0.928 & 0.921 & 0.860 & 1.030 \\
\bottomrule
\end{tabular}
\end{table}

\section*{Comparison with Established Bias Correction against ERA5}
The emulator replicating \gls{era5} ($\gamma = 1$), whose ensemble we also call guided (\emph{Materials and Methods}, \emph{Selecting the Guidance Strength}), is competitive with the two statistical corrections that guidance would replace, when all three are scored against observations. We compare it with \gls{mvt} and \gls{qdm} applied to the raw output of the seventeen \gls{cmip6} models (Table~\ref{tab:baselines} and Figs.~\ref{fig:si-baseline-maps}--\ref{fig:si-clouds-eur}). All four columns are scored against \gls{era5} \citep{si:hersbach-2020-era5} over 2001--2010, a decade that lies outside both the corrections' calibration windows (1979--2000 and 2011--2024) and the emulator's observational training years, on every day of the year and with seventeen members each (\emph{Sampling Designs}, \emph{The Emulator Replicating ERA5}). This historical decade is the only period in which \gls{era5} allows the comparison. The held-out MIROC6 comparison above scores the same corrections beyond their calibration period, where they are applied in practice (Table~\ref{tab:loocv-miroc6-baselines}).

On the aggregate diagnostics the three corrections are close, and guidance improves on the mean field and on spatial structure (Table~\ref{tab:baselines}). All three reduce the mean absolute bias of raw \gls{cmip6} output, and guidance has the smallest mean absolute bias and \gls{rmse}. The clearest separation is in the spatial texture of the residual bias: the Laplacian roughness of the mean-bias field is smallest for guidance, larger and similar for \gls{mvt} and \gls{qdm}, and largest for raw \gls{cmip6}. \Gls{qdm} is the best of the four on the quantities it is constructed to fix, the 99th-percentile bias and the standard-deviation ratio deviation. The Wasserstein distance marginally favors \gls{qdm}.

\paragraph*{Per-Cell Correction Displaces the Gradient between Neighboring Cells}
Aggregate scores are spatial averages of per-cell quantities, so they do not reveal how a correction treats the relation between neighboring cells. \Gls{mvt} and \gls{qdm} are fitted independently at each grid cell, and neither constrains the difference between adjacent cells. Figures~\ref{fig:si-clouds-hma} and \ref{fig:si-clouds-eur} and Table~\ref{tab:spatial-consistency} examine this relation at the cell pair of each of two regional boxes (\emph{Evaluation Metrics}, \emph{Spatial-Consistency Metrics}). Over High Mountain Asia the two cells are about 300\,km apart, and their \gls{era5} mean temperatures differ by $14.2\,\degC$. The displacement of each method's joint daily distribution from the principal axis of the \gls{era5} distribution is smallest for guidance, larger for raw \gls{cmip6} and largest for \gls{mvt} and \gls{qdm}, all in the same direction. Both corrections therefore leave the inter-cell gradient further from \gls{era5} than raw \gls{cmip6} does. The Europe example behaves the same way. A per-cell correction acts on each cell's marginal distribution without improving the spatial coherence of the map to which it belongs, a limitation reported independently for precipitation \citep{si:hess-2022-physically,si:aich-2026-conditional}. Our generative emulator instead generates each field as a whole, and we designed the two-region test to detect this difference.

Measurements on individual daily fields point the same way. The mean-bias maps of \gls{mvt} and \gls{qdm} show a speckled residual (Figs.~\ref{fig:si-clouds-hma}\emph{A} and \ref{fig:si-clouds-eur}\emph{A}), which could in principle be an artifact of averaging over time. The daily fields show that their texture itself departs from the observed one. By the roughness ratio, the corrected daily fields of \gls{mvt} and \gls{qdm} are $2$--$10\%$ too smooth in both regions, the raw \gls{cmip6} fields are too rough, and the guided fields are within a few percent of the observed texture, although over Europe the \gls{mvt} fields are closer to it (Table~\ref{tab:spatial-consistency}).

The corrections adjust the daily fields of each \gls{cmip6} model, whereas the guided column draws its seventeen members from the emulator conditioned on \gls{era5}, so the guided column is not a day-by-day correction of a particular simulation. We compare the methods this way because we condition the emulator on \gls{era5} alone ($\gamma = 1$) for the past, which observations describe best, and blend simulations with observations only in projections. We also compare them over the long-term future, with MIROC6 as the synthetic observation (\emph{Perfect-Model Validation with a Scheduled Guidance Strength}, \emph{Comparison with Two Bias Corrections and Classifier Guidance beyond the Calibration Period}).

\begin{table}[b!]
\centering
\caption{\textbf{Distributional error against ERA5, 2001--2010.} Raw CMIP6 output, guided samples at $\gamma = 1$, and the raw output after MVT and QDM bias correction, each with seventeen members pooled, on every day of the decade. Every metric is computed at each grid cell against ERA5 and reduced to one scalar by area weighting. Lower is better, and the smallest value of each row is in bold. The RI floor is $0.0003$. The metrics are defined in \emph{Evaluation Metrics}, \emph{Distributional Metrics}.}\label{tab:baselines}
\begin{tabular}{lrrrr}
 & Raw & Guided & MVT & QDM \\
\midrule
Mean absolute bias [$\degC$]   & 0.738 & \textbf{0.207} & 0.229 & 0.226 \\
RMSE [$\degC$]                 & 0.952 & \textbf{0.305} & 0.345 & 0.341 \\
1st-percentile bias [$\degC$]  & 1.543 & 0.549 & 0.542 & \textbf{0.439} \\
99th-percentile bias [$\degC$] & 2.254 & 0.443 & 0.438 & \textbf{0.357} \\
Standard-deviation ratio deviation & 0.280 & 0.055 & 0.043 & \textbf{0.035} \\
TX90p bias [percentage points] & 13.20 & 2.61 & 2.64 & \textbf{2.37} \\
Wasserstein distance [$\degC$] & 0.955 & 0.286 & 0.292 & \textbf{0.278} \\
Laplacian roughness [$\degC$]  & 0.646 & \textbf{0.245} & 0.360 & 0.355 \\
RI, per cell                   & 0.0251 & \textbf{0.0026} & 0.0030 & 0.0028 \\
\bottomrule
\end{tabular}
\end{table}

\begin{table}[b!]
\centering
\caption{\textbf{Spatial consistency at two adjacent grid-cell pairs.} The cell pairs (\emph{Evaluation Metrics}, \emph{Spatial-Consistency Metrics}) are $37.7^{\circ}$N and $40.5^{\circ}$N at $87.2^{\circ}$E over High Mountain Asia, whose observed mean temperatures differ by $14.2\,\degC$, and the same two latitudes at $39.4^{\circ}$E, in eastern Anatolia inside the Europe box, where they differ by $6.6\,\degC$. \emph{Displacement} is given as an absolute value in $\degC$, and smaller is better. \emph{Roughness ratio}, over each pair's regional box, is relative to ERA5, where $1$ is a perfect match. Figures~\ref{fig:si-clouds-hma} and \ref{fig:si-clouds-eur} show the joint distributions of the two pairs. Guided samples at a fixed $\gamma = 1$ over 2001--2010, every day of the year, seventeen members, as in Table~\ref{tab:baselines}. The metrics are defined in \emph{Evaluation Metrics}, \emph{Spatial-Consistency Metrics}.}
\label{tab:spatial-consistency}
\begin{tabular}{llrrrr}
region & quantity & raw CMIP6 & guided & MVT & QDM \\
\midrule
High Mountain Asia & displacement [$\degC$] & 0.856 & \textbf{0.334} & 2.516 & 2.480 \\
 & roughness ratio (ERA5 $1.00$) & 1.097 & \textbf{1.026} & 0.900 & 0.896 \\
\midrule
Europe & displacement [$\degC$] & 0.398 & \textbf{0.075} & 1.597 & 1.579 \\
 & roughness ratio (ERA5 $1.00$) & 1.310 & 1.032 & \textbf{0.979} & 0.937 \\
\bottomrule
\end{tabular}
\end{table}

\begin{table}[b!]\centering
\caption{\textbf{Inter-model spread retained by the guided ensemble, 2081--2100.} The metrics are defined in \emph{Evaluation Metrics}, \emph{Projection and Extreme-Heat Metrics}.}
\label{tab:retention}
\begin{tabular}{lrrr}
 & SSP2-4.5 & SSP3-7.0 & SSP5-8.5 \\
\midrule
Raw CMIP6 inter-model range [$\degC$]   & 1.149 & 1.618 & 2.103 \\
Guided inter-model range [$\degC$]      & 0.259 & 0.332 & 0.482 \\
\textbf{Retained inter-model spread}              & \textbf{22.5\%} & \textbf{20.5\%} & \textbf{22.9\%} \\
\bottomrule
\end{tabular}
\end{table}

\begin{table}\centering
\caption{\textbf{Years in which the guided ensemble mean first exceeds $1.5\,\degC$ and $2\,\degC$ above 1850--1900.} The first 20-year period above each level, the crossing year and its 5--95\% range across the 170 guided series, with the assessed warming of $0.85\,\degC$ from 1850--1900 to 1995--2014. The next two columns give the crossing year with that warming at its very likely bounds, $0.98$ and $0.67\,\degC$. The last column gives the first 20-year period above each level assessed by IPCC AR6 WGI (Table 4.5 of the report), with its very likely (5--95\%) range in brackets \citep{si:ipcc-2021-wg1-ar6}. The crossing year is defined in \emph{Evaluation Metrics}, \emph{Warming-Level Crossing Year}.}
 \label{tab:gwl-crossings}
\footnotesize
\begin{tabular}{llcrrrrc}
& & & & & \multicolumn{2}{c}{offset of} & \\
Level & Scenario & First 20-year period & Crossing year & 5--95\% range & $0.98\,\degC$ & $0.67\,\degC$ & IPCC AR6 \\
\midrule
$1.5\,\degC$ & SSP2-4.5 & 2020--2039 & 2029 & 2028--2032 & 2024 & 2036 & 2021--2040 [2012--2031 to 2037--2056] \\
            & SSP3-7.0 & 2020--2039 & 2029 & 2028--2031 & 2025 & 2035 & 2021--2040 [2013--2032 to 2033--2052] \\
            & SSP5-8.5 & 2019--2038 & 2028 & 2027--2030 & 2024 & 2033 & 2018--2037 [2011--2030 to 2029--2048] \\
\midrule
$2\,\degC$   & SSP2-4.5 & 2038--2057 & 2047 & 2045--2051 & 2043 & 2054 & 2043--2062 [2028--2047 to 2075--2094] \\
            & SSP3-7.0 & 2036--2055 & 2045 & 2042--2048 & 2040 & 2051 & 2037--2056 [2026--2045 to 2053--2072] \\
            & SSP5-8.5 & 2032--2051 & 2041 & 2039--2043 & 2038 & 2045 & 2032--2051 [2023--2042 to 2044--2063] \\
\bottomrule
\end{tabular}
\end{table}

\begin{table}
\centering
\caption{\textbf{\glsentryshort{cmip6} models used in this study.} The column \emph{Reference} gives the dataset references of the historical and \glsentryshort{ssp} simulations of each model.}\label{tab:cmip6-models}
\begin{tabular}{ll}
Model & Reference \\
\midrule
AWI-CM-1-1-MR & \citep{si:cmip6-cmip-awi-cm-1-1-mr,si:cmip6-scenariomip-awi-cm-1-1-mr} \\
BCC-CSM2-MR & \citep{si:cmip6-cmip-bcc-csm2-mr,si:cmip6-scenariomip-bcc-csm2-mr} \\
CanESM5 & \citep{si:cmip6-cmip-canesm5,si:cmip6-scenariomip-canesm5} \\
CESM2-WACCM & \citep{si:cmip6-cmip-cesm2-waccm,si:cmip6-scenariomip-cesm2-waccm} \\
CMCC-CM2-SR5 & \citep{si:cmip6-cmip-cmcc-cm2-sr5,si:cmip6-scenariomip-cmcc-cm2-sr5} \\
CMCC-ESM2 & \citep{si:cmip6-cmip-cmcc-esm2,si:cmip6-scenariomip-cmcc-esm2} \\
EC-Earth3 & \citep{si:cmip6-cmip-ec-earth3,si:cmip6-scenariomip-ec-earth3} \\
EC-Earth3-Veg & \citep{si:cmip6-cmip-ec-earth3-veg,si:cmip6-scenariomip-ec-earth3-veg} \\
EC-Earth3-Veg-LR & \citep{si:cmip6-cmip-ec-earth3-veg-lr,si:cmip6-scenariomip-ec-earth3-veg-lr} \\
FGOALS-g3 & \citep{si:cmip6-cmip-fgoals-g3,si:cmip6-scenariomip-fgoals-g3} \\
IPSL-CM6A-LR & \citep{si:cmip6-cmip-ipsl-cm6a-lr,si:cmip6-scenariomip-ipsl-cm6a-lr} \\
MIROC6 & \citep{si:cmip6-cmip-miroc6,si:cmip6-scenariomip-miroc6} \\
MPI-ESM1-2-HR & \citep{si:cmip6-cmip-mpi-esm1-2-hr,si:cmip6-scenariomip-mpi-esm1-2-hr} \\
MPI-ESM1-2-LR & \citep{si:cmip6-cmip-mpi-esm1-2-lr,si:cmip6-scenariomip-mpi-esm1-2-lr} \\
MRI-ESM2-0 & \citep{si:cmip6-cmip-mri-esm2-0,si:cmip6-scenariomip-mri-esm2-0} \\
NorESM2-LM & \citep{si:cmip6-cmip-noresm2-lm,si:cmip6-scenariomip-noresm2-lm} \\
NorESM2-MM & \citep{si:cmip6-cmip-noresm2-mm,si:cmip6-scenariomip-noresm2-mm} \\
\bottomrule
\end{tabular}
\end{table}
\begin{table}[p]\centering
\caption{\textbf{Held-out MPI-ESM1-2-LR: condition-to-condition guidance against raw CMIP6.} The metrics, samples and conventions of Table~\ref{tab:loocv-miroc6-baselines}, for raw CMIP6 and the emulator guided from the sixteen remaining models. The RI floor is $0.0001$. The metrics are defined in \emph{Evaluation Metrics}, \emph{Trajectory Metrics} and \emph{Distributional Metrics}.}\label{tab:loocv-mpi}
\footnotesize
\begin{tabular}{llrr}
Metric & Scenario & Raw CMIP6 & Guided \\
\midrule
\multicolumn{4}{l}{\textit{Annual global-mean warming, 2025--2100}} \\
Trajectory RMSE [$\degC$] & SSP2-4.5 & 0.392 & \textbf{0.257} \\
 & SSP3-7.0 & 0.424 & \textbf{0.397} \\
 & SSP5-8.5 & 0.621 & \textbf{0.394} \\
\addlinespace[1pt]
Band coverage & SSP2-4.5 & 0.84 & 0.34 \\
 & SSP3-7.0 & 0.79 & 0.17 \\
 & SSP5-8.5 & 0.63 & 0.38 \\
\midrule
\multicolumn{4}{l}{\textit{Daily fields, 2081--2100}} \\
Mean absolute bias [$\degC$] & SSP2-4.5 & 1.119 & \textbf{0.391} \\
 & SSP3-7.0 & 1.135 & \textbf{0.613} \\
 & SSP5-8.5 & 1.361 & \textbf{0.651} \\
\addlinespace[1pt]
RMSE [$\degC$] & SSP2-4.5 & 1.396 & \textbf{0.583} \\
 & SSP3-7.0 & 1.422 & \textbf{0.839} \\
 & SSP5-8.5 & 1.659 & \textbf{0.865} \\
\addlinespace[1pt]
Standard-deviation ratio deviation & SSP2-4.5 & 0.217 & \textbf{0.062} \\
 & SSP3-7.0 & 0.240 & \textbf{0.075} \\
 & SSP5-8.5 & 0.245 & \textbf{0.076} \\
\addlinespace[1pt]
Skewness bias & SSP2-4.5 & 0.216 & \textbf{0.111} \\
 & SSP3-7.0 & 0.231 & \textbf{0.124} \\
 & SSP5-8.5 & 0.219 & \textbf{0.120} \\
\addlinespace[1pt]
Excess-kurtosis bias & SSP2-4.5 & 0.543 & \textbf{0.261} \\
 & SSP3-7.0 & 0.576 & \textbf{0.292} \\
 & SSP5-8.5 & 0.542 & \textbf{0.294} \\
\addlinespace[1pt]
TX90p bias [percentage points] & SSP2-4.5 & 13.97 & \textbf{2.42} \\
 & SSP3-7.0 & 13.97 & \textbf{3.70} \\
 & SSP5-8.5 & 16.73 & \textbf{4.04} \\
\addlinespace[1pt]
TN10p bias [percentage points] & SSP2-4.5 & 5.02 & \textbf{3.73} \\
 & SSP3-7.0 & \textbf{4.91} & 5.87 \\
 & SSP5-8.5 & \textbf{5.05} & 6.45 \\
\addlinespace[1pt]
1st-percentile bias [$\degC$] & SSP2-4.5 & 1.737 & \textbf{0.906} \\
 & SSP3-7.0 & 1.654 & \textbf{1.146} \\
 & SSP5-8.5 & 1.547 & \textbf{1.154} \\
\addlinespace[1pt]
99th-percentile bias [$\degC$] & SSP2-4.5 & 2.605 & \textbf{0.438} \\
 & SSP3-7.0 & 2.790 & \textbf{0.566} \\
 & SSP5-8.5 & 3.210 & \textbf{0.644} \\
\addlinespace[1pt]
CRPS [$\degC$] & SSP2-4.5 & 1.297 & \textbf{1.016} \\
 & SSP3-7.0 & 1.291 & \textbf{1.060} \\
 & SSP5-8.5 & 1.368 & \textbf{1.061} \\
\addlinespace[1pt]
RI, per cell & SSP2-4.5 & 0.0392 & \textbf{0.0049} \\
 & SSP3-7.0 & 0.0368 & \textbf{0.0109} \\
 & SSP5-8.5 & 0.0489 & \textbf{0.0128} \\
\addlinespace[1pt]
Laplacian roughness [$\degC$] & SSP2-4.5 & 1.178 & \textbf{0.356} \\
 & SSP3-7.0 & 1.173 & \textbf{0.423} \\
 & SSP5-8.5 & 1.184 & \textbf{0.475} \\
\addlinespace[1pt]
Wasserstein distance [$\degC$] & SSP2-4.5 & 1.263 & \textbf{0.440} \\
 & SSP3-7.0 & 1.268 & \textbf{0.650} \\
 & SSP5-8.5 & 1.478 & \textbf{0.690} \\
\bottomrule
\end{tabular}
\end{table}

\FloatBarrier

\section*{LOOCV across Seventeen Folds}
\begin{figure}[p]
\centering
\includegraphics[width=\textwidth]{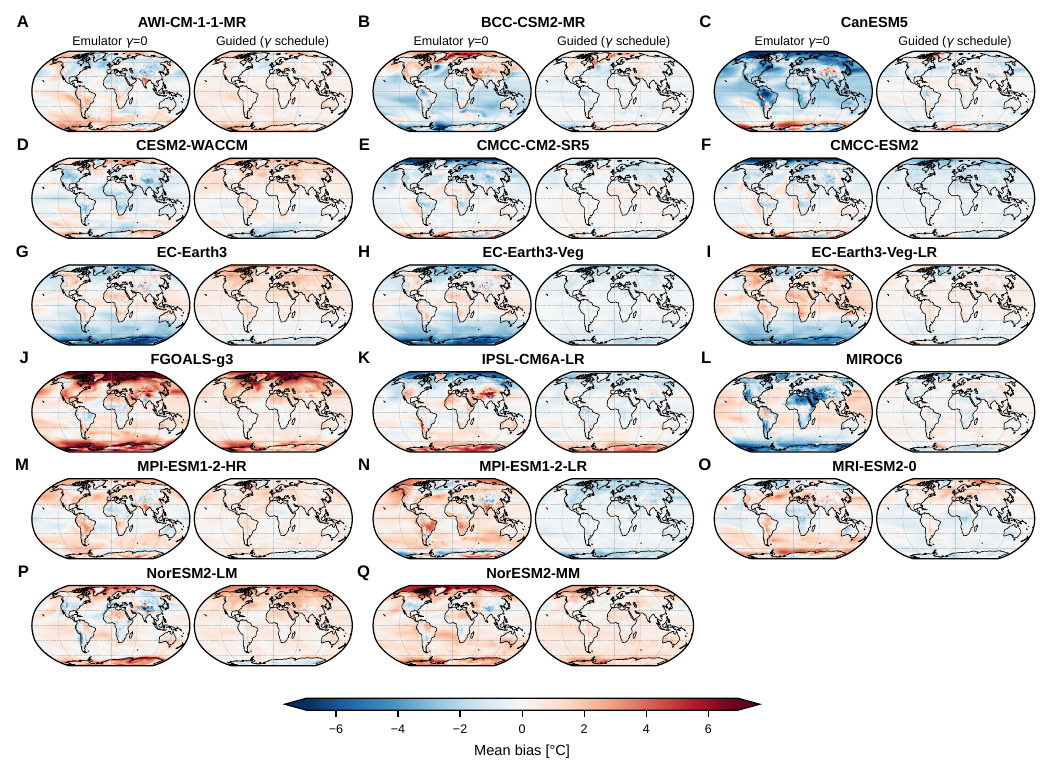}
\caption{\textbf{LOOCV performance across seventeen folds. Mean bias of the emulated ensemble against each held-out model, without and with guidance, SSP3-7.0, 2080--2100.} (\emph{A}--\emph{Q}) One LOOCV fold each, named by the held-out model. Left of each pair: the emulator conditioned on the sixteen source models without guidance ($\gamma = 0$). Right: the emulator guided toward the held-out model with the $\gamma$ schedule rounded to the nearest sampled scale. Each map is the mean at every grid cell minus that of the held-out model. Samples from the scale sweep, with one sample per source model on twelve days per year, the sixteen source models pooled and the held-out model's run on the same days (\emph{Sampling Designs}).}
\label{fig:si-folds-mean-370}
\end{figure}

\begin{figure}[p]
\centering
\includegraphics[width=\textwidth]{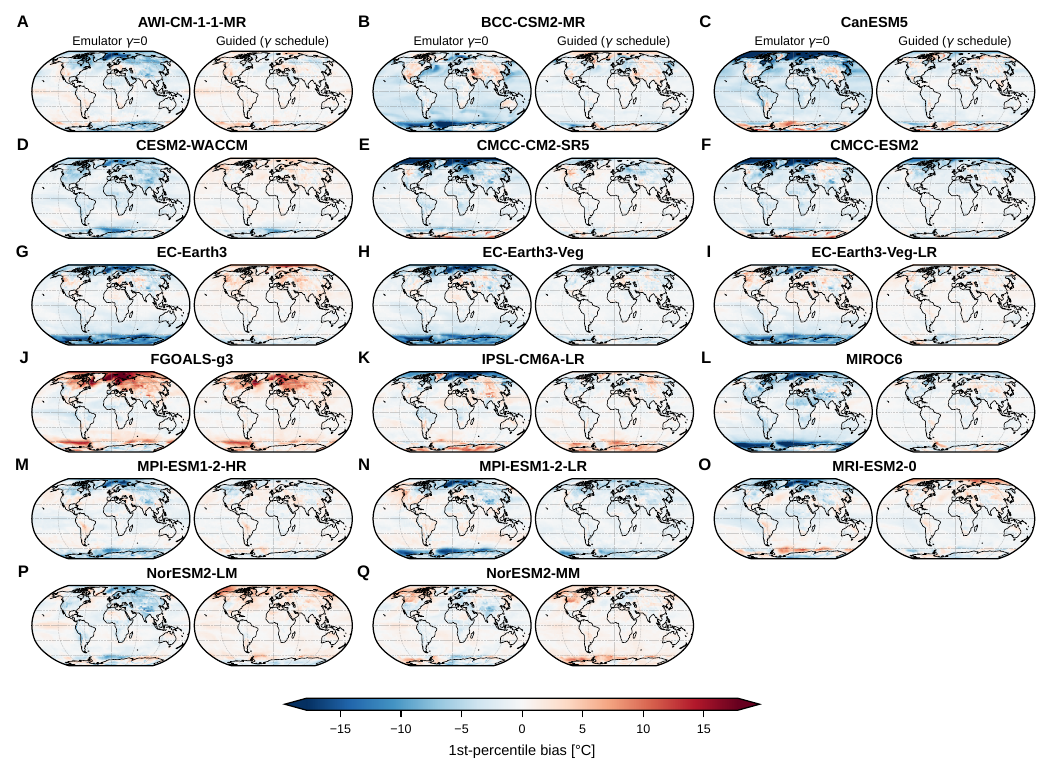}
\caption{\textbf{LOOCV performance across seventeen folds. 1st-percentile bias of the emulated ensemble against each held-out model, without and with guidance, SSP3-7.0, 2080--2100.} Folds, pairs, $\gamma$ and sampling as in Fig.~\ref{fig:si-folds-mean-370}. Each map is the 1st percentile of daily temperature at every grid cell minus that of the held-out model. One color scale for all panels.}
\label{fig:si-folds-q01-370}
\end{figure}

\begin{figure}[p]
\centering
\includegraphics[width=\textwidth]{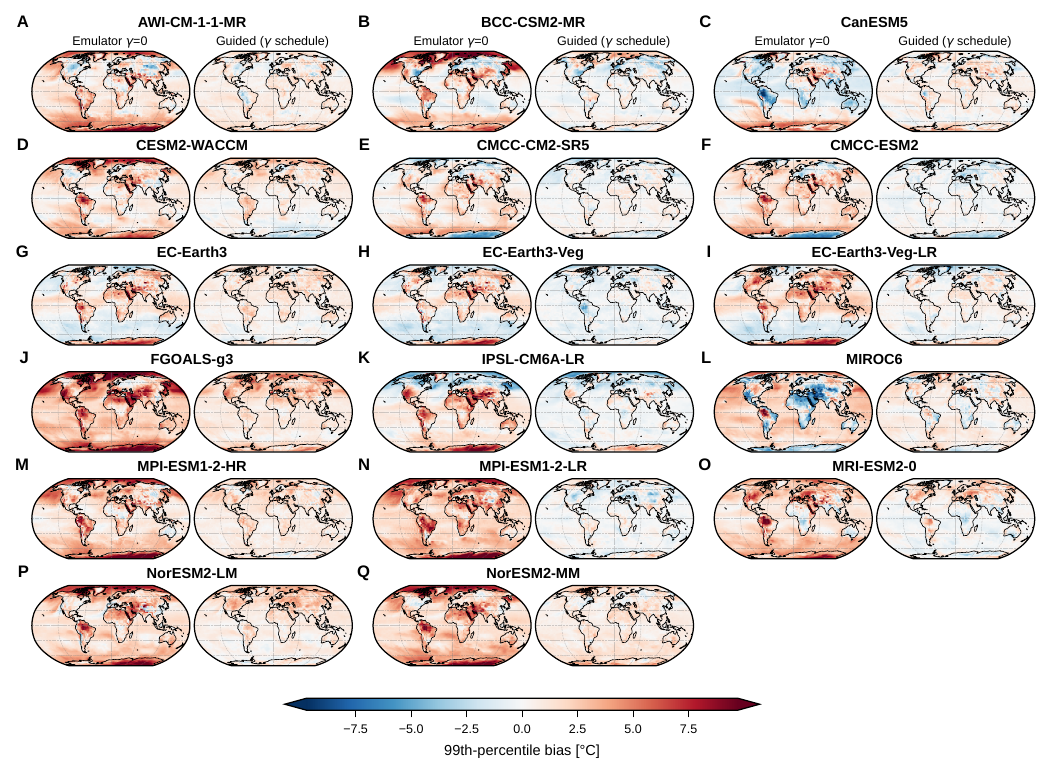}
\caption{\textbf{LOOCV performance across seventeen folds. 99th-percentile bias of the emulated ensemble against each held-out model, without and with guidance, SSP3-7.0, 2080--2100.} As Fig.~\ref{fig:si-folds-q01-370}, for the 99th percentile.}
\label{fig:si-folds-q99-370}
\end{figure}

\begin{figure}[p]
\centering
\includegraphics[width=\textwidth]{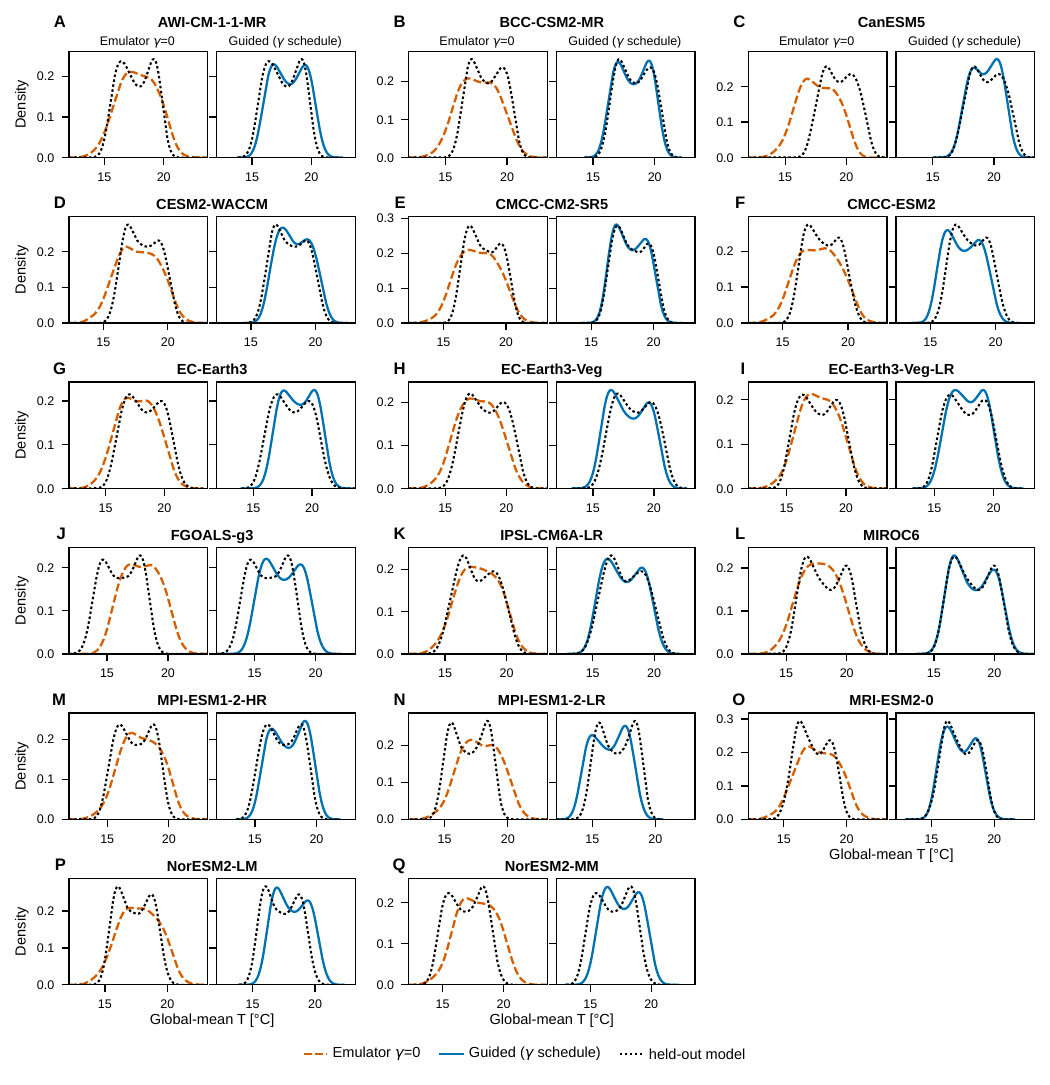}
\caption{\textbf{LOOCV performance across seventeen folds. Distribution of the daily global-mean temperature for each held-out model, without and with guidance, SSP3-7.0, 2080--2100.} Folds, pairs, $\gamma$ and sampling as in Fig.~\ref{fig:si-folds-mean-370}. Kernel density of the daily area-weighted global-mean temperature for the emulator without guidance (dashed), the guided emulator (solid) and the held-out model on the same days (dotted).}
\label{fig:si-folds-kde-370}
\end{figure}

\begin{figure}[p]
\centering
\includegraphics[width=\textwidth]{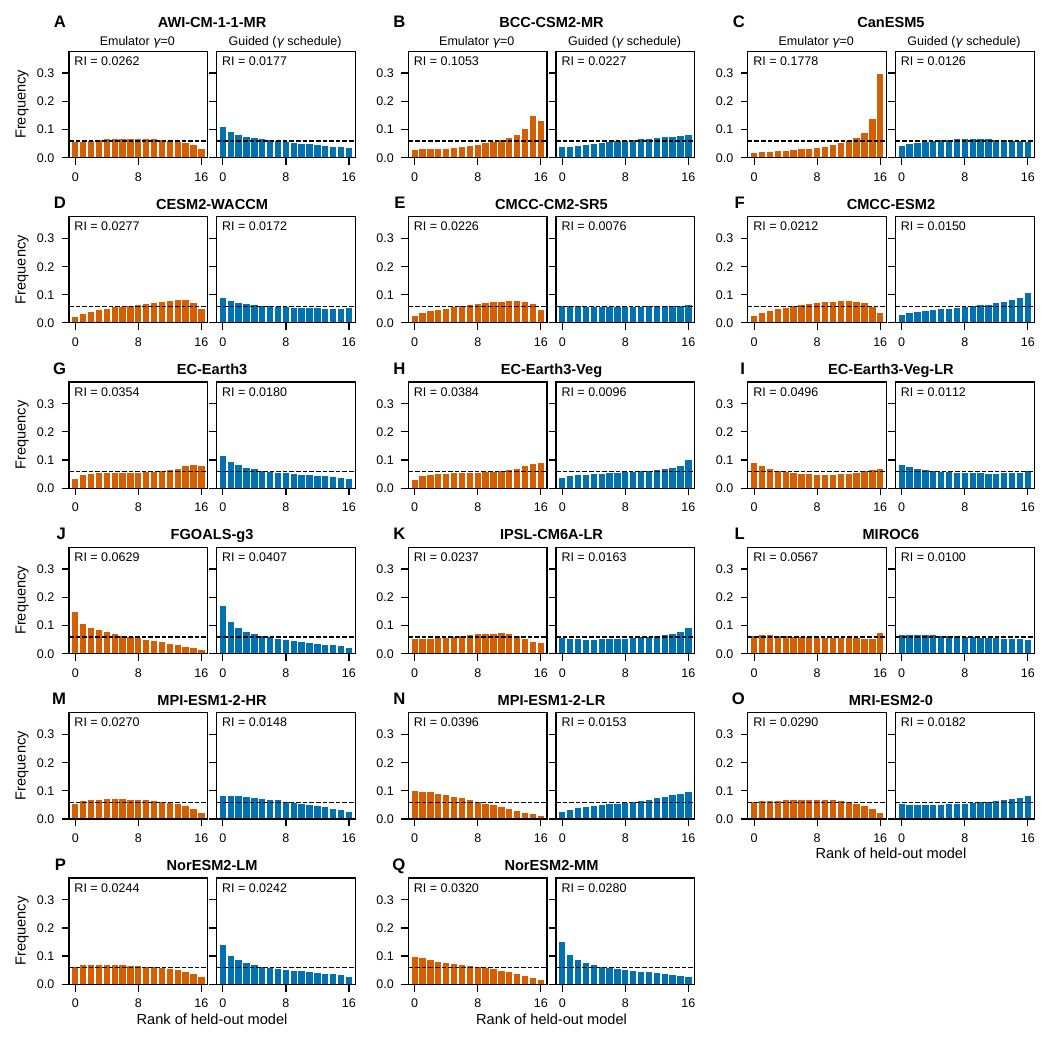}
\caption{\textbf{LOOCV performance across seventeen folds. Ensemble calibration against each held-out model, without and with guidance, SSP3-7.0, 2080--2100.} Folds, pairs, $\gamma$ and sampling as in Fig.~\ref{fig:si-folds-mean-370}. Area-weighted rank histogram of the held-out model among the sixteen ensemble members at each grid cell and day. The dashed line marks the uniform frequency $1/17$ expected of a calibrated sixteen-member ensemble. The RI floor is $0.0037$.}
\label{fig:si-folds-rank-370}
\end{figure}

The figures of this section show the perfect-model test in each of the seventeen \gls{loocv} folds under SSP3-7.0 over 2080--2100. Each figure compares the emulator conditioned on the sixteen source models without guidance ($\gamma = 0$) with the same emulator guided toward the held-out model. Figs.~\ref{fig:si-folds-mean-370}--\ref{fig:si-folds-q99-370} map the mean, 1st-percentile and 99th-percentile biases, Fig.~\ref{fig:si-folds-kde-370} shows the distribution of the daily global-mean temperature, and Fig.~\ref{fig:si-folds-rank-370} the rank histograms. Table~\ref{tab:loocv-metrics} gives the metrics of every fold, and \emph{All Seventeen Folds under SSP3-7.0} discusses the figures. The same figures for SSP2-4.5 and SSP5-8.5 over 2080--2100, and for the three scenarios over 2030--2050, which we use as an additional mid-term validation period, are available at \url{https://github.com/mandt-lab/gdso-figures} for completeness.

\FloatBarrier

\section*{LOOCV for MIROC6 (Considered as Synthetic Observation) -- Comparison of Condition-to-Condition Guidance, MVT, QDM (Applied to Raw CMIP6 Data), and Classifier Guidance}
\begin{figure}[p]
\centering
\includegraphics[width=\textwidth]{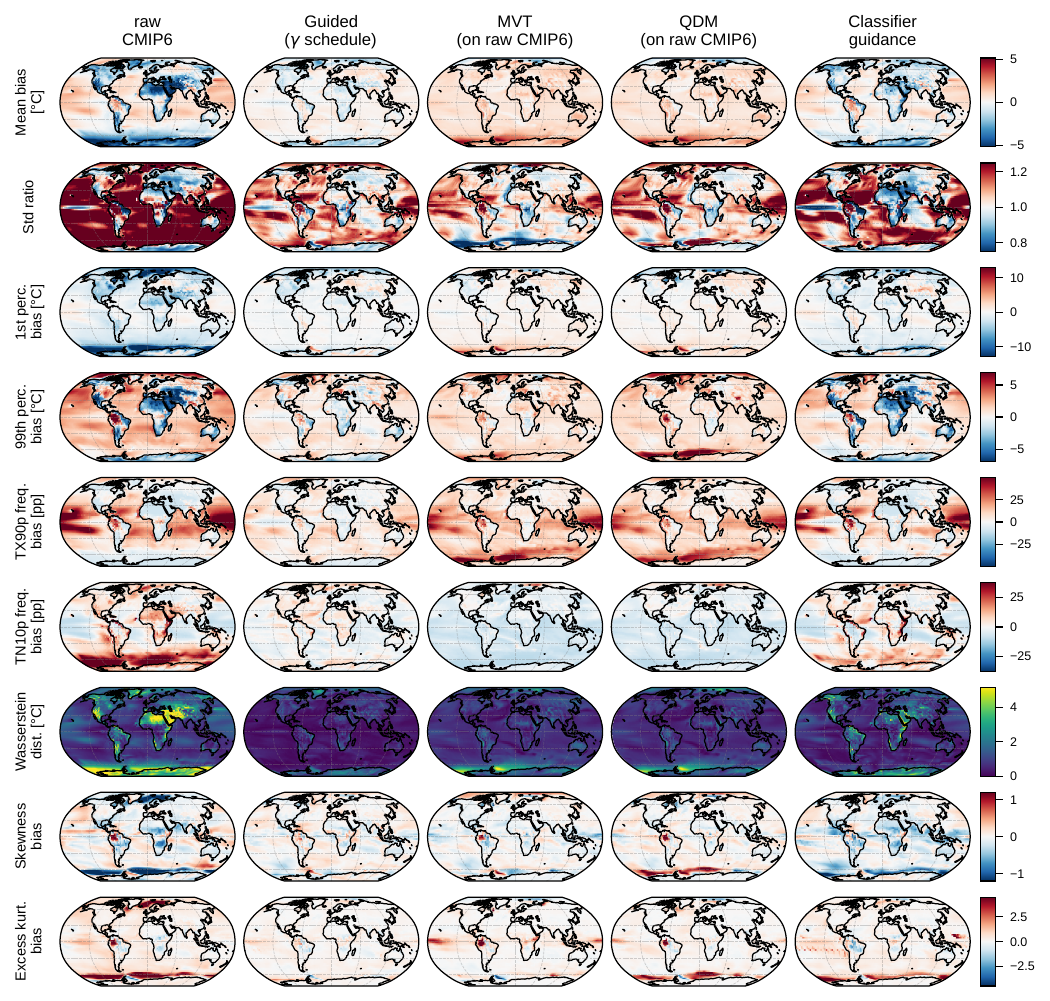}
\caption{\textbf{LOOCV performance comparison of condition-to-condition guidance, MVT, QDM (applied to raw CMIP6 data), and classifier guidance for the MIROC6 fold. Nine distributional error diagnostics against the held-out MIROC6, 2081--2100, SSP3-7.0.} Columns, left to right: the raw output of the sixteen CMIP6 source models; the emulator guided from the same sixteen models with the $\gamma$ schedule; the raw output after MVT and QDM bias correction; and classifier guidance. Rows, top to bottom: mean bias; ratio of modeled to MIROC6 daily standard deviation; 1st- and 99th-percentile bias; bias in the exceedance frequency of each cell's MIROC6 90th-percentile warm-day (TX90p) and 10th-percentile cold-day (TN10p) thresholds, in percentage points; 1-Wasserstein distance between the daily distributions; skewness bias; and excess-kurtosis bias. Computed with one sample per source model and day of 2081--2100 for the emulator and for classifier guidance.}
\label{fig:si-m6-maps-raw-370}
\end{figure}

\begin{figure}[p]
\centering
\includegraphics[width=\textwidth]{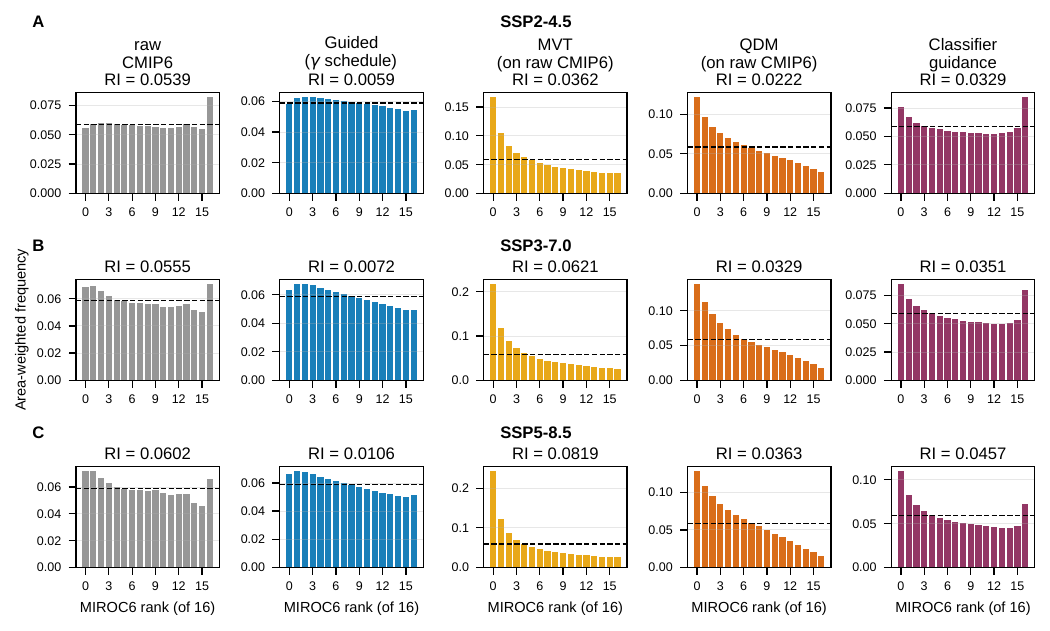}
\caption{\textbf{LOOCV performance comparison of condition-to-condition guidance, MVT, QDM (applied to raw CMIP6 data), and classifier guidance for the MIROC6 fold. Ensemble calibration against the held-out MIROC6, 2081--2100.} Area-weighted rank histogram of MIROC6 among the sixteen ensemble members at each grid cell and day, for (\emph{A}) SSP2-4.5, (\emph{B}) SSP3-7.0 and (\emph{C}) SSP5-8.5. Columns, left to right: raw output of the sixteen CMIP6 source models; the emulator guided from the same sixteen models with the $\gamma$ schedule; the raw output after MVT and QDM bias correction; and classifier guidance. The dashed line marks the uniform frequency $1/17$ expected of a calibrated sixteen-member ensemble. Every day of 2081--2100, with one sample per source model and day for the emulator and for classifier guidance.}
\label{fig:si-m6-rank-raw}
\end{figure}

\begin{figure}[p]
\centering
\includegraphics[width=\textwidth]{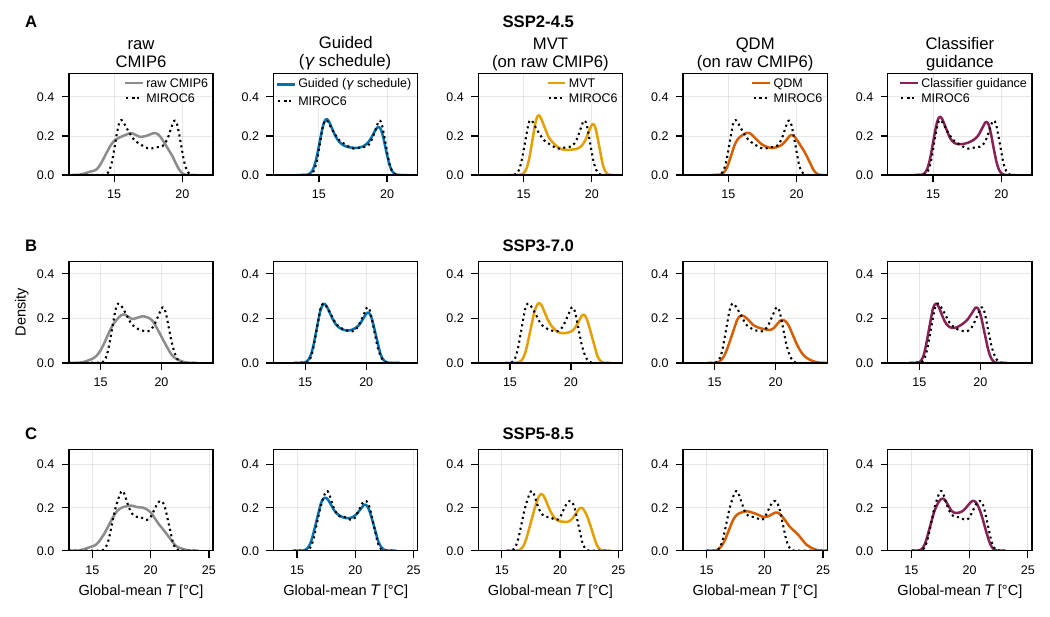}
\caption{\textbf{LOOCV performance comparison of condition-to-condition guidance, MVT, QDM (applied to raw CMIP6 data), and classifier guidance for the MIROC6 fold. Distribution of the daily global-mean temperature, held-out MIROC6, 2081--2100.} Kernel density of the daily area-weighted global-mean temperature for each method (color) and for MIROC6 (dotted), for (\emph{A}) SSP2-4.5, (\emph{B}) SSP3-7.0 and (\emph{C}) SSP5-8.5.}
\label{fig:si-m6-kde-raw}
\end{figure}

\begin{figure}[p]
\centering
\includegraphics[width=\textwidth]{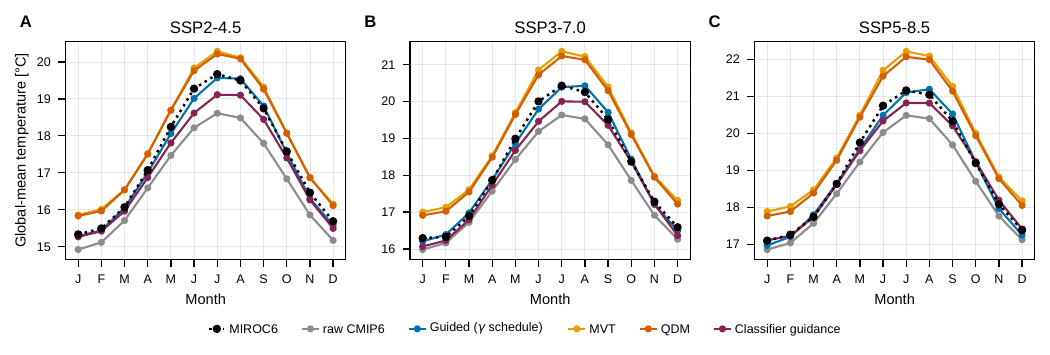}
\caption{\textbf{LOOCV performance comparison of condition-to-condition guidance, MVT, QDM (applied to raw CMIP6 data), and classifier guidance for the MIROC6 fold. Seasonal cycle of the global-mean temperature, held-out MIROC6, 2081--2100.} Global-mean temperature by calendar month for MIROC6 (black, dotted) and each method (color), for (\emph{A}) SSP2-4.5, (\emph{B}) SSP3-7.0 and (\emph{C}) SSP5-8.5.}
\label{fig:si-m6-cycle-raw}
\end{figure}

The figures of this section compare condition-to-condition guidance with \gls{mvt} and \gls{qdm} applied to the raw output of the sixteen source models, and with classifier guidance, in the \gls{loocv} fold in which MIROC6 is held out, over 2081--2100. Fig.~\ref{fig:si-m6-maps-raw-370} maps nine distributional errors under SSP3-7.0. Figs.~\ref{fig:si-m6-rank-raw}--\ref{fig:si-m6-cycle-raw} show, under the three scenarios, the rank histograms, the distribution of the daily global-mean temperature and its seasonal cycle. Table~\ref{tab:loocv-miroc6-baselines} gives the metrics, and \emph{Comparison with Two Bias Corrections and Classifier Guidance beyond the Calibration Period} discusses them. The figures for SSP2-4.5 and SSP5-8.5 over 2081--2100, and all four figures for the three scenarios over 2030--2050, which we use as an additional mid-term validation period, are available at \url{https://github.com/mandt-lab/gdso-figures} for completeness.

\FloatBarrier

\section*{LOOCV for MIROC6 (Considered as Synthetic Observation) -- Comparison of Condition-to-Condition Guidance, MVT, and QDM (Applied to the Unguided Emulator as an Ablation)}
\begin{figure}[p]
\centering
\includegraphics[width=\textwidth]{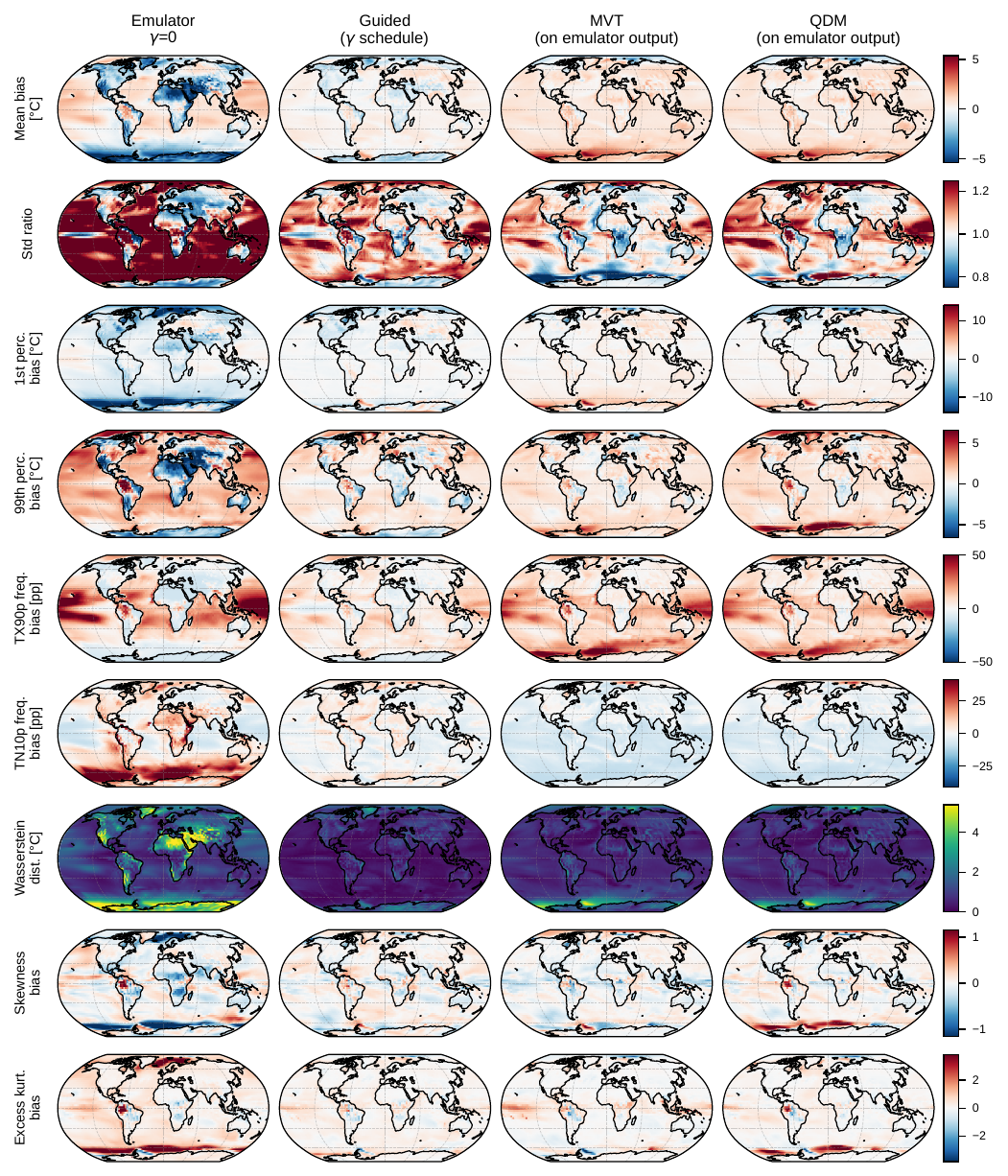}
\caption{\textbf{LOOCV performance comparison of condition-to-condition guidance, MVT, and QDM (applied to the unguided emulator output) for the MIROC6 fold. Nine distributional error diagnostics against the held-out MIROC6, 2081--2100, SSP3-7.0.} Columns, left to right: the emulator conditioned on the sixteen CMIP6 source models without guidance ($\gamma = 0$); the emulator guided from the same sixteen models with the $\gamma$ schedule; and the unguided output after MVT and QDM bias correction. Rows, top to bottom: mean bias; ratio of modeled to MIROC6 daily standard deviation; 1st- and 99th-percentile bias; bias in the exceedance frequency of each cell's MIROC6 90th-percentile warm-day (TX90p) and 10th-percentile cold-day (TN10p) thresholds, in percentage points; 1-Wasserstein distance between the daily distributions; skewness bias; and excess-kurtosis bias. Computed with one sample per source model and day of 2081--2100 for the emulator.}
\label{fig:si-m6-maps-diff-370}
\end{figure}

\begin{figure}[p]
\centering
\includegraphics[width=\textwidth]{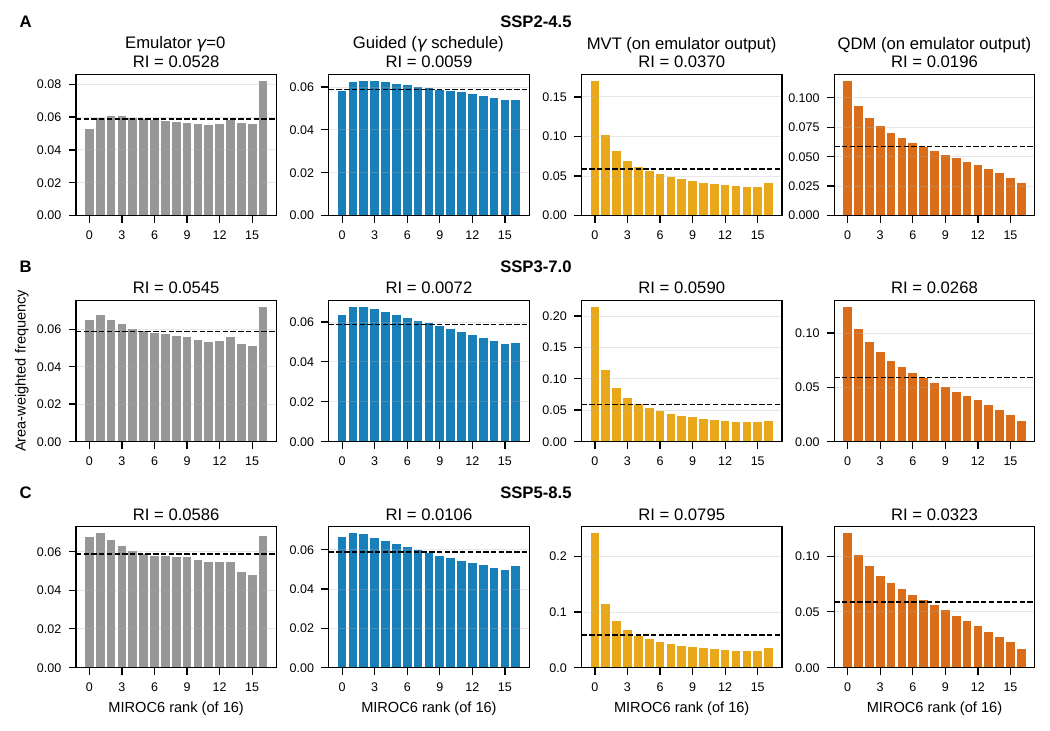}
\caption{\textbf{LOOCV performance comparison of condition-to-condition guidance, MVT, and QDM (applied to the unguided emulator output) for the MIROC6 fold. Ensemble calibration against the held-out MIROC6, 2081--2100.} As Fig.~\ref{fig:si-m6-rank-raw}, with the output of the emulator conditioned on the same sixteen models without guidance ($\gamma = 0$) in the leftmost column and as the input to MVT and QDM, and without classifier guidance.}
\label{fig:si-m6-rank-diff}
\end{figure}

\begin{figure}[p]
\centering
\includegraphics[width=\textwidth]{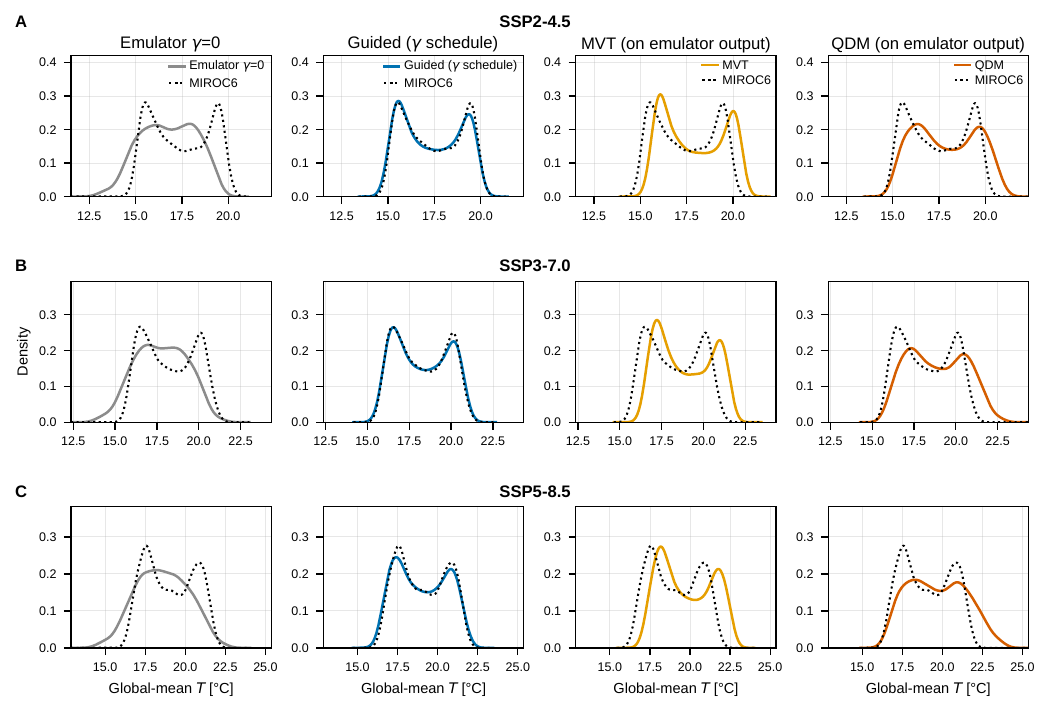}
\caption{\textbf{LOOCV performance comparison of condition-to-condition guidance, MVT, and QDM (applied to the unguided emulator output) for the MIROC6 fold. Distribution of the daily global-mean temperature, held-out MIROC6, 2081--2100.} As Fig.~\ref{fig:si-m6-kde-raw}, with the columns of Fig.~\ref{fig:si-m6-rank-diff}.}
\label{fig:si-m6-kde-diff}
\end{figure}

\begin{figure}[p]
\centering
\includegraphics[width=\textwidth]{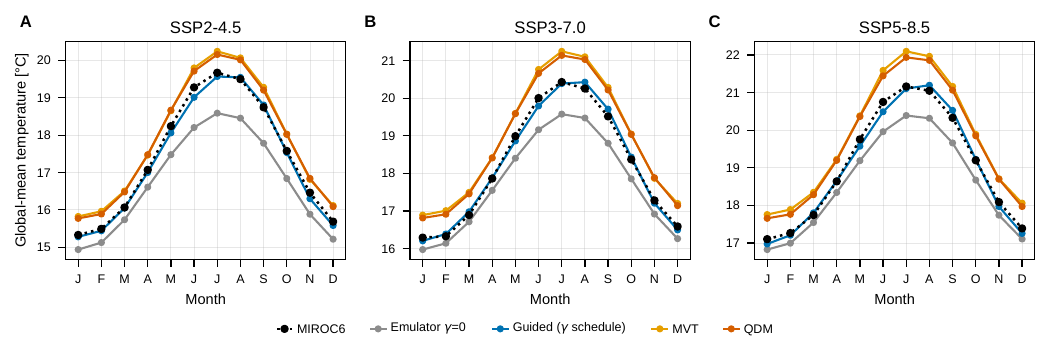}
\caption{\textbf{LOOCV performance comparison of condition-to-condition guidance, MVT, and QDM (applied to the unguided emulator output) for the MIROC6 fold. Seasonal cycle of the global-mean temperature, held-out MIROC6, 2081--2100.} As Fig.~\ref{fig:si-m6-cycle-raw}, with the methods of Fig.~\ref{fig:si-m6-rank-diff}.}
\label{fig:si-m6-cycle-diff}
\end{figure}

The figures of this section show the ablation of the MIROC6 comparison. The output of the emulator conditioned on the sixteen source models without guidance ($\gamma = 0$) replaces raw \gls{cmip6}, both in the leftmost column and as the input to \gls{mvt} and \gls{qdm}. Fig.~\ref{fig:si-m6-maps-diff-370} maps the nine distributional errors under SSP3-7.0, and Figs.~\ref{fig:si-m6-rank-diff}--\ref{fig:si-m6-cycle-diff} show the rank histograms, the distribution of the daily global-mean temperature and its seasonal cycle. Table~\ref{tab:loocv-miroc6-baselines} gives the metrics. The figures for SSP2-4.5 and SSP5-8.5 over 2081--2100, and all four figures for the three scenarios over 2030--2050, which we use as an additional mid-term validation period, are available at \url{https://github.com/mandt-lab/gdso-figures} for completeness.

\FloatBarrier

\section*{Comparison of Condition-to-Condition Guidance, MVT, and QDM against ERA5 over 2001--2010}
\begin{figure}[p]
\centering
\includegraphics[width=\textwidth]{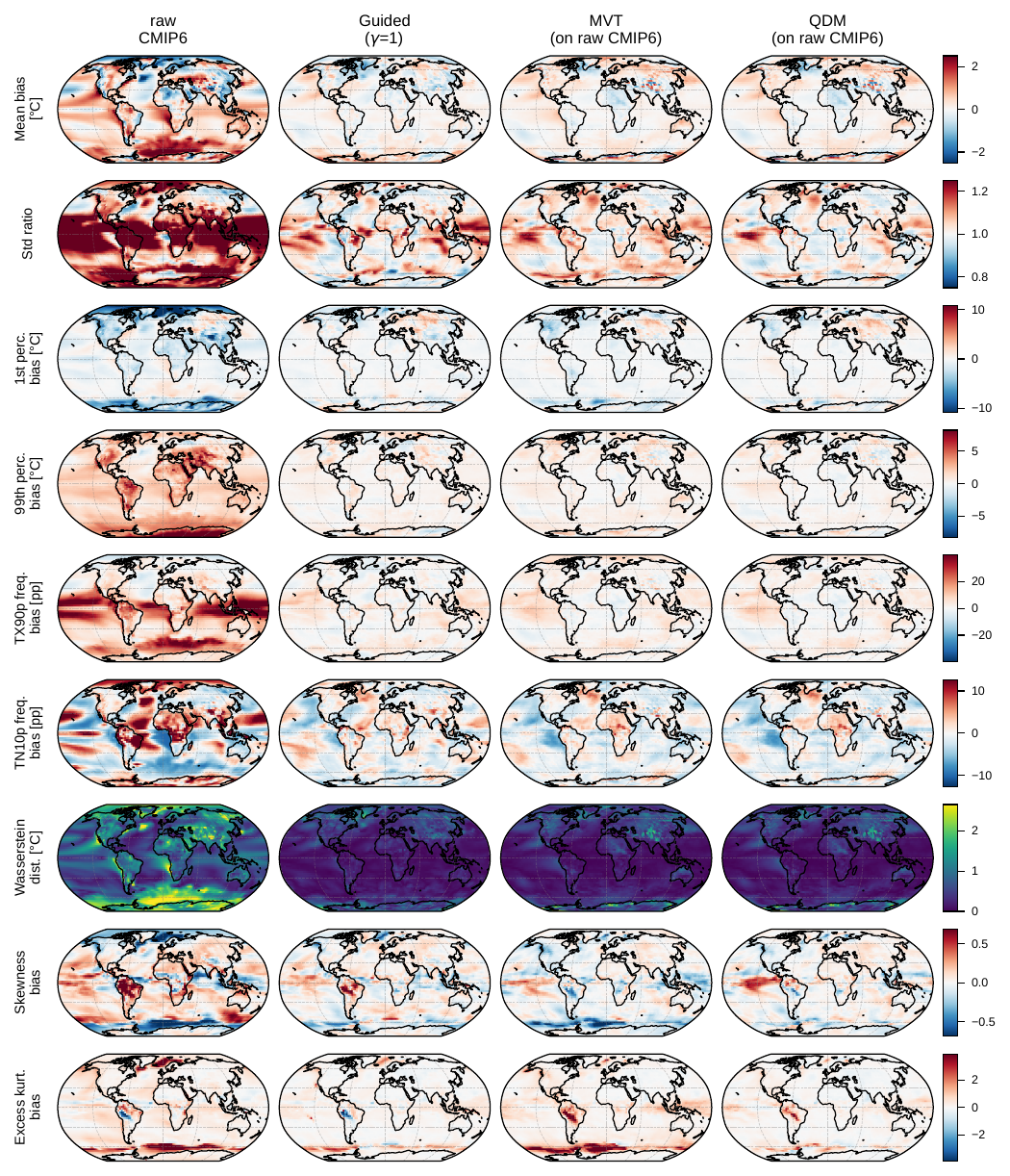}
\caption{\textbf{Performance comparison of condition-to-condition guidance, MVT and QDM (applied to the raw CMIP6 output), evaluated against ERA5 over 2001--2010. Nine distributional error diagnostics.} Columns, left to right: raw CMIP6 output; guided diffusion samples ($\gamma = 1$); and the same raw output after mean--variance--trend (MVT) and quantile delta mapping (QDM) bias correction. Rows, top to bottom: mean bias; ratio of modeled to ERA5 daily standard deviation; 1st- and 99th-percentile bias; bias in the exceedance frequency of each cell's ERA5 90th-percentile warm-day (TX90p) and 10th-percentile cold-day (TN10p) thresholds, in percentage points; 1-Wasserstein distance between the daily distributions; skewness bias; and excess-kurtosis bias.}
\label{fig:si-baseline-maps}
\end{figure}

\begin{figure}[p]
\centering
\includegraphics[width=\textwidth]{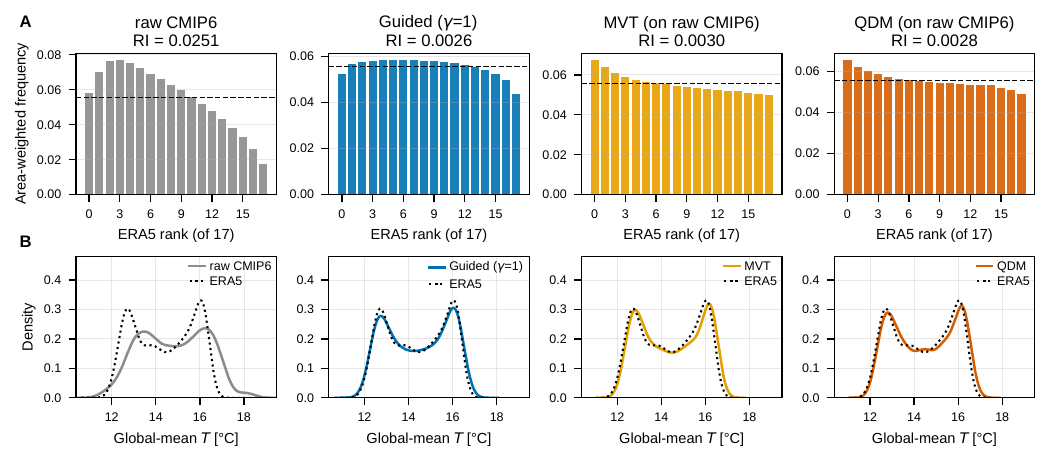}
\caption{\textbf{Performance comparison of condition-to-condition guidance, MVT and QDM (applied to the raw CMIP6 output), evaluated against ERA5 over 2001--2010. Ensemble calibration and the distribution of global-mean daily temperature for each method.} (\emph{A}) Area-weighted rank histogram: the rank of ERA5 among the seventeen ensemble members at each grid cell, weighted by cell area; the dashed line marks the uniform frequency $1/18$ expected of a calibrated seventeen-member ensemble. (\emph{B}) Kernel density of the daily global-mean temperature for each method (color) and for ERA5 (dotted).}
\label{fig:si-baseline-dist}
\end{figure}

\begin{figure}[p]
\centering
\includegraphics[width=\textwidth]{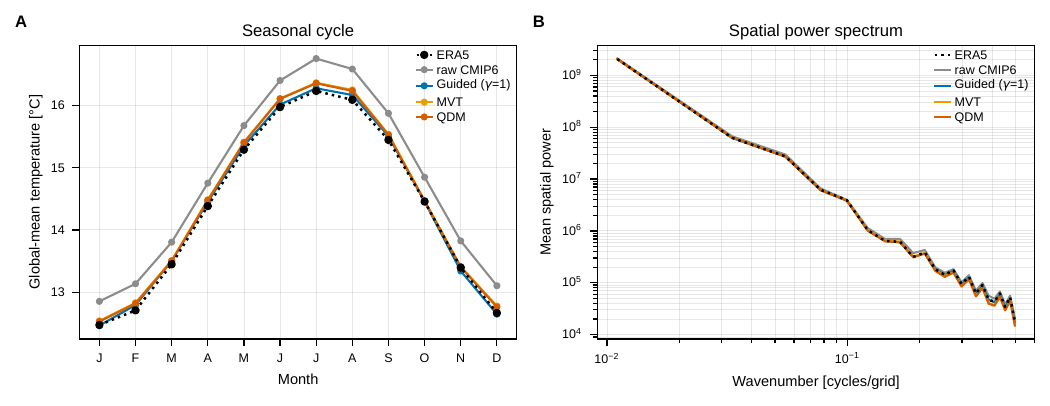}
\caption{\textbf{Performance comparison of condition-to-condition guidance, MVT and QDM (applied to the raw CMIP6 output), evaluated against ERA5 over 2001--2010. Seasonal cycle and spatial power spectrum of the daily temperature fields.} (\emph{A}) Global-mean temperature by calendar month for ERA5 (black, dotted) and for each method (color). (\emph{B}) Radially averaged spatial power spectrum of the daily fields against wavenumber, on logarithmic axes, truncated at $0.5$ cycles per grid interval, the Nyquist wavenumber of each axis.}
\label{fig:si-baseline-structure}
\end{figure}

\begin{figure}[p]
\centering
\includegraphics[width=\textwidth]{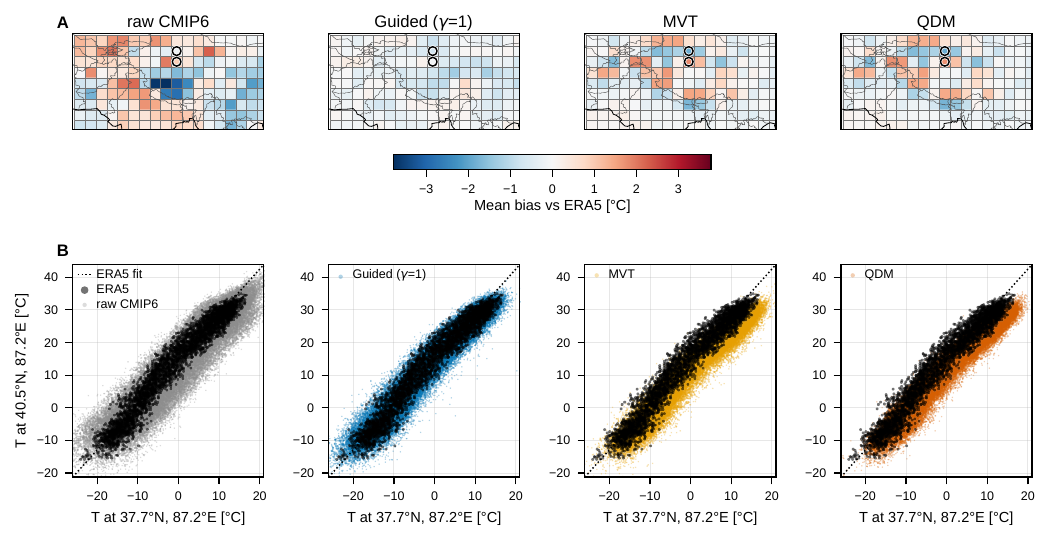}
\caption{\textbf{Performance comparison of condition-to-condition guidance, MVT and QDM (applied to the raw CMIP6 output), evaluated against ERA5 over 2001--2010. Mean bias and the joint distribution of two adjacent grid cells over High Mountain Asia.} (\emph{A}) Mean bias against ERA5 within $20^{\circ}$--$45^{\circ}$N, $60^{\circ}$--$110^{\circ}$E, cell edges drawn; columns are raw CMIP6 output, guided diffusion samples ($\gamma = 1$), MVT and QDM. Circles mark the two north--south adjacent cells, at $37.7^{\circ}$N and $40.5^{\circ}$N and both at $87.2^{\circ}$E. (\emph{B}) Every sampled day and ensemble member at those two cells, plotted as temperature at $40.5^{\circ}$N against temperature at $37.7^{\circ}$N, with those for ERA5 shown as black dots. The dotted line is the principal axis of the ERA5 joint distribution. Every day of the year, seventeen ensemble members.}
\label{fig:si-clouds-hma}
\end{figure}

\begin{figure}[p]
\centering
\includegraphics[width=\textwidth]{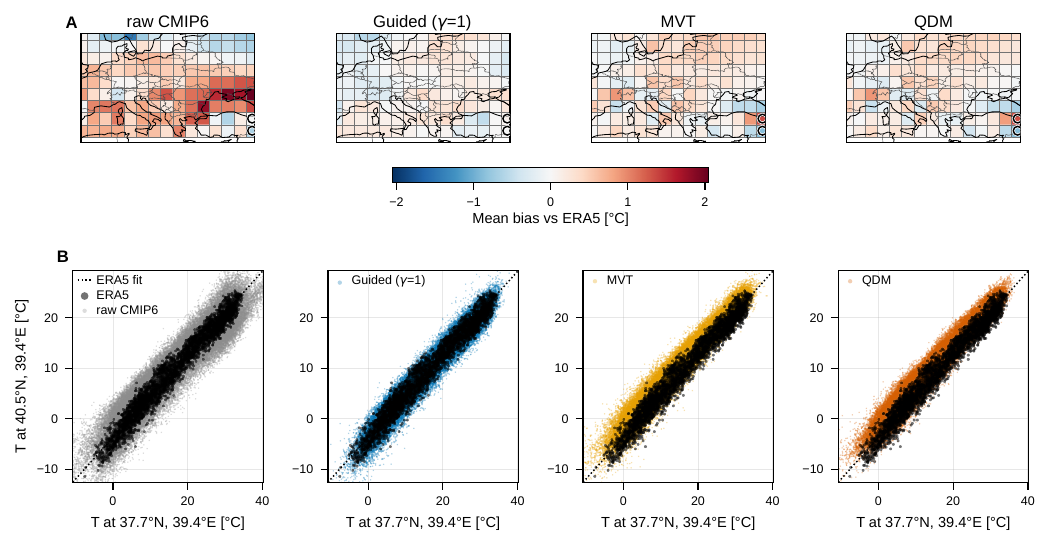}
\caption{\textbf{Performance comparison of condition-to-condition guidance, MVT and QDM (applied to the raw CMIP6 output), evaluated against ERA5 over 2001--2010. Mean bias and the joint distribution of two adjacent grid cells over Europe, 2001--2010.} As Fig.~\ref{fig:si-clouds-hma}, for the box $35^{\circ}$--$60^{\circ}$N, $0^{\circ}$--$40^{\circ}$E and its cell pair: $37.7^{\circ}$N and $40.5^{\circ}$N, both at $39.4^{\circ}$E, whose ERA5 mean temperatures differ by $6.6\,\degC$. The dotted reference line is the principal axis of the ERA5 joint distribution for this pair.}
\label{fig:si-clouds-eur}
\end{figure}

The figures of this section compare guidance at full scale ($\gamma = 1$) with \gls{mvt} and \gls{qdm} applied to the raw output of the seventeen \gls{cmip6} models, all scored against \gls{era5} over 2001--2010. Fig.~\ref{fig:si-baseline-maps} maps nine distributional errors. Fig.~\ref{fig:si-baseline-dist} shows the ensemble calibration and the distribution of the daily global-mean temperature, and Fig.~\ref{fig:si-baseline-structure} the seasonal cycle and the spatial power spectrum. Figs.~\ref{fig:si-clouds-hma} and \ref{fig:si-clouds-eur} show the joint distribution of two adjacent grid cells over High Mountain Asia and over Europe. Tables~\ref{tab:baselines} and \ref{tab:spatial-consistency} give the metrics, and \emph{Comparison with Established Bias Correction against ERA5} discusses them.

\FloatBarrier

\section*{LOOCV for MPI-ESM1-2-LR (Considered as Synthetic Observation)}
\begin{figure}[p]
\centering
\includegraphics[width=8.7cm]{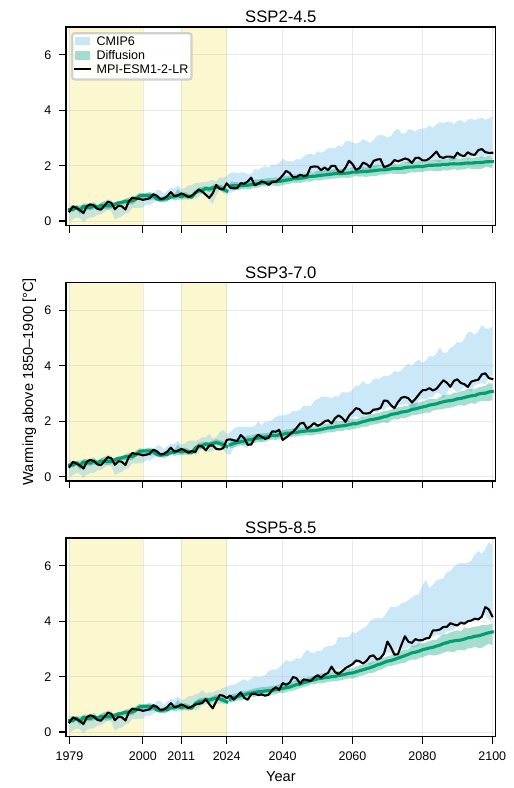}
\caption{\textbf{LOOCV performance of condition-to-condition guidance for the MPI-ESM1-2-LR fold. Global-mean warming of the held-out MPI-ESM1-2-LR and of the guided ensemble, 1979--2100.} Annual global-mean warming above 1850--1900 under SSP2-4.5, SSP3-7.0 and SSP5-8.5 (top to bottom) for MPI-ESM1-2-LR (black), the sixteen remaining CMIP6 models (blue, 5--95\% band) and the diffusion emulator (green, ensemble mean and 5--95\% band). The emulator replicates MPI-ESM1-2-LR ($\gamma = 1$) over 1979--2024 with ten samples per day, and is guided from the sixteen source models with the $\gamma$ schedule over 2025--2100 with ten samples per source model and day. Yellow shading marks the years of MPI-ESM1-2-LR simulation in the emulator's training data. Warming and bands are constructed as in Fig.~\ref{fig:projections} of the main text.}
\label{fig:si-mpi-warming}
\end{figure}

\begin{figure}[p]
\centering
\includegraphics[width=8.7cm]{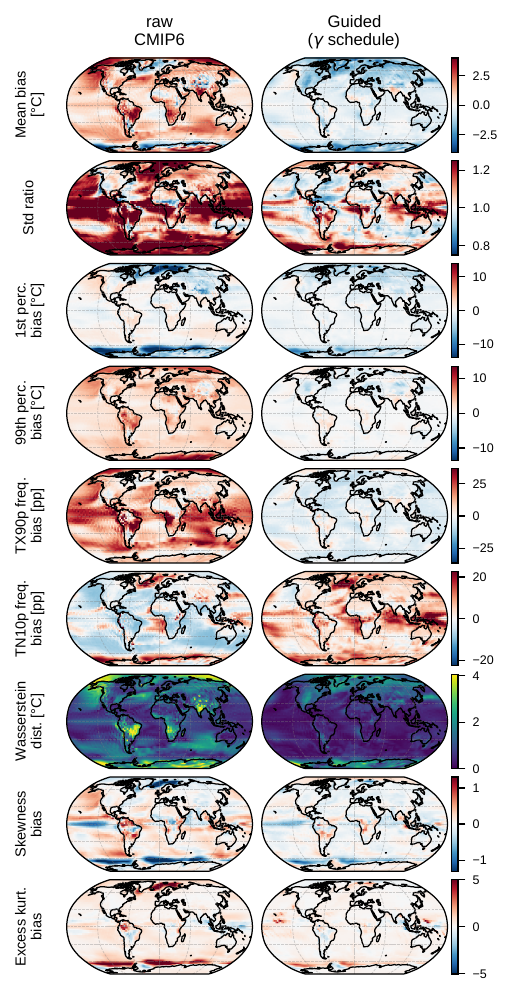}
\caption{\textbf{LOOCV performance of condition-to-condition guidance for the MPI-ESM1-2-LR fold. Nine distributional error diagnostics against the held-out MPI-ESM1-2-LR, SSP3-7.0, 2080--2100.} Columns, left to right: the raw output of the sixteen CMIP6 source models; the emulator guided from the same sixteen models with the $\gamma$ schedule. Rows, top to bottom: mean bias; ratio of modeled to MPI-ESM1-2-LR daily standard deviation; 1st- and 99th-percentile bias; bias in the exceedance frequency of each cell's MPI-ESM1-2-LR 90th-percentile warm-day (TX90p) and 10th-percentile cold-day (TN10p) thresholds, in percentage points; 1-Wasserstein distance between the daily distributions; skewness bias; and excess-kurtosis bias. Native $64\times128$ grid, Robinson projection, every day of 2080--2100, with one sample per source model and day for the emulator. The sixteen members are pooled into a single sample of 122{,}640 days per cell and each diagnostic is computed once on that sample; MPI-ESM1-2-LR contributes its own 7{,}665 days. Color scales are shared along each row. They are centered on zero, except for the standard-deviation ratio, centered on $1$ between $0.75$ and $1.25$, and the Wasserstein distance, which starts at zero.}
\label{fig:si-mpi-maps-370}
\end{figure}

\begin{figure}[p]
\centering
\includegraphics[width=\textwidth]{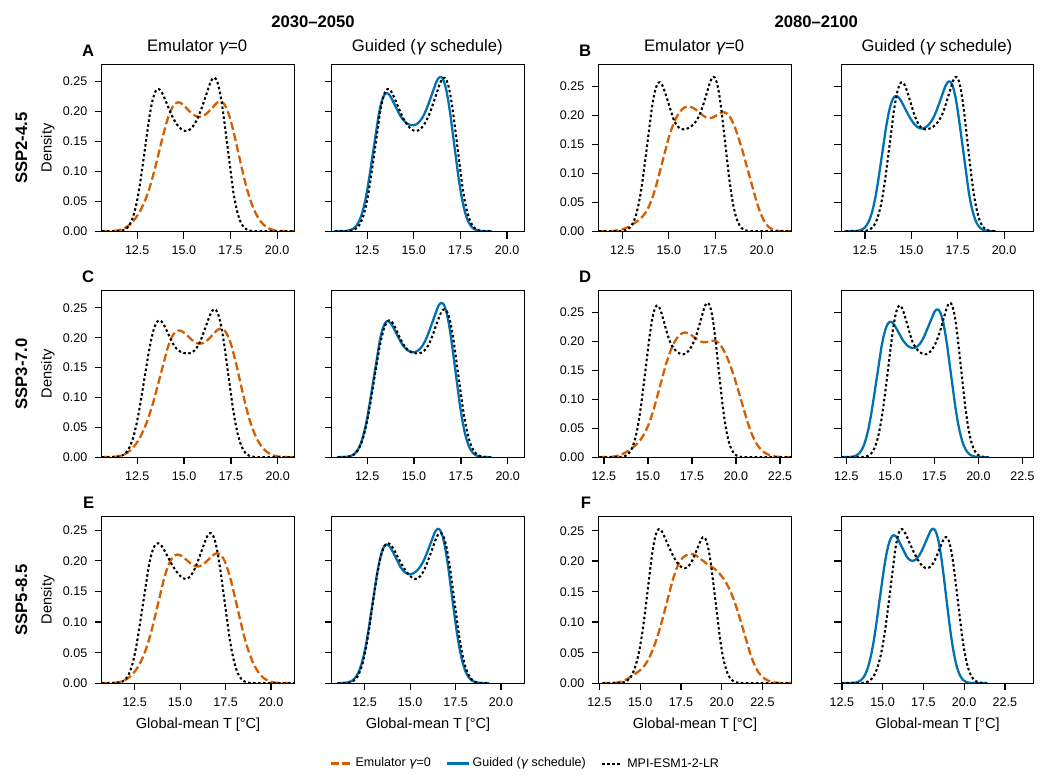}
\caption{\textbf{LOOCV performance of condition-to-condition guidance for the MPI-ESM1-2-LR fold. Distribution of the daily global-mean temperature, held-out MPI-ESM1-2-LR.} Kernel density of the daily area-weighted global-mean temperature under SSP2-4.5 (\emph{A}, \emph{B}), SSP3-7.0 (\emph{C}, \emph{D}) and SSP5-8.5 (\emph{E}, \emph{F}), over 2030--2050 (\emph{A}, \emph{C}, \emph{E}) and 2080--2100 (\emph{B}, \emph{D}, \emph{F}). In each panel, left: the emulator conditioned on the sixteen source models without guidance ($\gamma = 0$); right: the emulator guided from the same models with the $\gamma$ schedule. Both halves use one sample per source model on the same twelve days per year, and MPI-ESM1-2-LR (dotted) is shown on those days (\emph{Sampling Designs}). The emulator curve pools the sixteen source models.}
\label{fig:si-mpi-kde}
\end{figure}

\begin{figure}[p]
\centering
\includegraphics[width=\textwidth]{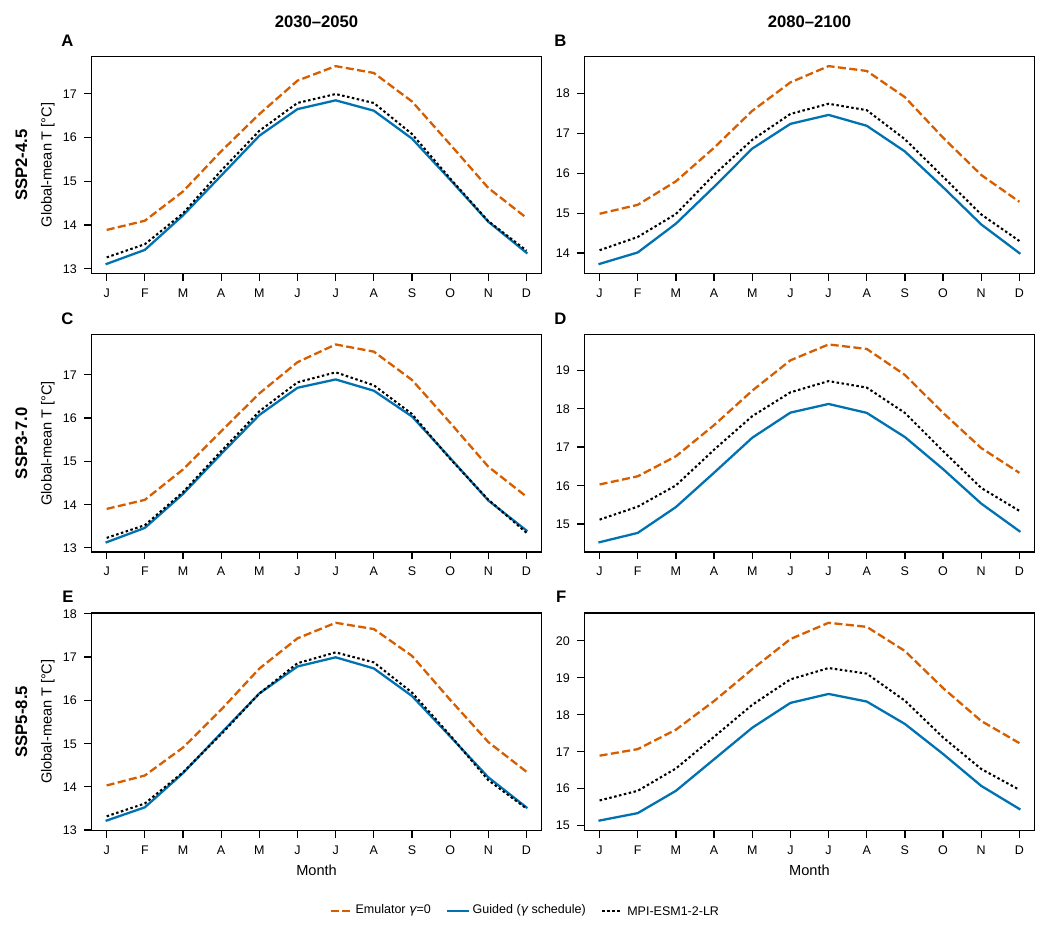}
\caption{\textbf{LOOCV performance of condition-to-condition guidance for the MPI-ESM1-2-LR fold. Seasonal cycle of the global-mean temperature, held-out MPI-ESM1-2-LR.} Global-mean temperature by calendar month under SSP2-4.5 (\emph{A}, \emph{B}), SSP3-7.0 (\emph{C}, \emph{D}) and SSP5-8.5 (\emph{E}, \emph{F}), over 2030--2050 (\emph{A}, \emph{C}, \emph{E}) and 2080--2100 (\emph{B}, \emph{D}, \emph{F}), for MPI-ESM1-2-LR (black, dotted), the emulator without guidance (dashed) and the guided emulator (solid). Both emulators use one sample per source model on the same twelve days per year, and MPI-ESM1-2-LR is shown on those days (\emph{Sampling Designs}). The source models are pooled before the monthly means are computed.}
\label{fig:si-mpi-cycle}
\end{figure}

\begin{figure}[p]
\centering
\includegraphics[width=\textwidth]{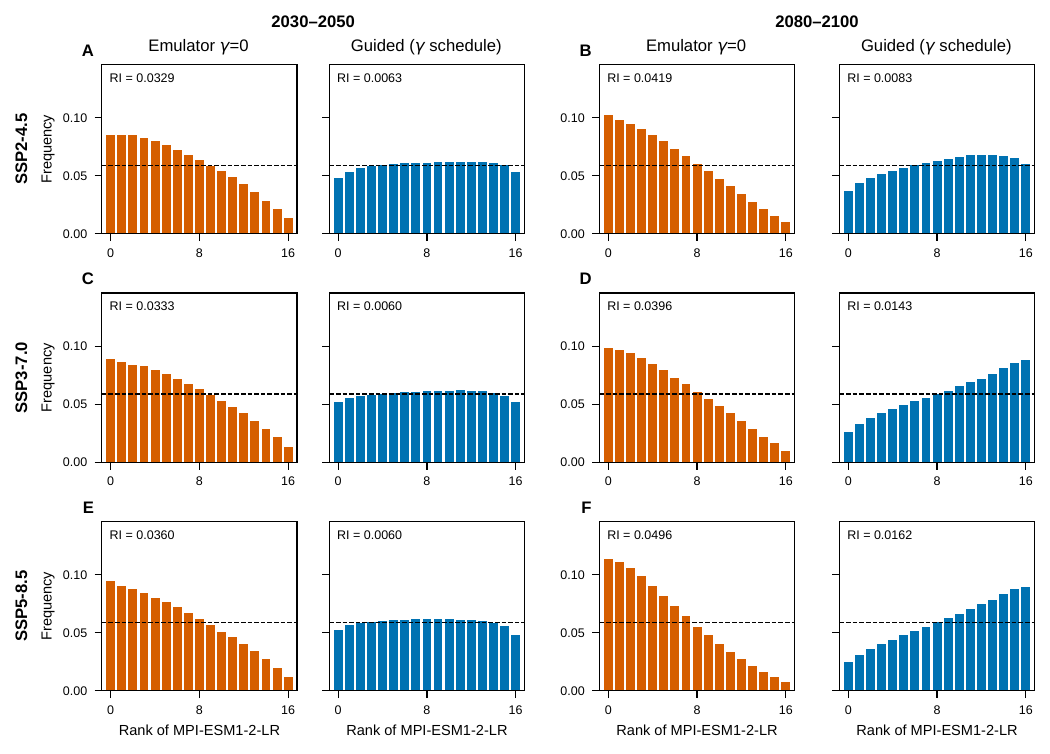}
\caption{\textbf{LOOCV performance of condition-to-condition guidance for the MPI-ESM1-2-LR fold. Ensemble calibration against the held-out MPI-ESM1-2-LR.} Area-weighted rank histogram of MPI-ESM1-2-LR among the sixteen ensemble members at each grid cell and day under SSP2-4.5 (\emph{A}, \emph{B}), SSP3-7.0 (\emph{C}, \emph{D}) and SSP5-8.5 (\emph{E}, \emph{F}), over 2030--2050 (\emph{A}, \emph{C}, \emph{E}) and 2080--2100 (\emph{B}, \emph{D}, \emph{F}). In each panel, left: the emulator conditioned on the sixteen source models without guidance ($\gamma = 0$); right: the emulator guided from the same models with the $\gamma$ schedule. Both halves use one sample per source model on the same twelve days per year (\emph{Sampling Designs}). The dashed line marks the uniform frequency $1/17$ expected of a calibrated sixteen-member ensemble. The RI floor is $0.0037$.}
\label{fig:si-mpi-rank}
\end{figure}

The figures of this section show the perfect-model test in the \gls{loocv} fold in which MPI-ESM1-2-LR is held out. Fig.~\ref{fig:si-mpi-warming} shows the annual global-mean warming over 1979--2100. Fig.~\ref{fig:si-mpi-maps-370} maps nine distributional errors of raw \gls{cmip6} and of the guided ensemble under SSP3-7.0 over 2080--2100. Figs.~\ref{fig:si-mpi-kde}--\ref{fig:si-mpi-rank} compare the emulator without and with guidance over 2030--2050 and 2080--2100, through the distribution and seasonal cycle of the daily global-mean temperature and the rank histograms. Table~\ref{tab:loocv-mpi} gives the metrics, and \emph{Held-Out MPI-ESM1-2-LR} discusses them. The figures for SSP2-4.5 and SSP5-8.5 over 2080--2100, and for the three scenarios over 2030--2050, the mid-term period of Figs.~\ref{fig:si-mpi-kde}--\ref{fig:si-mpi-rank}, are available at \url{https://github.com/mandt-lab/gdso-figures} for completeness.

\FloatBarrier

\section*{Evaluation of the Unguided Emulator}
\begin{figure}[p]
\centering
\includegraphics[width=\textwidth]{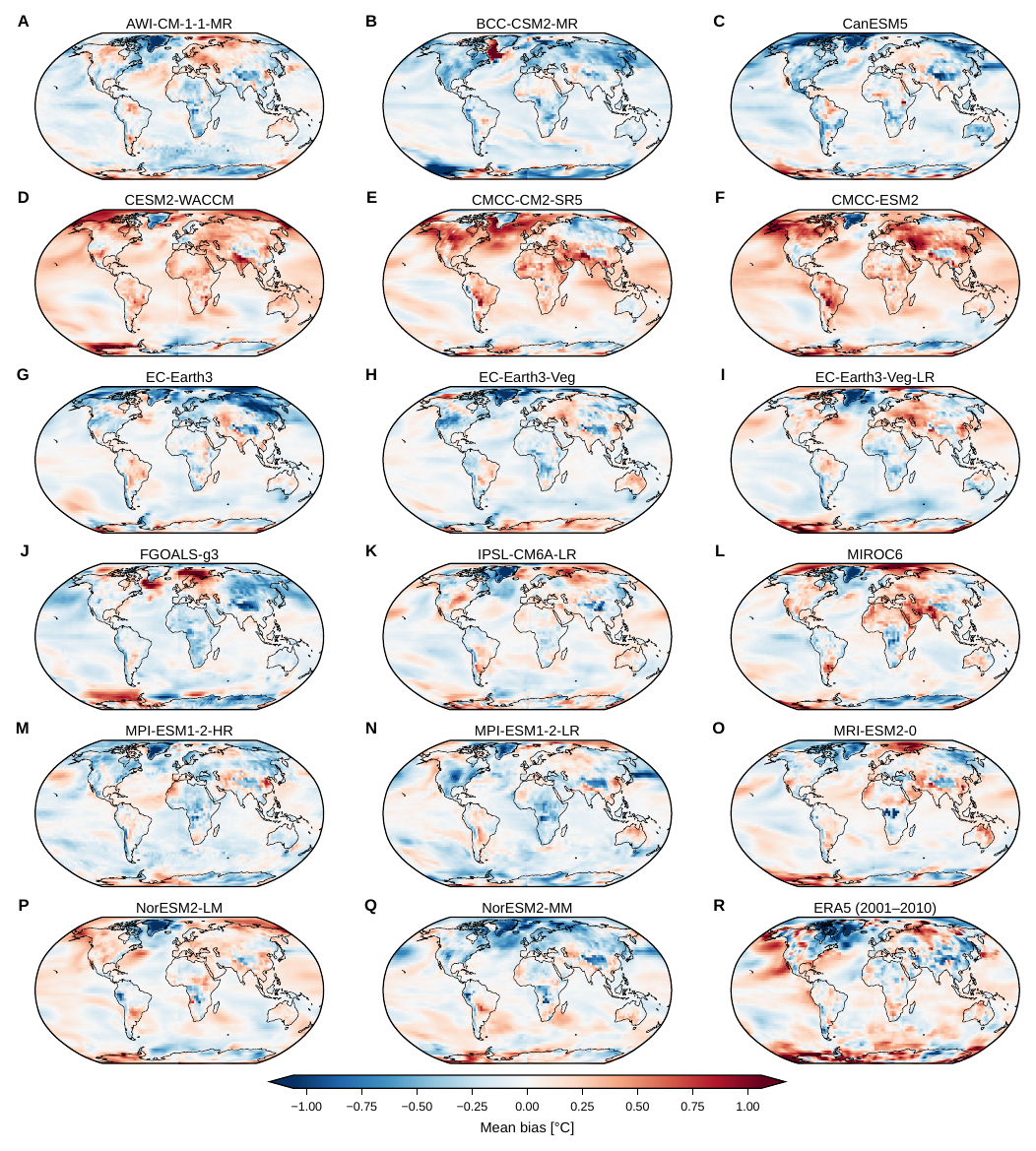}
\caption{\textbf{Performance of the diffusion emulator. Mean bias of the diffusion emulator conditioned on each of eighteen model identities, SSP3-7.0, 2080--2100.} The emulator is trained jointly on seventeen CMIP6 models over 1979--2100 and on ERA5 over 1979--2000 and 2011--2024. In (\emph{A}--\emph{Q}) it is conditioned on one CMIP6 model without guidance, the $\gamma = 0$ endpoint of condition-to-condition guidance, sampled ten times for every day of 2080--2100 (years of CMIP6 record included in the training set) under SSP3-7.0 and compared with that model's own daily temperature on the same days. In (\emph{R}) it is conditioned on ERA5 without guidance, the $\gamma = 1$ endpoint, sampled ten times for every day of 2001--2010 (decade of ERA5 record withheld from training set), and compared with ERA5 daily temperature. Each panel shows, at each grid cell, the emulator's mean daily temperature over the window minus the one of the target.}
\label{fig:si-prior-mean-370}
\end{figure}

\begin{figure}[p]
\centering
\includegraphics[width=\textwidth]{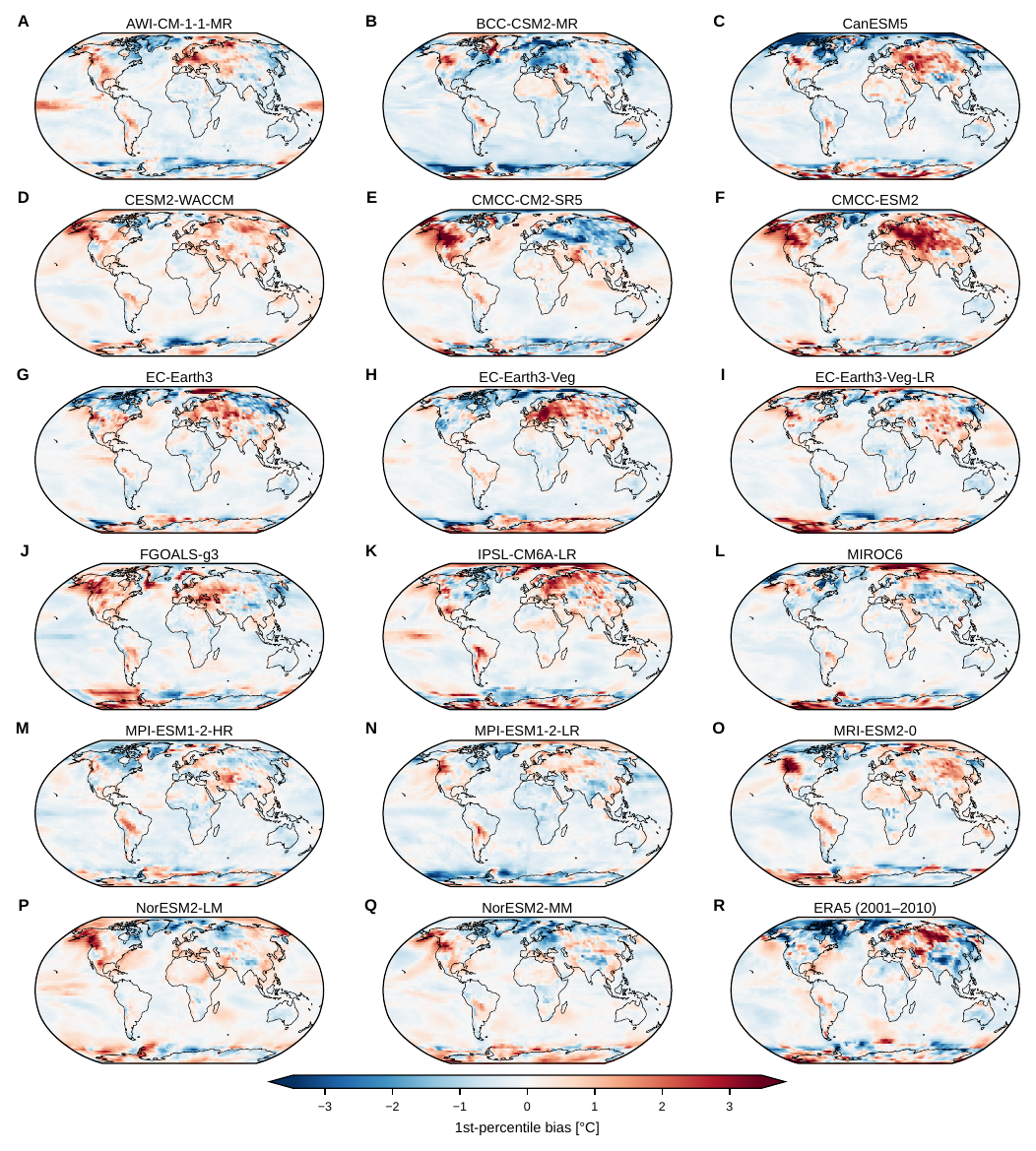}
\caption{\textbf{Performance of the diffusion emulator. 1st-percentile bias of the diffusion emulator conditioned on each of eighteen model identities, SSP3-7.0, 2080--2100.} Setup, panels, sampling and windows as in Fig.~\ref{fig:si-prior-mean-370}. Each panel shows the 1st percentile of the emulator's daily temperature distribution at each grid cell, over the window's days and with the ten samples pooled, minus the same percentile of the target's.}
\label{fig:si-prior-q01-370}
\end{figure}

\begin{figure}[p]
\centering
\includegraphics[width=\textwidth]{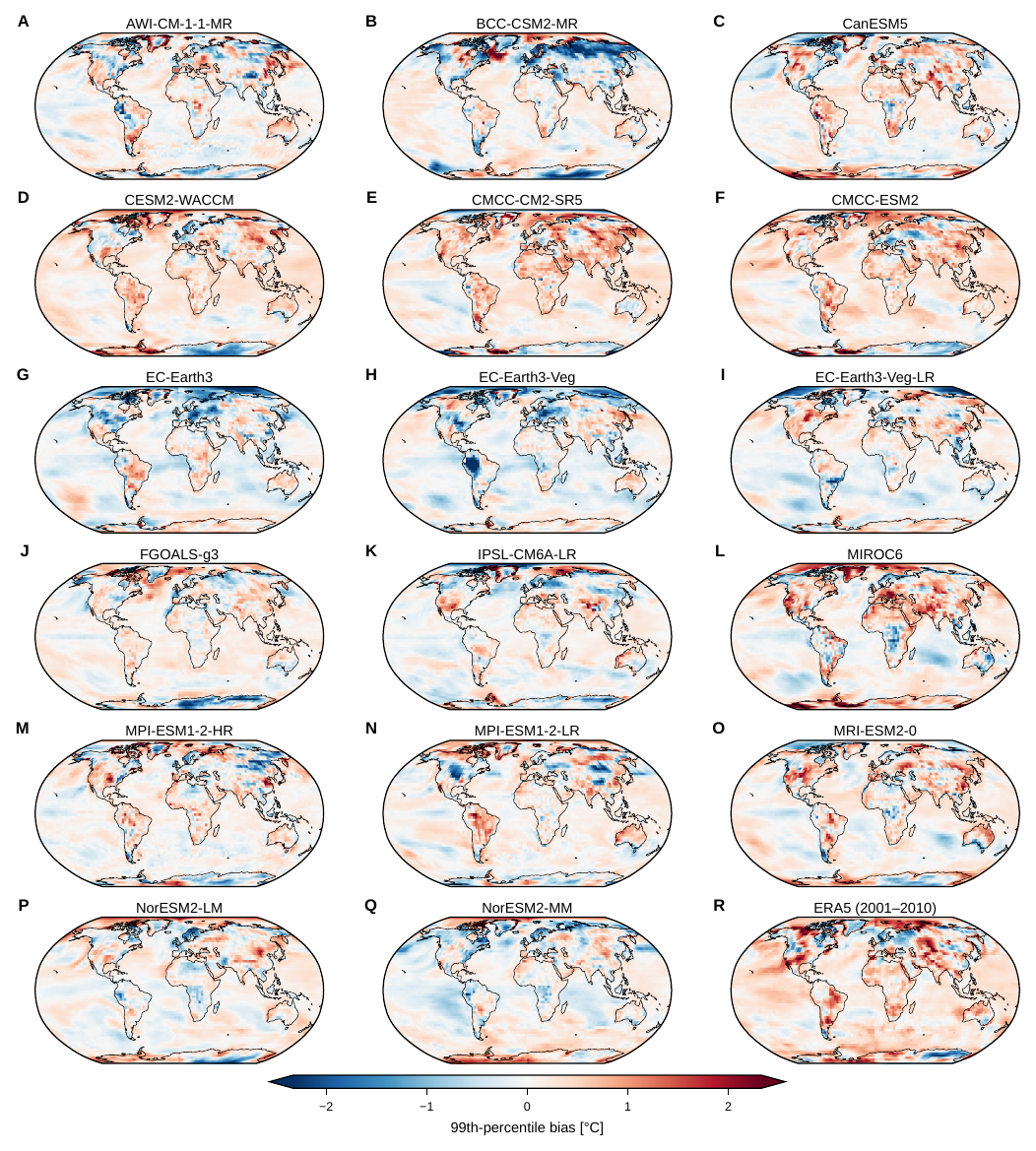}
\caption{\textbf{Performance of the diffusion emulator. 99th-percentile bias of the diffusion emulator conditioned on each of eighteen model identities, SSP3-7.0, 2080--2100.} As Fig.~\ref{fig:si-prior-q01-370}, for the 99th percentile.}
\label{fig:si-prior-q99-370}
\end{figure}

\begin{figure}[p]
\centering
\includegraphics[width=\textwidth]{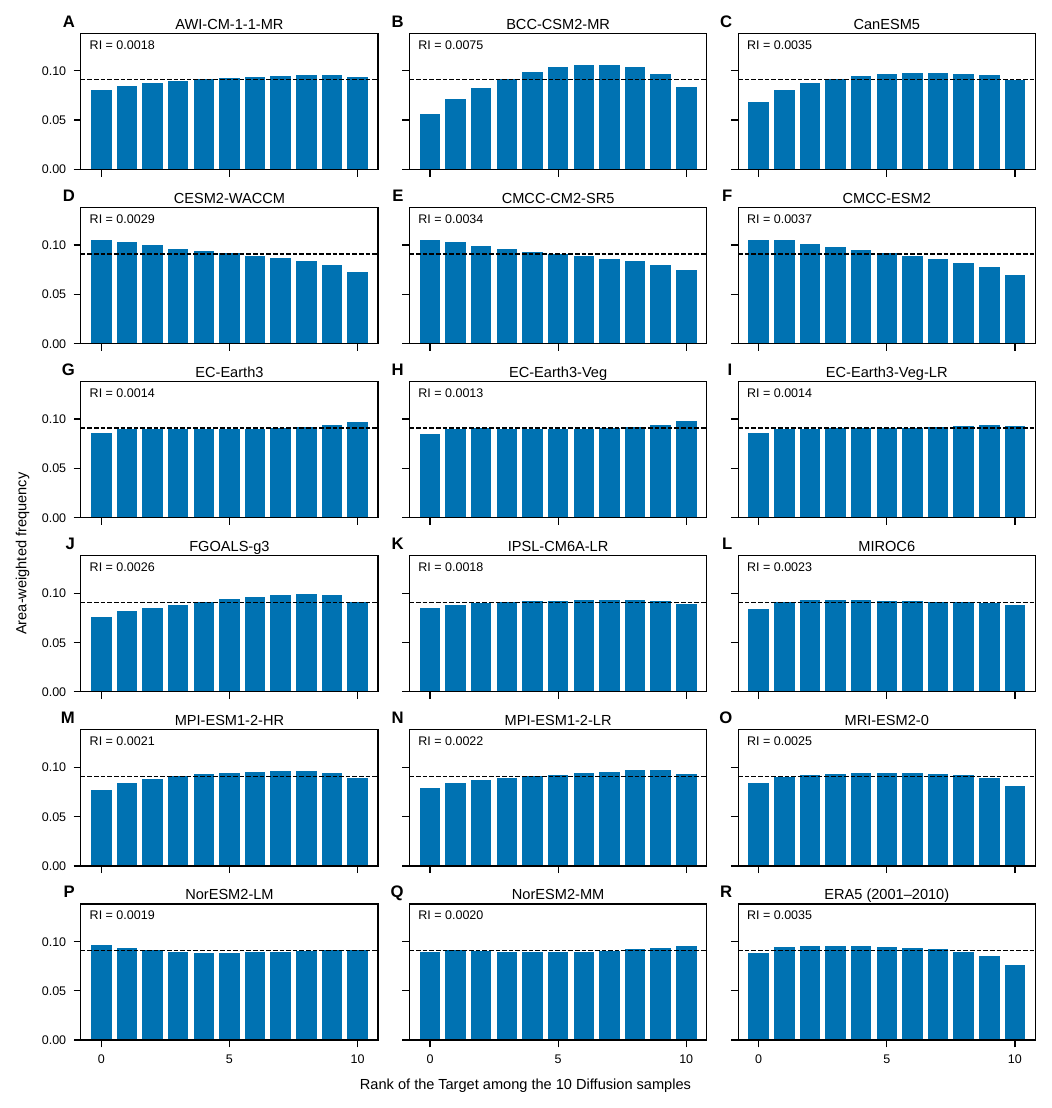}
\caption{\textbf{Performance of the diffusion emulator. Rank histograms of the diffusion emulator conditioned on each of eighteen model identities, SSP3-7.0, 2080--2100.} Setup, panels, sampling and windows as in Fig.~\ref{fig:si-prior-mean-370}. Each panel shows the area-weighted frequency of the rank of the target's daily temperature among the ten emulator samples of the same day and grid cell. The dashed line marks the uniform frequency $1/11$ expected of a calibrated ten-member ensemble.}
\label{fig:si-prior-rank-370}
\end{figure}

\begin{figure}[p]
\centering
\includegraphics[width=\textwidth]{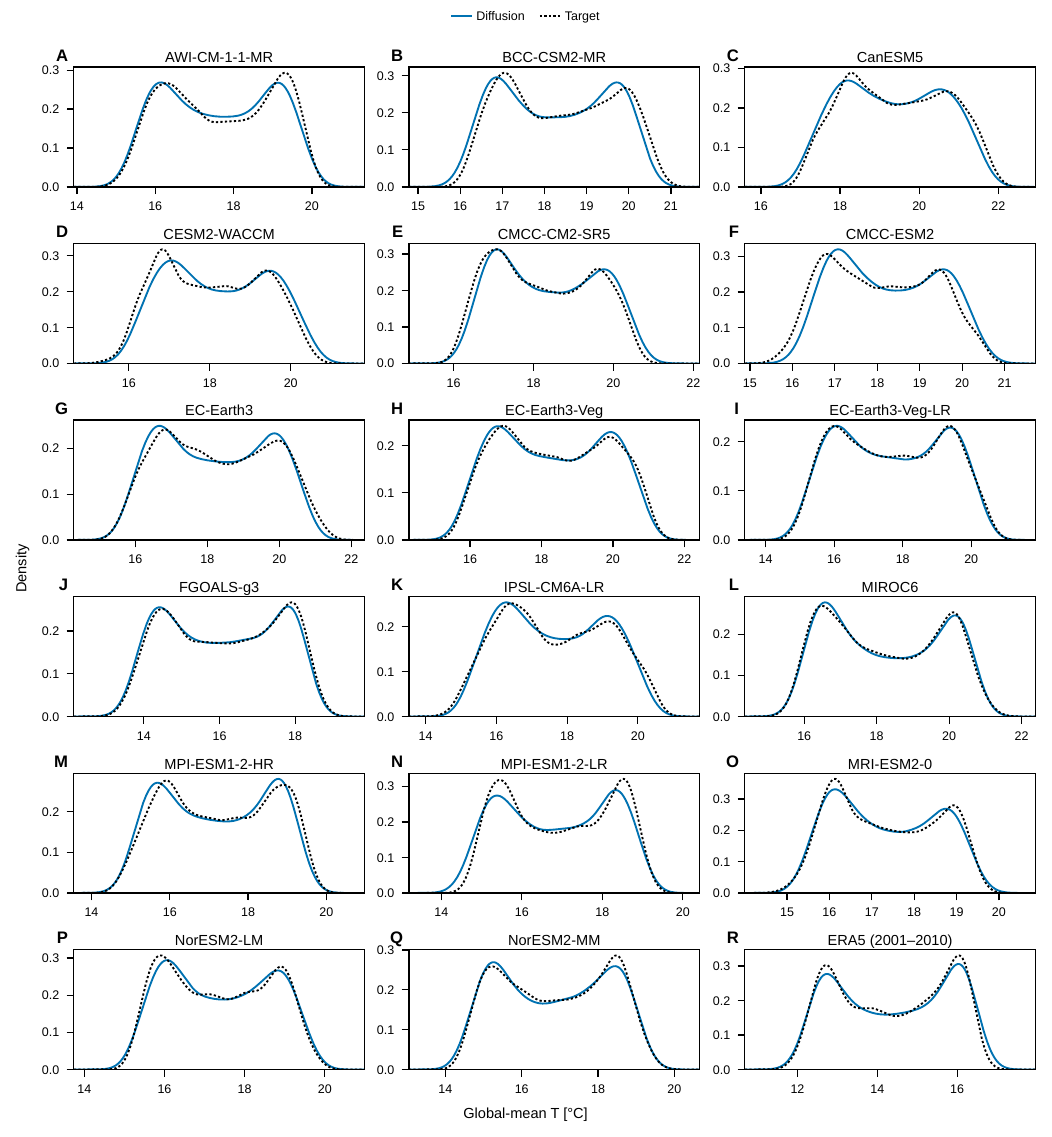}
\caption{\textbf{Performance of the diffusion emulator. Distribution of the daily global-mean temperature of the diffusion emulator conditioned on each of eighteen model identities, SSP3-7.0, 2080--2100.} Setup, panels, sampling and windows as in Fig.~\ref{fig:si-prior-mean-370}. Kernel density of the area-weighted daily global-mean temperature for the emulator (blue), with the ten samples pooled, and for the target (black, dotted).}
\label{fig:si-prior-kde-370}
\end{figure}

\begin{figure}[p]
\centering
\includegraphics[width=\textwidth]{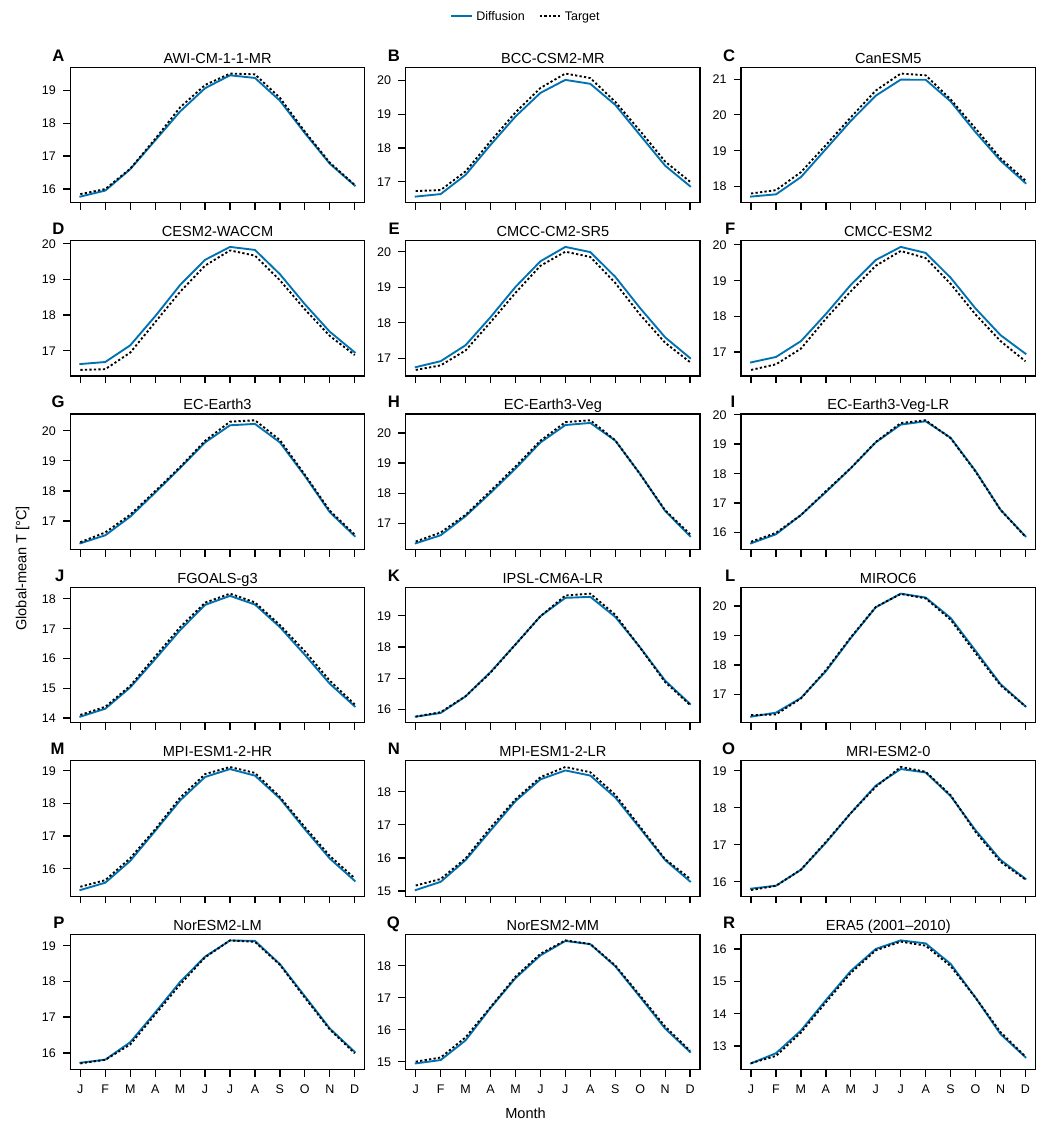}
\caption{\textbf{Performance of the diffusion emulator. Seasonal cycle of the global-mean temperature of the diffusion emulator conditioned on each of eighteen model identities, SSP3-7.0, 2080--2100.} Setup, panels, sampling and windows as in Fig.~\ref{fig:si-prior-mean-370}. Area-weighted global-mean temperature by calendar month, averaged over the years of the window, for the emulator (blue), with the ten samples pooled, and for the target (black, dotted).}
\label{fig:si-prior-cycle-370}
\end{figure}

The figures of this section evaluate the emulator without guidance, conditioned on each of the seventeen \gls{cmip6} models under SSP3-7.0 over 2080--2100 ($\gamma = 0$) and on \gls{era5} over 2001--2010 ($\gamma = 1$). Figs.~\ref{fig:si-prior-mean-370}--\ref{fig:si-prior-q99-370} map the mean, 1st-percentile and 99th-percentile biases, and Fig.~\ref{fig:si-prior-rank-370} shows the rank histograms. Figs.~\ref{fig:si-prior-kde-370} and \ref{fig:si-prior-cycle-370} show the distribution and the seasonal cycle of the global-mean temperature. \emph{Fidelity of the Unguided Emulator} discusses them. The same figures for SSP2-4.5 and SSP5-8.5 over 2080--2100, and for the three scenarios over 2030--2050, a mid-term period that is also part of the training record, are available at \url{https://github.com/mandt-lab/gdso-figures} for completeness; panel \emph{R} (\gls{era5}, 2001--2010) shows the same values in all of them, since it depends on neither the scenario nor the projection period.

\FloatBarrier

\end{document}